\documentclass[apj,onecolumn]{openjournal}

\usepackage[utf8]{inputenc}

\usepackage{graphicx}   
\usepackage{amsmath}    
\usepackage{natbib}     
\usepackage{hyperref}   
\usepackage{nameref}
\usepackage{float}      

\floatstyle{plain}
\restylefloat{table}

\makeatletter
\renewcommand{\fnum@table}{Table~\thetable}
\makeatother

\hypersetup{colorlinks=true,linkcolor=blue,citecolor=blue,filecolor=blue,urlcolor=blue}
\usepackage{cleveref}
\usepackage{natbib}
\usepackage{chngcntr}
\usepackage{xcolor}

\usepackage[T1]{fontenc}
\usepackage{ae,aecompl}
\usepackage{newtxtext,newtxmath}
\usepackage{comment}

\usepackage{bbding}
\newcommand{\angstrom}{\textup{\AA}}

\begin{document}

\title[Simulating realistic Lyman-$\alpha$ emitting galaxies including the effect of radiative transfer]{Simulating realistic Lyman-$\alpha$ emitters including the effect of radiative transfer}

\author{Hasti Khoraminezhad$^{1}$, Shun Saito$^{1,2}$, Max Gronke$^{3,4}$, Chris Byrohl$^{2,5}$}
\affil{
$^{1}$ Institute for Multi-messenger Astrophysics and Cosmology, Department of Physics, Missouri University of Science and Technology,\\ 1315 N Pine Street, Rolla, MO 65409, USA\\
$^2$Kavli Institute for the Physics and Mathematics of the Universe (Kavli IPMU, WPI), University of Tokyo, Chiba 277-8582, Japan\\
$^3$ Astronomisches Rechen-Institut, Zentrum für Astronomie, Universität Heidelberg, Mönchhofstraße 12-14, 69120 Heidelberg, Germany\\
$^4$Max-Planck-Institut f\"{u}r Astrophysik, Karl-Schwarzschild-Str. 1, 85741 Garching, Germany\\
$^5$Institut für Theoretische Astrophysik, Universität Heidelberg, ZAH, Albert-Ueberle-Str. 2, 69120 Heidelberg, Germany
}

\begin{abstract}

We present an empirical yet physically motivated simulation of realistic Lyman-$\alpha$ emitters (LAEs) at $z\sim2\!-\!3$, crucial for ongoing and forthcoming cosmological LAE surveys. 
We combine an empirical \texttt{UniverseMachine} galaxy-halo model with a simple spherical expanding shell model for the Lyman-$\alpha$ radiative transfer, calibrating only three free parameters to simultaneously reproduce the observed Lyman-$\alpha$ luminosity function and the angular clustering. 
Our LAE model is further supported by its consistency with other observables such as the Lyman-$\alpha$ equivalent width distribution, the Lyman-$\alpha$ escape fraction as a function of stellar mass and dust reddening, and the systemic velocity offsets. Our LAE model provides predictions for the halo occupation distributions for LAEs and relationship between Ly$\alpha$ luminosity and halo mass, including the distribution of satellite LAEs. Our work provides a crucial first step towards creating a high-fidelity LAE synthetic catalog for the LAE cosmology surveys.
Our LAE catalog and spectra are publicly available at \url{https://scholarsmine.mst.edu/research_data/14}.\\ 
\end{abstract}


\section{Introduction}
\label{sec:intro}
Realistic high-fidelity mock galaxy catalogs play a crucial role in precision cosmology from large-scale structure surveys \citep[e.g.,][]{Kitaura:2016MN,Saito:2016MN,Rossi:2021MN,Prada:2023ar,Euclid:2025AA}.
In particular, there is currently a strong demand for a high-fidelity catalog for Lyman-$\alpha$ (Ly$\alpha$) emitting galaxies (LAEs).
LAEs offer powerful probes of the high-redshift Universe \citep[e.g.,][]{1967ApJ...147..868P,Chiang:2013JCAP,Ebina:2024JCAP} and thus LAEs at $z\sim 2\!-\!3$ are targeted by various optical and near-infrared galaxy surveys; ongoing ones such as the Hobby-Eberly Telescope Dark Energy Experiment (HETDEX, \citealt{2021ApJ...923..217G, 2021AJ....162..298H}) and the Javalambre Physics of the Accelerating Universe Astrophysical Survey (J-PAS, \citealt{Benitez:2014ar}) as well as forthcoming ones such as the second stage of the Dark Energy Spectroscopic Instrument (DESI-II, \citealt{Schlegel:2022ar}), the MUltiplexed Survey Telescope (MUST, \citealt{Zhao:2024ar}), the Wide-field Spectroscopic Telescope (WST, \citealt{Mainieri:2024ar}), and the proposed Spectroscopic Stage-5 experiment (Spec-S5, \citealt{2025arXiv250307923B}). 
A realistic LAE mock catalog is crucial to optimize their survey strategies, calibrate LAE selection functions, and to interpret cosmological analyses, since any systematic uncertainties could bias the inferred cosmological parameters.\\ 

A realistic mock LAE catalog shall reproduce two essential observables: (i) the observed LAE luminosity function and (ii) the LAE clustering on large scales. 
The clustering strength of galaxies on large scales, $\gtrsim \mathcal{O}(10\,{\rm Mpc})$, is the main target of the aforementioned surveys as a source of cosmological information such as the Baryon Acoustic Oscillations (BAO) and the Redshift Space Distortion (RSD). 
In addition, the clustering at intermediate scales, $\mathcal{O}(0.1$-$10\,{\rm Mpc})$, carries the important information on how the galaxies are connected with underlying dark matter halos and thus how galaxies have evolved in the context of the large-scale structure \citep[see e.g.,][for a review]{Wechsler:2018AR}. 
The LAE luminosity function controls the Poisson shot noise of the clustering measurements at a given survey flux threshold. 
Beyond these two primary requirements, it is beneficial for the catalog to also reproduce additional observed properties such as the distribution of Ly$\alpha$ equivalent width (EW) that characterizes the strength of the Ly$\alpha$ emission line with respect to the continuum level. The EW distribution and additional observational constraints \citep[see e.g.,][for a review]{Ouchi:2020AR} further improve the fidelity of the observed sample selection.\\

Cosmological galaxy formation simulations are an important tool in the creation of mock catalogs.
Yet, their limited dynamical spatial range and modeling uncertainties down to individual HII regions, sourcing most of an LAEs' photons, impede a straightforward use to create LAE mocks. In addition, Ly$\alpha$ photons are resonantly scattered by neutral hydrogen atoms on any scale that spans the interstellar medium (ISM), the circumgalactic medium (CGM), and the intergalactic medium (IGM). In dense HI regions, each Ly$\alpha$ photon could undergo an extremely large number of scatterings, often up to millions, on the ISM and CGM scales \citep{Laursen_2009, Dijkstra:2017ar} and be absorbed by dust grains \citep{2008A&A...488..491A, 2014A&A...562A..52D}.  
Resonant scattering also changes the spectral line profile of Ly$\alpha$: photons diffusing in frequency to the line wings can escape more easily. 
The Ly$\alpha$ line is typically redshifted and broadened compared to the intrinsic emission \citep{Verhamme2006, Yang_2016}. 
Each scattering imparts a frequency shift, and photons that eventually escape have often shifted to the red side of the line or even emerged as double-peaked profiles bracketing the systemic wavelength \citep[see e.g.,][]{Steidel:2010ApJ,Leclercq:2017AA,Runnholm:2021PASP}. 
The impact on Ly$\alpha$ radiative transfer on the spectral profile was theoretically investigated by analytic methods \citep[e.g.,][]{Harrington:1973mn,Neufeld:1990ApJ,Nebrin:2025MN,Smith:2025MN} and simulation methods \citep[e.g.,][]{Zheng_2002,Dijkstra_2006,Tasitsiomi_2006, Verhamme2006,Semelin:2007AA,Zheng:2010ApJ,Yajima:2012MN,Gronke_2015,Gronke:2016ApJ,Song:2020ApJ,Smith:2022mn,Xu:2023mn,AlmadaMonter2024MN,Chang:2024MN}. 
In addition, even in an optically thin regime on IGM scales, the redshifted Ly$\alpha$ photon on the blue side of the spectrum could be attenuated by HI in the IGM \citep[e.g.,][]{Laursen:2011ApJ,Byrohl:2020AA}. 
All of these processes complicate the interpretation of LAE observations. This means that the presence or absence of Ly$\alpha$ emission is not a simple indicator of the intrinsic properties of the galaxy but rather a convoluted result of Ly$\alpha$ radiative transfer (RT) through density and kinematics of HI gas and dust at a wide range of scales, $\mathcal{O}(10\,{\rm pc}$-$10\,{\rm Mpc})$ (see \cite{Dijkstra:2017ar} for a comprehensive review of Ly$\alpha$ RT). Despite extensive surveys and large samples, connecting the observed Ly$\alpha$ luminosities and spectra to the underlying star formation activity and dark-matter halos remains non-trivial. 
For example, it is unclear whether the LAE-halo connection can be modeled by a simple framework that works for a mass-selected galaxy sample such as the Halo Occupation Distribution (HOD) approach \citep[e.g.,][]{Hamana:2004mn,Ouchi:2018PA,Hong:2019mn,Herrero-Alons2021AA}.\\ 

One may wonder if these complications are relevant only for small-scale galaxy formation physics. 
However, some previous work pointed out that complications in Ly$\alpha$ RT could have an impact on the cosmological interpretation of large-scale LAE clustering. 
The IGM attenuation on the blue side of the Ly$\alpha$ spectrum could potentially correlate with the large-scale density field and introduces the anisotropic selection effect, leading to suppression of clustering along the line of sight \citep{Zheng:2011ApJ,Behrens:2018AA,Gurung-Lopez:2019mn_b}. 
Furthermore, the peak shift and smearing of the Ly$\alpha$ emission could be regarded as a velocity offset that leads to an effect similar to the Finger-of-God suppression \citep{Byrohl:2019mn,Gurung-Lopez:2021mn,LujanNiemeyer2025ApJ}. 
Thus, we argue that \textit{one needs to model the Ly$\alpha$ spectra} for a realistic LAE mock catalog, since Ly$\alpha$ RT shapes the Ly$\alpha$ line profile and the escape fraction of photons, which ultimately impact the selection of LAEs and their inferred large-scale distribution \citep[see e.g.,][for a recent comparison of LAE spectra between HETDEX and DESI]{Landriau:2025ar}.\\ 

It is not computationally feasible to fully solve the physics of Ly$\alpha$ RT in a hydrodynamical simulation at all relevant scales in a self-consistent manner and to scale it up to the cosmological volume required for the LAE surveys. 
There are recent promising progress along this line, but they still require calibration of their subgrid physics such as dust attenuation, and it is unclear how well they can reproduce the radiative transfer processes on the sub-parsec scale \citep{Smith:2022mn,Xu:2023mn,Byrohl:2023mn}. 
Alternatively, one could empirically model the Ly$\alpha$ escape fraction for LAEs without incorporating Ly$\alpha$ RT physics, but adopt a physical model for all galaxy populations from a hydrodynamical simulation \citep{Nagamine:2010PA,Shimizu:2010mn,Shimizu:2011mn,Im:2024ApJ,Ravi:2024Ph,Sullivan:2025ar} or from a semi-analytical model \citep{LeDelliou:2006mn,Kobayashi:2008as}. 
Notably, there are only a few exceptions in the literature that model LAEs at $z=2\!-\!3$ by incorporating Ly$\alpha$ RT \citep{Garel:2015mn,Gurung-Lopez:2019mn_a,Gurung-Lopez:2019mn_b,Byrohl:2023mn} \citep[see also e.g.,][for reionization studies at higher redshifts]{Hutter:2015mn,Inoue:2018PA,Gronke:2021mn}.
We will discuss the differences in these previous works in more detail in Sec.~\ref{subsec:comparison}.\\

In this work, we advance the modeling of LAEs by developing a framework to construct a high-fidelity mock LAE catalog that satisfies the following conditions: 
\begin{enumerate}
    \item it is applicable to a dark-matter only $N$-body simulation so that it can be readily scaled to a Gpc-scale simulation in the future, 
    \item it includes a model for global galaxy properties such as the stellar mass and star formation rate (SFR), which allows us to model the intrinsic Ly$\alpha$ luminosity and predict the relation between LAEs and other population of galaxies, 
    \item it incorporates a Ly$\alpha$ RT model that allows us to predict the Ly$\alpha$ spectra, and 
    \item the model is calibrated such that it simultaneously reproduces the observed luminosity function and clustering of LAEs as well as other LAE diagnostics, when possible, such as the EW distribution, systemic velocity offset, and the dust reddening strength.
\end{enumerate}

Our approach is built on an empirical galaxy formation model, \texttt{UniverseMachine} \citep{Behroozi:2019mn}, which is calibrated to a broad range of galaxy observables and thus provides a realistic (albeit imperfect) population of galaxies, satisfying conditions 1) and 2).
To meet condition 3), we then link Ly$\alpha$ emission to these galaxies via the direct results of the Monte Carlo Ly$\alpha$ RT code with a spherical expanding shell model on the ISM/CGM scales \citep{Gurung-Lopez:2021zELDA} and the mean IGM transmission modified from \cite{Laursen:2011ApJ}. 
As we shall show, our calibrated model successfully reproduces various LAE observables (condition 4)).\\ 

This paper is organized as follows.
In Sec.~\ref{sec:initial-phase} we describe $N$-body simulation data and observational data.
Sec.~\ref{sec:lae_model} details our approach to modeling LAEs. 
In Sec.~\ref{sec:result}, we present our main fitting results as well as implications of our simulated LAE mock catalogs.
We clarify the difference between our work and others and the limitations of our approach in Sec.~\ref{sec:discussion} before we conclude this work in Sec.~\ref{sec:conclusion}.
Supplementary information on model validation and further technical considerations is provided in the appendices.
Throughout this paper, we denote the comoving and physical distance units as cMpc/$h$ and pMpc/$h$ etc., respectively. 
The rest frame wavelength is denoted without any superscripts, while the observed one is clarified as $\lambda^{\rm obs}$.
For example, the rest frame Ly$\alpha$ wavelength is denoted by $\lambda_{\rm Ly\alpha}=1215.67\,\angstrom$.
We define galaxies with Ly$\alpha$ luminosity of $\mathrm{log}_{10}(L_{\rm Ly\alpha}/\,\rm erg\,s^{-1}) \geq 41.7$ and equivalent width of $\mathrm{EW}_0 > 40 \, \text{\AA}$ as our fiducial ``LAEs'' when we compare our model with observational data unless it has different specific criteria, and call galaxies that do not satisfy the criteria as ``others''.

\section{Simulation and observational dataset}
\label{sec:initial-phase} 

In this section, we describe the external datasets including both cosmological simulations and observations. 
Note that we only include the observational measurements adopted in our fitting procedure here, while we will compare predictions from our LAE mock with other measurements in the literature in later sections. 

\subsection{\texorpdfstring{$N$-body simulation}{N-body simulation}}
\label{sec:nbody}

The goal of this work is to model LAEs on top of a dark-matter only $N$-body simulation such that it simultaneously reproduces both the LAE luminosity function and the spatial clustering at $\mathcal{O}(0.1\text{-}10\,{\rm cMpc}/h)$. 
This requires a simulation with as large a box size and as high a mass resolution as possible.
Taking into account that LAEs are typically hosted by dark-matter halos with a mass of $\gtrsim 10^{11}\,\rm{M}_{\odot}$ at $2\lesssim z\lesssim 3$ \citep[see e.g.,][]{Ouchi:2020AR}, we adopt the \texttt{Small MultiDark Planck (SMDPL)} simulation \citep{10.1093/mnras/stw248} in this work.
We justify our choice by comparing the mass resolution in Appendix \ref{sec:app-nbody}.
 \texttt{SMDPL} has $3840^3$ dark matter particles with particle mass of $m_{\rm p} = 9.63 \times 10^{7}\, {\rm M_{\odot}}/h$ in a periodic box with a comoving length of $400\, {\rm cMpc}/h$. 
The simulation is run using the \texttt{L-Gadget-2} code, a performance-optimized version of the publicly available cosmological code  \texttt{Gadget-2} \citep{10.1111/j.1365-2966.2005.09655.x} with a force resolution of $1.5\, {\rm ckpc}/h$. The cosmological parameters for these simulations are $h=0.6777$, $\Omega_{\Lambda}=0.692885$, $\Omega_{m}=0.307115$, $\Omega_{b}=0.048206$, $n=0.96$ and $\sigma_{8}=0.8228$, closely aligned with \citet{refId0}.
Dark matter halos and subhalos are identified using the \texttt{Rockstar} halo finder \citep{Behroozi_2013} and halo merger trees are constructed using the \texttt{Consistent Trees} algorithm \citep{Behroozi_2013_2}.

\subsection{The UniverseMachine galaxy-halo model}
\label{sec:halos}

Since LAEs are observationally thought to be young star-forming galaxies \citep{Dijkstra:2017ar, Kusakabe:2018PASJ, Ouchi:2020AR, 2022ApJ...936..131M, 2025arXiv250108568F} and intrinsic Lyman-$\alpha$ luminosity is primarily correlated with the star formation rate (SFR) \citep{Kennicutt:1998ARAA, 2009ApJ...704..724T, 2010MNRAS.401.2343D, 2019A&A...623A.157S, Runnholm_2020}, we need a prescription to predict a SFR for each individual galaxy or dark matter halo in the simulation. 
To this end, we adopt an empirical framework \texttt{UniverseMachine} \citep{Behroozi:2019mn} to connect the properties of galaxies and dark matter halos. 
In this framework, the assembly history of the stellar mass is parameterized as a function of the assembly history of the halo mass on a given dark-matter-only $N$-body simulation.
Then, the free parameters are determined so that they reproduce observational data across $0<z<10$ including stellar mass functions, cosmic and specific SFRs, Ultraviolet (UV) luminosity functions, quenched fractions, correlation functions for all, quenched, and star-forming galaxies, measurements of the environmental dependence of central galaxy quenching, and UV-stellar mass relations.\\ 

\begin{figure*}
\centering
\includegraphics[scale=0.39]{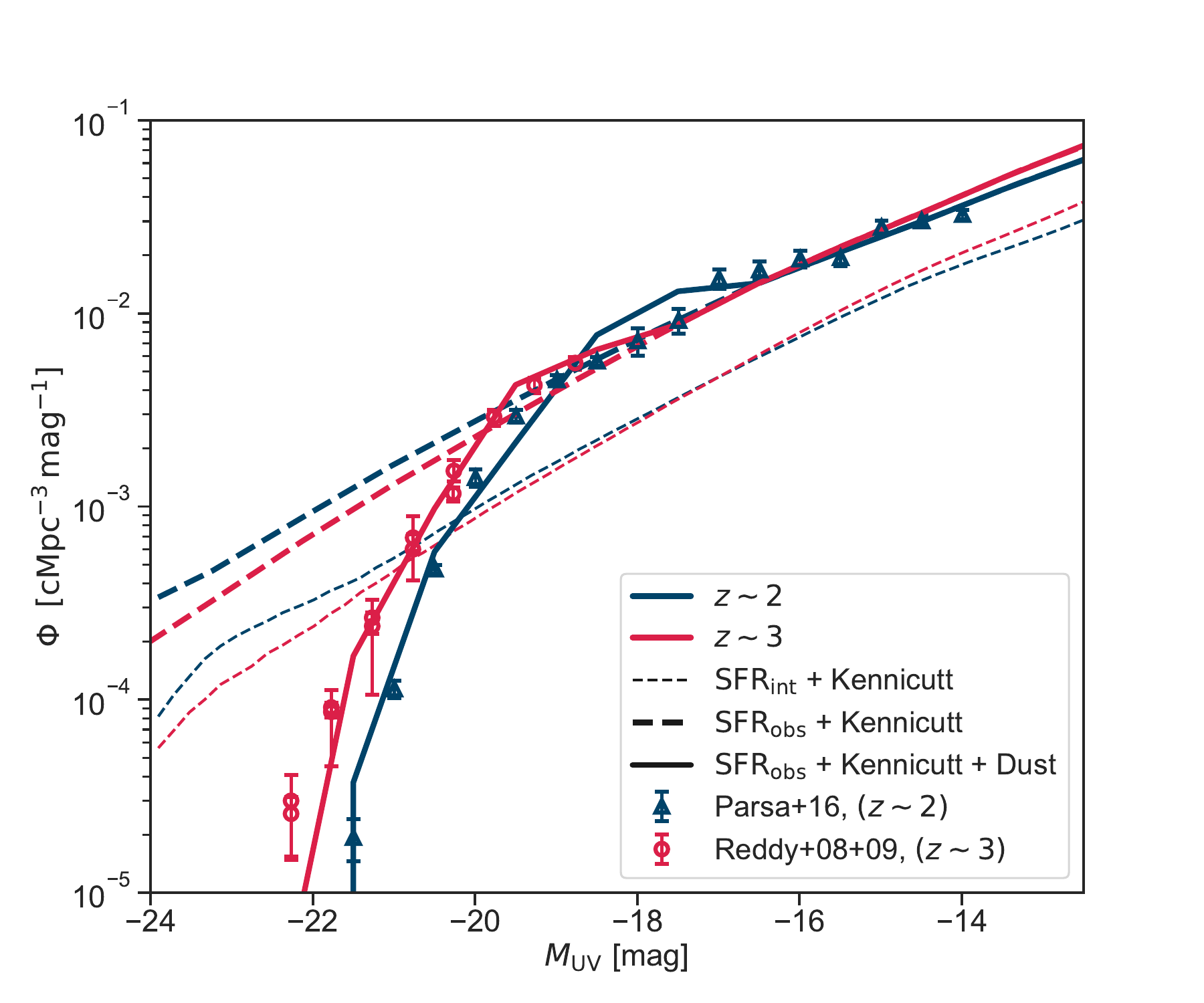}
\caption{UV luminosity function at $z\sim2$ (blue) and $z\sim3$ (red) compared to observational data from \cite{10.1093/mnras/stv2857} at $z\sim2$ and \cite{Reddy_2008, Reddy_2009} at $z\sim3$. The model (thick solid curves) calibrates the SFR to reproduce the UV luminosity function incorporating the empirical dust model from \cite{2020MNRAS.492.5167V}. The dust model is not included in dashed lines. The thin dashed lines assume the $\rm SFR_{\rm int}$ from the \texttt{UniverseMachine} model assuming Kennicutt relation, $L_{\rm cont,\lambda} = 1.4 \times 10^{40}\,{\rm erg\,s^{-1}\,\text{\AA}^{-1}}\times(\rm SFR/M_{\odot}yr^{-1})$ \citep{Kennicutt:1998ARAA}.}
\label{fig:UV_LF}
\end{figure*}
 
In this work, we adopt a publicly available \texttt{UniverseMachine} model run on the \texttt{SMDPL} simulation (\textsc{SMDPL-UM} hereafter) and use two snapshots at $z=2.42$ and $3.03$ which hereafter we will denote $z\sim 2$ and $z\sim 3$, respectively. We do not use a light-cone realization of SMDPL-UM, which is left for future work. 
We use (sub)halo properties of the three-dimensional position, peculiar velocity, the maximum circular velocity of a halo, $V_{\rm max} \equiv \text{max} (\sqrt{GM_{\rm halo}(<R)/R})$), the virial mass $M_{\rm vir}$, and the virial radius $R_{\rm vir}$ as well as galaxy properties of stellar mass $M_{*}$, and the $\mathrm{SFR_{int}}$ in this work. 
We do not adopt the observed $\mathrm{SFR_{obs}}$ and $M_{\rm UV,obs}$ available in the SMDPL-UM catalog and instead recalibrate these quantities in order to ensure consistency with the observed UV luminosity function at $z \sim 2$ and $3$, incorporating an empirical dust model. 
This allows us to obtain a more realistic prediction for the observed equivalent width and the observed escape fraction (see Sec~\ref{sec:lae_properties} for a comparison of our model with observations).
Following \citet{Behroozi:2019mn}, the observed $\mathrm{SFR_{obs}}$ is modeled by
\begin{align}\label{eq:SFR_obs}
    \log_{10}\left(\frac{\rm SFR_{obs}}{\rm M_{\odot}\,yr^{-1}}\right) = \log_{10}\left(\frac{\rm SFR_{int}}{\rm M_{\odot}\,yr^{-1}}\right) + \mu + \kappa\exp\left\{-\frac{(z-2)^{2}}{2}\right\},
\end{align}
where the mean shift $\mu$ and the scatter $\kappa$ are added to the UM-intrinsic $\mathrm{SFR_{int}}$ mainly to take into account the uncertainties in inferring SFR from observations.
Note that our $\mathrm{SFR_{obs}}$ is not exactly equivalent to the one in \citet{Behroozi:2019mn}, since we do not use SFR derived from a spectral fitting. 
Rather, we account for the amount of ionized photon budget by $\mathrm{SFR_{obs}}$ to explain the UV luminosity function. We assume the Kennicutt relation that relates ${\rm SFR_{obs}}$ to the UV luminosity density at $\lambda=1500\,$\AA, $L^{\rm int}_{\rm \lambda, UV} = 1.4 \times 10^{40}\,{\rm erg\,s^{-1}\,\text{\AA}^{-1}}\times(\rm SFR/M_{\odot}yr^{-1})$ \citep{Kennicutt:1998ARAA}. 
Fig.~\ref{fig:UV_LF} shows that $\mathrm{SFR_{int}}$ is not sufficient to explain the faint-end UV luminosity function in ${\rm M_{UV}}\gtrsim -20\,\,\rm mag$. 
We then recalibrate the mean shift value $\mu$ such that it reproduces the faint end of the UV luminosity function, producing $\mu=0.5$, i.e., $0.5\, \rm dex$ higher SFR.\\

The 0.5 dex addition is still consistent with the observed SFR data points within 2$\sigma$ errors at $z\sim 2$ and $3$ (see Fig.~3 in \citet{Behroozi:2019mn}).
We do not recalibrate the scatter $\kappa$, since the shift at faint end appears constant.
This SFR addition is not fully convincing and satisfying but is necessary to account for the insufficient amount of UV photons. 
This discrepancy may be attributed to starburst activities \citep[e.g.,][]{Sun:2023ApJL,Sun:2024ar,Haskell:2024MNL} or tensions between infrared and UV SFR indicators \citep[e.g.,][]{Fang:2018ApJ}.
Note that the original $M_{\rm UV,obs}$ available in the SMDPL-UM catalog is not reliable, as they are calibrated with UV luminosity functions only at $z>4$ \citep{Behroozi:2019mn}.
We also checked the redshift evolution of UV luminosity function at $z=2\!-\!3$. In Appendix~\ref{sec:app-UVLF-zevol}, we demonstrate that, once Eq.~(\ref{eq:SFR_obs}) is applied, the predicted rest-frame UV luminosity function of the SMDPL-UM catalog exhibits negligible evolution over $1.9 \leq z \leq 3.2$.\\ 

The mismatch at the bright end can be attributed to dust attenuation for UV photons. 
To model the dust attenuation, we adopt the `model A' in \cite{2020MNRAS.492.5167V} which empirically calibrates the UV dust optical depth as a function of ${\rm M_{UV,int}}$ and redshift (see Eq.~(7) in \citealp{2020MNRAS.492.5167V} but with different parameters; $C_0=4.43$, $C_1=1.99$, $M_0=-19.5$, $\sigma_{\beta}=0.34$ for constant values, for the redshift dependent part, we used $\beta_{M_0}=-0.97$ and $d\beta /dM_{\rm UV}^{\rm obs}=-0.71$ at $z=2.42$ and $\beta_{M_0}=-1.78$ and $d\beta /dM_{\rm UV}^{\rm obs}=-0.73$ at $z=3.03$).
This operation gives us the dust optical depth for UV photons, $e^{-\tau_{\rm UV}}=L_{\rm \lambda, UV}^{\rm obs}/L_{\rm \lambda, UV}^{\rm int}$.
This simple approach to recovering a realistic UV luminosity function is sufficient for our purpose to model the equivalent width of LAEs.\\

A potential caveat in the UM framework could be ``orphan'' subhalos/galaxies which are no longer associated with any subhalo or halo in the halo catalog but originate from subhalos in earlier snapshots.
We refer the reader to \citet{10.1093/mnras/stw439} and \citet{Behroozi:2019mn} for a specific implementation of the orphans. 
In Appendix \ref{sec:app-orphans}, we quantify the impact of the orphans on our results.
We also note that the measurement of the stellar mass function relevant to our LAE sample is not available particularly at $\log_{10}(M_{*}/M_{\odot})< 11$ at $z\sim 3$, and thus it does not contribute to constraining the \textsc{SMDPL-UM} model.\\

\subsection{Observational measurements of the LAE summary statistics }
\label{sec:obs}
We use recent measurements of the luminosity function and the angular correlation function to calibrate our LAE model at both $z\sim 2$ and $z\sim 3$.
In summary, we use the $z\sim 2.2$ luminosity function from \citet{Konno_2016}, the $z\sim 3.1$ luminosity function from \citet{Ouchi_2008}, and the angular correlation function from \citet{White:2024JC} as our fiducial choice.
In Appendices~\ref{sec:app-umeda} and \ref{sec:app-khostovan}, we show our results with different measurements from \cite{Umeda:2025ApJS} and \citet{Khostovan:2019mn}.
All of these measurements are based on LAEs detected with narrow-band surveys. 
We will briefly describe these default datasets below.\\

The $z\sim 3.1$ luminosity function from \citet{Ouchi_2008} is derived from a $1\,\rm deg^{2}$ sky of the Subaru/\textit{XMM-Newton} Deep Survey field.
Their $z\sim 3.1$ sample consists of 356 LAEs detected through a narrow-band filter, \textsc{NB503}, with a central wavelength of $\lambda^{\rm obs}_c = 5029 \text{\AA}$ and a full width at half maximum (FWHM) of $\Delta \lambda^{\rm obs} = 74 \text{\AA}$.
Their limiting magnitude roughly corresponds to $\log_{10}(L_{\rm Ly\alpha}\,[{\rm erg\,s^{-1}}])\simeq 42$. A spectroscopic follow-up campaign shows that the contamination fraction of the photometric LAE sample at $z = 3.1$ is estimated to be $< 13\%$. 
The $z\sim 2.2$ luminosity function from \citet{Konno_2016} is measured in the luminosity range of $\log_{10}(L_{\rm Ly\alpha}\,/{\rm erg\,s^{-1}}) = 41.7 - 44.4$ with a total of $3137$ LAEs detected by a narrow-band filter, \textsc{NB387}, with a central wavelength of $\lambda^{\rm obs}_c = 3870 \text{\AA}$ and a FWHM of $\Delta \lambda^{\rm obs} = 94 \text{\AA}$ in five independent fields that span a total of $1.43\,{\rm deg^2}$. 
They estimated that the contamination fraction from, e.g., low-redshift [OII] emitters is less than a few percent. 
This is in part because they applied an EW cut of $\gtrsim 70\, \text{\AA}$ to mitigate contamination from low-$z$ $\rm [O II]$ emitters. 
In \citet{Konno_2016}, they also reported that the bright-end LAE luminosity function is dominated by active galactic nuclei (AGNs) \citep[see also e.g.,][]{Zheng:2016ApJS,Spinoso:2020AA,Zhang:2021ApJ,Liu:2022ApJS,Liu:2022ApJ,Torralba-Torregrosa:2023A&A}. 
Therefore, we limit ourselves to the maximum luminosity of $\log_{10}(L_{\rm Ly\alpha}\,/{\rm erg\,s^{-1}}) = 43.0$ in this work, since we do not include AGNs in our model. 
In addition, we ignore interloper contamination and the impact of an incomplete sample selection (an EW cut, for instance) in these luminosity function measurements \citep{Konno_2016}.
A comparison of various LAE luminosity function measurements at $z=2\!-\!3$ in the literature is presented in \citet{Blanc:2011ApJ,Konno_2016,Herenz:2019AA,Torralba-Torregrosa:2023AA,Umeda:2025ApJS}.\\

We also use both $z\sim 2.4$ and $z\sim 3.1$ angular correlation functions in \citet{White:2024JC}, based on LAEs selected from the One-hundred-$\rm deg^2$ DECam Imaging in Narrowbands (\textsc{ODIN}; \citet{2023ApJ...951..119R,2023arXiv231216075F,2024ApJ...962...36L}). 
The \textsc{ODIN} survey is a wide-field, deep imaging survey targeting seven fields with three narrow-band filters, designed to select LAEs at redshifts $z \sim 2.4, 3.1$, and $4.5$.
The 822 (1099) LAEs at $z\sim 2.4\,(3.1)$ are detected with \textsc{N419} and \textsc{N501} filters over the \textsc{COSMOS} field with limiting magnitude of $25.5$ and $25.7$, corresponding to Ly$\alpha$ line flux limits of $1.547 \times 10^{42}$ and $1.584 \times 10^{42}\, \rm erg\,s^{-1}$. 
A subset of these objects was subjected to spectroscopic follow-up with the Dark Energy Spectroscopic Instrument (DESI; \citet{2013arXiv1308.0847L,2016arXiv161100036D}), and these data were used to refine the narrow-band selection criteria, as well as to constrain the interloper fraction and the redshift distribution. 
With these follow-up observations, the contamination fractions are estimated between $0.02$ and $0.08$ at $z\sim 2.4$ and between $0.04$ and $0.11$ at $z\sim 3.1$, respectively. 
The clustering amplitude is directly affected by the contamination fraction $f_{\rm cont}$ by a factor of $(1-f_{\rm cont})^{2}$ (assuming that the contaminating objects do not cluster).
We will adopt the most conservative values, i.e., $f_{\rm cont}=0.08$ at $z\sim 2.4$ and $f_{\rm cont}=0.11$ at $z\sim 3.1$ in our fitting processes.

\section{Modeling the Lyman-\texorpdfstring{$\alpha$}{alpha} emitters}
\label{sec:lae_model}

\subsection{Overview}
\label{subsec:overview}

We model the Ly$\alpha$ \textit{spectrum} for each individual galaxy in the \textsc{SMDPL-UM} model, which can affect the LAE observability due to RT effects. 
We incorporate the Ly$\alpha$ RT effect on the Ly$\alpha$ spectrum across the ISM/CGM and IGM scales rather than assuming a simple scaling relation between the observed Ly$\alpha$ luminosity and other galaxy properties.
We give an overview of our model here and then will discuss each ingredient in more detail in the subsequent subsections.\\

The starting point of our model is the linear relation between the intrinsic Ly$\alpha$ luminosity and SFR, given by \citep[see e.g.,][]{Dijkstra:2017ar}
\begin{align}\label{eq:L_int}
    L_{\rm Ly\alpha}^{\rm int} = 1.0\times 10^{42} \,\rm{erg\,s^{-1}}\times \left(\frac{\rm SFR_{obs}}{\rm M_{\odot}\,yr^{-1}}\right), 
\end{align}
where the case-B recombination at the ISM gas temperature of $10^{4}\,{\rm K}$ is assumed, and the relation between SFR and the ionizing photon emission rate is calibrated assuming the Salpeter Initial Mass Function (IMF) \citep{Kennicutt:1998ARAA}.  
Although it is well known that different assumptions on the IMF shape and metallicity can lead to different normalizations of the relationship between the $\rm SFR_{obs}$ and $L_{\rm Ly\alpha}^{\rm int}$ \citep{Schaerer:2003AA}, this normalization still represents a conservative Ly$\alpha$ luminosity when compared to other literature values \citep{Kennicutt:1998ARAA, 2001MNRAS.322..231K, 2003PASP..115..763C, 2011ApJ...742..108F, Dijkstra:2017ar,     2019A&A...623A.157S}.
We assume a Gaussian line profile with intrinsic line width $W_{\rm in}$ and a flat continuum with intrinsic equivalent width (in rest frame) $\mathrm{EW}_{\rm in}$, both of which will be specified in Sec.~\ref{subsec:ism}.\\  

We model the rest-frame spectrum $\phi(\lambda)=\phi_{\rm Ly\alpha}(\lambda)+\phi_{\rm cont}(\lambda)$ that consists of the Ly$\alpha$ emission, $\phi_{\rm Ly\alpha}(\lambda)$, and the continuum $\phi_{\rm cont}(\lambda)$, as follows. 
The Ly$\alpha$ RT on the ISM and CGM scales is effectively modeled by the expanding gas shell model, closely following \citet{Gurung-Lopez:2019mn_a} as we discuss in Sec.~\ref{subsec:ism} in detail. 
This provides the line profile $\phi^{\rm ISM}_{\rm Ly\alpha}(\lambda;\vec{\theta}_{g}(\alpha,\beta,\gamma))$ for each individual galaxy. 
$\vec{\theta}_{g}(\alpha,\beta,\gamma)$ denotes a set of galaxy parameters necessary for the shell model with three dimensionless free parameters $(\alpha,\beta,\gamma)$, and will be specified in Sec.~\ref{subsec:ism}. 
Then, the Ly$\alpha$ luminosity $ L^{\rm ISM}_{\rm Ly\alpha}$ after escaping from the ISM/CGM scales is given by 
\begin{align}\label{eq:L_ism}
    L^{\rm ISM}_{\rm Ly\alpha} = \int \phi^{\rm ISM}_{\rm Ly\alpha}(\lambda;\vec{\theta}_{g}(\alpha,\beta,\gamma)) d\lambda = f_{\rm esc}^{\rm ISM}L_{\rm Ly\alpha}^{\rm int}, 
\end{align}
where $f_{\rm esc}^{\rm ISM}$ is the ISM escape fraction of Ly$\alpha$ photons, and the integration is evaluated within $\pm 10\,\angstrom$ from the Ly$\alpha$ line center. 
We also account for the impact of IGM transmission $T_{\rm IGM}(\lambda)$.
As we discuss in Sec.~\ref{subsec:igm}, we will simply assume that the IGM transmission $T_{\rm IGM}(\lambda;z)$ depends only on redshift, neglecting line-of-sight variations. Thus, we model the observed Ly$\alpha$ luminosity by 
\begin{align}\label{eq:f_esc_igm}
 L^{\rm obs}_{\rm Ly\alpha} = \int \phi^{\rm ISM}_{\rm Ly\alpha}(\lambda;\vec{\theta}_{g}(\alpha,\beta,\gamma)) T^{\rm IGM}(\lambda;z) \, d\lambda = f^{\rm IGM}_{\rm esc}L^{\rm ISM}_{\rm Ly\alpha} = f^{\rm IGM}_{\rm esc}f^{\rm ISM}_{\rm esc}L_{\rm Ly\alpha}^{\rm int}, 
\end{align}
where the fraction of Ly$\alpha$ luminosity, $f^{\rm IGM}_{\rm esc}\equiv L^{\rm obs}_{\rm Ly\alpha}/L^{\rm ISM}_{\rm Ly\alpha}$, depends on each galaxy property through the observed line profile $\phi^{\rm obs}(\lambda)=\phi^{\rm ISM}(\lambda;\vec{\theta}_{g})\times T^{\rm IGM}(\lambda;z)$, and the observed Ly$\alpha$ escape fraction corresponds to 
$f^{\rm obs}_{\rm esc, Ly\alpha}=f^{\rm ISM}_{\rm esc}f^{\rm IGM}_{\rm esc}$ in this notation.
We will discuss later in more detail how to estimate the rest-frame equivalent width of each LAE.
In summary, we attempt to model the LAEs with the three dimensionless free parameters $(\alpha,\beta,\gamma)$ so that they reproduce both the Ly$\alpha$ luminosity function and the angular correlation function at the same time. 
The advantage of our work is that it predicts the observed Ly$\alpha$ luminosity and line profile for each individual galaxy, which allows us to study statistical correlation between Ly$\alpha$ emission diagnostics and other galaxy properties. 

\begin{figure*}
\centering
\includegraphics[scale=0.38]{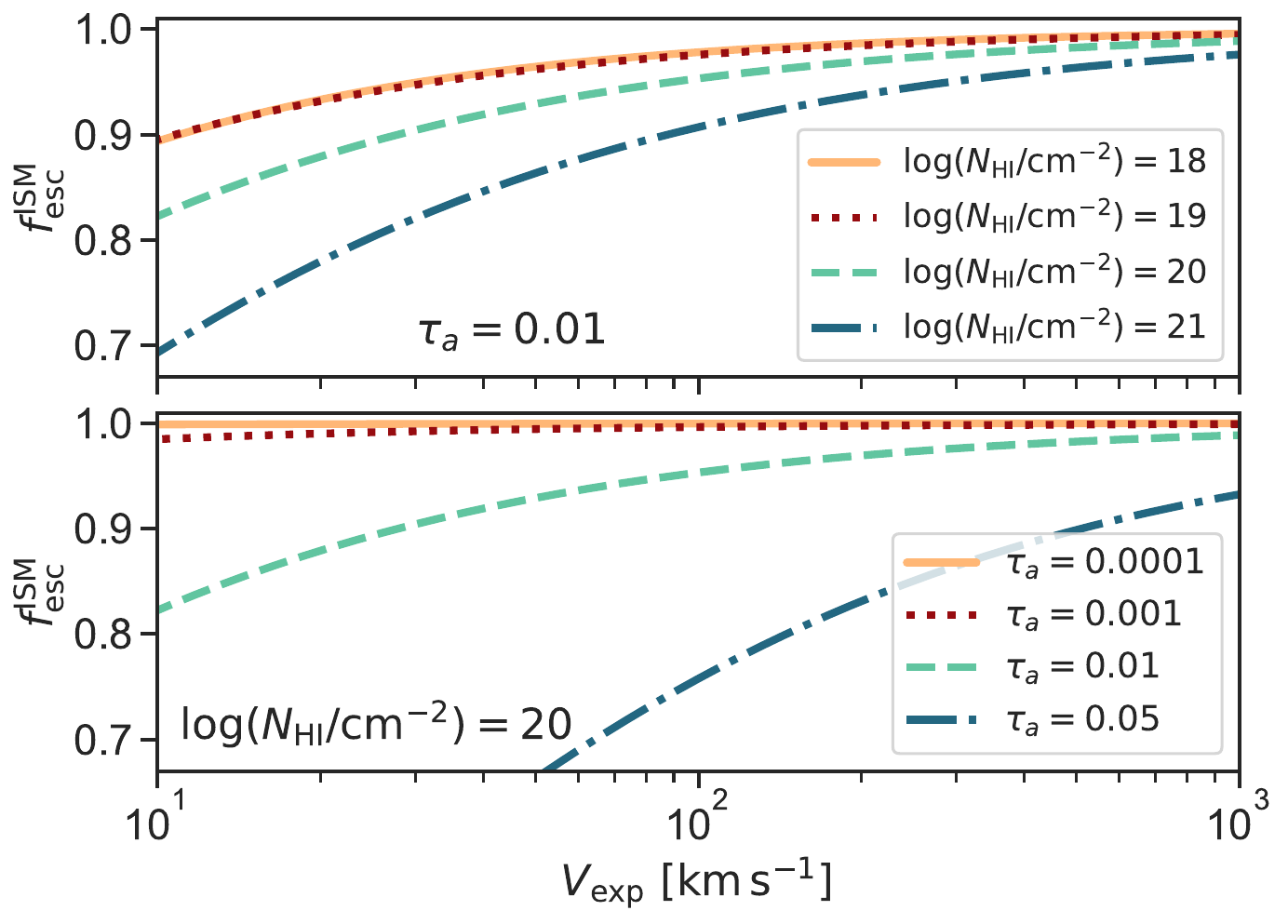}
\includegraphics[scale=0.38]{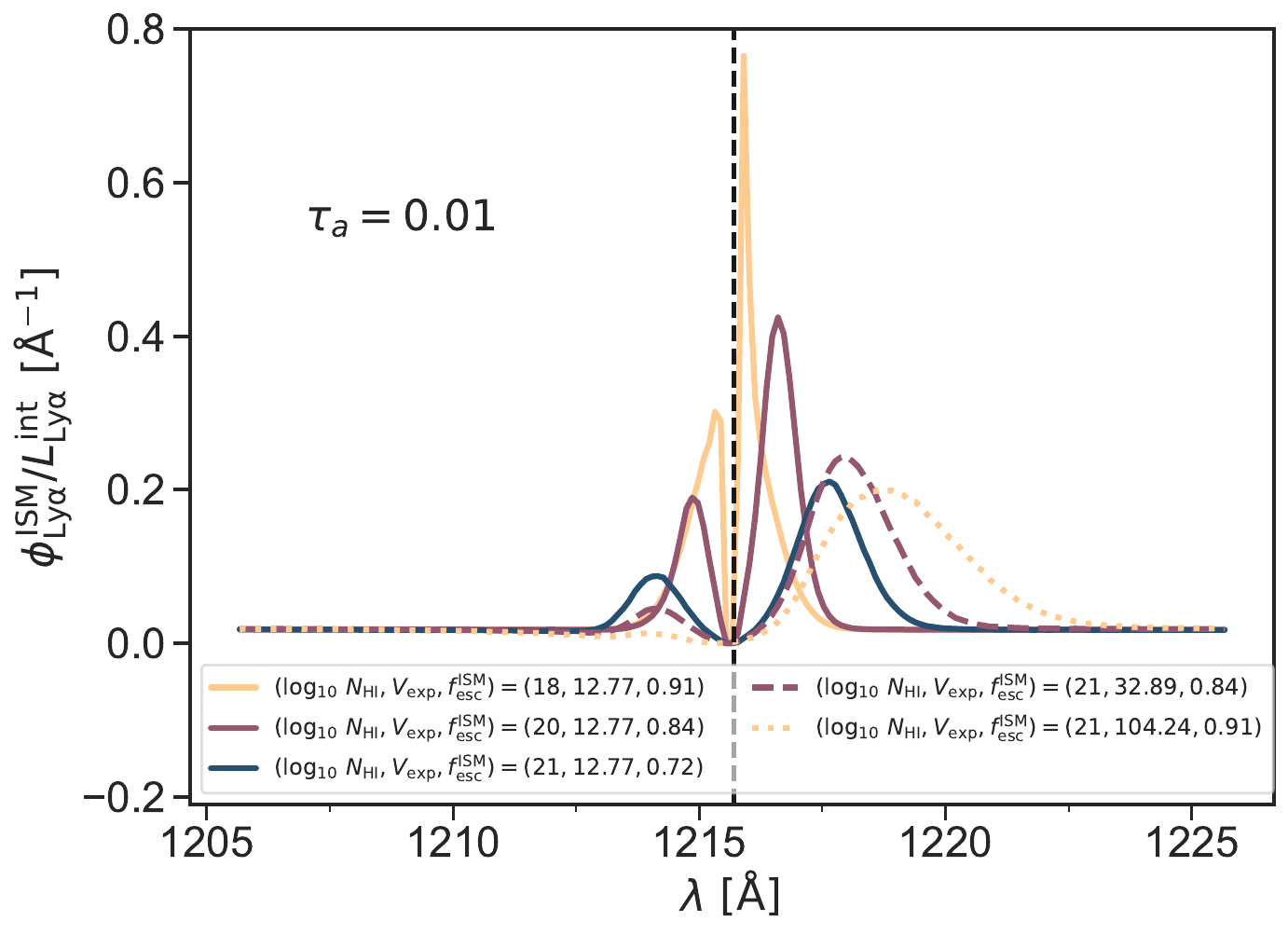}
\caption{
{\it Left}: ISM escape fraction, $f_{\rm esc}^{\rm ISM}$, as a function of expansion velocity, $V_{\rm exp}$, neutral hydrogen column density, $N_{\rm HI}$, and dust absorption optical depth, $\tau_a$. Top panel shows how $f_{\rm esc}^{\rm ISM}$ changes for various $N_{\rm HI}$ values at a fixed $\tau_a=0.01$. The bottom panel presents the dependence on $\tau_a$ for a fixed $\log_{10}({\rm N}_{\rm HI}/{\rm cm}^{-2})=20$. {\it Right}: Example line profile of the galaxies with various shell model parameters. 
The three parameters in each legend corresponds to $\log_{10}({\rm N}_{\rm HI}/{\rm cm}^{-2})$, $V_{\rm exp}/({\rm km\,s^{-1}})$, and $f_{\rm esc}^{\rm ISM}$. The vertical dashed line shows the rest frame Ly$\alpha$ wavelength, $\lambda_{\rm Ly\alpha}=1215.67\,\angstrom$.
}
\label{fig:fISM}
\end{figure*}

\begin{figure}
\centering
\includegraphics[scale=0.45]{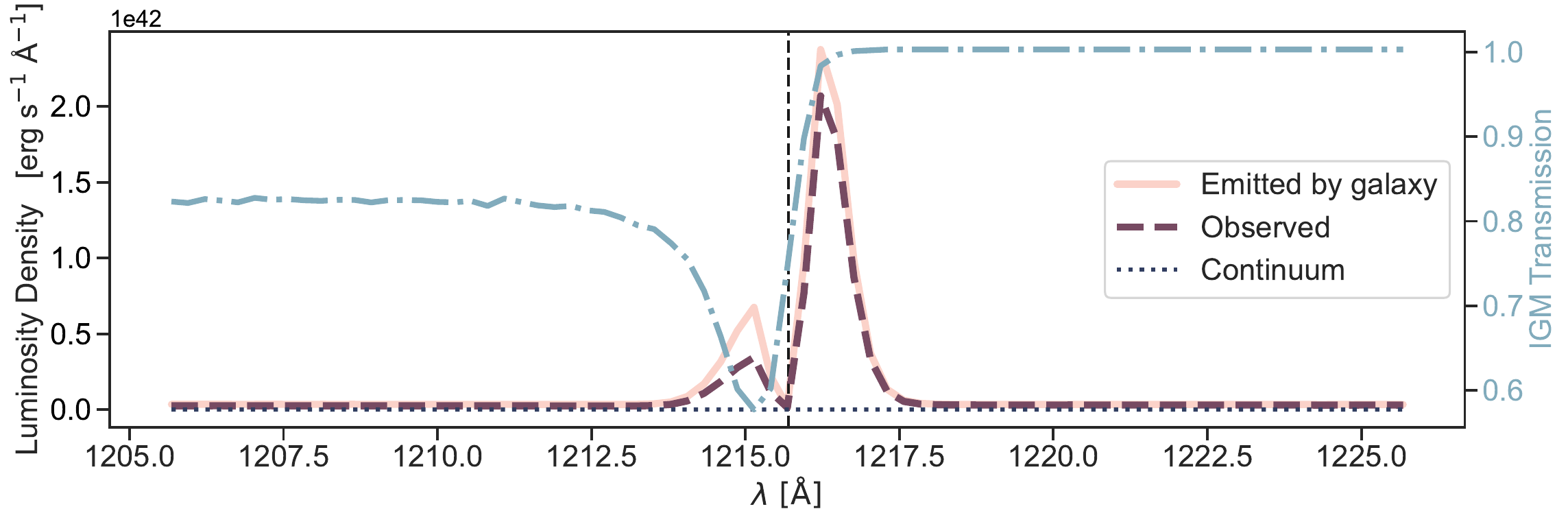}
\caption{The impact of Ly$\alpha$ RT on an example Ly$\alpha$ profile. This galaxy resides at $z=2.42$, with $M^{*}=4.59\times 10^{8}\, \rm M_{\odot}$, $\rm SFR=1.01\,\,M_{\odot}\, yr^{-1}$ and the shell model's key parameters of $V_{\rm exp} = 21.41 \,\rm \,\,km/s$, $\rm log_{10} \, N_{\rm HI}= 19.48\,\rm cm^{-1}$ and $\tau_a =0.03$. The light red solid curve represents the emitted Ly$\alpha$ luminosity density from the ISM/CGM. The dark dashed curve corresponds to the observed Ly$\alpha$ spectrum after incorporating both ISM/CGM and IGM effects. 
The blue dot-dashed curve (right y-axis) represents the IGM transmission. 
The vertical black dashed line marks the rest-frame Ly$\alpha$ line center wavelength, ($\lambda = 1215.67 \,\text{\AA}$).\\}
\label{fig:flux}
\end{figure}

\subsection{Modeling the Ly$\alpha$ Radiative Transfer on ISM/CGM scales}
\label{subsec:ism}

To effectively capture the impact of Ly$\alpha$ RT on ISM/CGM scales, we adopt the model with an expanding spherical gas shell, following \cite{Gurung-Lopez:2019mn_a}. 
The expanding spherical shell model has been widely used in the literature to study Ly$\alpha$ RT and understand how key parameters, such as outflow velocity and gas column density, shape the observed line profiles (e.g., \cite{Zheng_2002, Dijkstra_2006, Verhamme2006, Tasitsiomi_2006,Semelin:2007AA, Laursen_2009, Yajima:2012MN, Gronke_2015, 10.1093/mnras/stv565}).
In this model, an LAE galaxy is surrounded by an expanding homogeneous isothermal spherical shell of neutral hydrogen gas with a fixed temperature of ($T = 10^4\, \rm K$) and the inner-to-outer radius fraction of $0.9$. 
Even in this simplified geometry, it is computationally infeasible to perform a full Monte Carlo RT simulation for millions of individual simulated galaxies. 
For this reason, we adopt the publicly available Python package \texttt{zELDA} \citep{2019MNRAS.490..733G, Gurung-Lopez:2021zELDA} that provides a fast prediction of the Ly$\alpha$ line profile trained by precomputed grids of full Monte Carlo simulations with a simple neural network\footnote{We do not implement the most updated version, \texttt{zELDA II} in \citet{Gurung-Lopez:2025ar}, since it relies on the specific IGM model in the \texttt{IllustrisTNG} simulation.}. 
In \texttt{zELDA}, the Ly$\alpha$ line profile including both emission and continuum is calculated using five input parameters; expansion velocity $V_{\rm exp}$, neutral hydrogen column density $N_{\rm HI}$, optical depth of dust attenuation $\tau_{\rm a}$, intrinsic line width $W_{\rm in}$, and intrinsic equivalent width $\mathrm{EW}_{\rm in}$.
\begin{align}\label{eq:theta_g}
    \vec{\theta}_{g} = (V_{\rm exp},\;N_{\rm HI},\;\tau_{\rm a},\;W_{\rm in},\;\mathrm{EW}_{\rm in}).
\end{align}
In the left panel of Fig.~\ref{fig:fISM}, we show the ISM escape fraction $f^{\rm ISM}_{\rm esc}$ as a function of three key parameters, $V_{\rm exp}$, $N_{\rm HI}$, and $\tau_{\rm a}$. A technical note is that \texttt{zELDA} is trained only in a limited range of parameters $(V_{\rm exp} \,\,{\rm km\,s^{-1}} \in (0, 1000), \,\, \log_{10}(N_{\rm HI}/{\rm cm^{-2}}) \in (17, 21.5)\,\, \log_{10}(\tau_{\rm a}) \in (-4, 0) )$ and, in principle, there could be some galaxies that have values for these parameters outside of the trained range. 
In this case, we simply adopt the ISM escape fraction and the spectral profile from the closest boundary, as it physically makes sense to expect $f^{\rm ISM}_{\rm esc}=1$ for sufficiently large $V_{\rm exp}$, for example. 
In the right panel of Fig.~\ref{fig:fISM}, we further illustrate the impact of these parameters on the emergent Ly$\alpha$ line profiles by showing the corresponding spectra for galaxies with different $N_{\rm HI}$ and $V_{\rm exp}$ values. 
Three solid lines ($\log_{10}(N_{\rm HI}/{\rm cm}^{-2})=18,20,21$ for orange, purple, and blue, respectively) have different $N_{\rm HI}$ at a fixed $V_{\rm exp}$, showing that a larger neutral hydrogen density diffuses more Ly$\alpha$ photons and results in a smaller $f^{\rm ISM}_{\rm esc}$. 
As a consequence, the peak positions at both the red and blue sides of the spectra are more apart from the line center.  
In addition, the expansion velocity increases from the blue solid line ($V_{\rm exp}=12.77\,{\rm km\,s^{-1}}$), the purple dashed line ($V_{\rm exp}=32.89\,{\rm km\,s^{-1}}$), to the orange dotted line ($V_{\rm exp}=104.24\,{\rm km\,s^{-1}}$), 
at fixed $N_{\rm HI}$. 
This comparison shows that an outflow gas with higher velocity suppresses the blue side of photon's spectra more, while shifting the red peak position to a larger wavelength.  
Notice that the same values of $f^{\rm ISM}_{\rm esc}$ are shared by the orange solid and dotted lines, as well as the purple solid and dashed lines, highlighting that we cannot determine gas physics only from the escape fraction.\\  

The next question is then how to assign $\vec{\theta}_{g}$ for each simulated galaxy. 
Motivated by \cite{Gurung-Lopez:2019mn_a}, we characterize them as follows:
\begin{align}\label{eq:vexp}
    V_{\rm exp} = \alpha\,{\rm sSFR}\, r_{1/2} =  0.0978\alpha\,\left(\frac{\rm sSFR}{10^{-10}\rm yr^{-1}}\right)\left(\frac{r_{1/2}}{\rm pkpc}\right)\, {\rm km\,s^{-1}},
\end{align}
\begin{align}\label{eq:NHI}
    N_{\rm HI} = \beta \, \dfrac{M_{\rm HI}}{(4 \pi m_{\rm H}) r_{1/2}^{2}} = 
    9.94\times 10^{21}\beta \, \left(\frac{M_{\rm HI}}{\rm 10^{9}\,M_{\odot}}\right)\left(\frac{r_{1/2}}{\rm pkpc}\right)^{-2} \, {\rm cm^{-2}},
\end{align}
\begin{align}\label{eq:tau_a}
    \tau_{\rm a} = \Big(\dfrac{\gamma}{\beta}\Big)\,(1-A_{\rm{Lya}})\dfrac{E_{\odot}}{Z_{\odot}}N_{\rm HI} Z_{\rm cold}
    = 0.11\gamma \left(\frac{M_{\rm HI}}{\rm 10^{9}\,M_{\odot}}\right)\left(\frac{r_{1/2}}{\rm pkpc}\right)^{-2}\left(\frac{Z_{\rm cold}}{0.0002}\right).
\end{align}

Here, the expansion velocity $V_{\rm exp}$ is scaled by the specific SFR (sSFR$\equiv \mathrm{SFR}/M_*$) and the stellar-half mass radius $r_{1/2}$ \citep{Hopkins:2012MN,Arribas:2014AA}, while the neutral hydrogen column density, $N_{\rm HI}$, is scaled by the neutral hydrogen mass, $M_{\rm HI}$, and $r_{1/2}$. 
The optical depth of the dust attenuation $\tau_{\rm a}$ is assumed to be proportional to the amount of gas-phase metallicity $N_{\rm HI} Z_{\rm cold}$.
We assume a Ly$\alpha$ dust albedo of $A_{\rm Ly\alpha} \;=\; 0.39$ at solar metallicity, $Z_{\odot}=0.02$ (see \cite{1979MNRAS.187P..73S, 2000ApJ...542..710G, Calzetti_2000,  Verhamme2006, 2009ApJ...704L..98S}), and $E_{\odot}/Z_{\odot}=8.85\times 10^{-20}\,\rm cm^{2}$ refers to a reference dust opacity of the gas for solar metallicity. We introduced three constant free parameters $(\alpha,\beta,\gamma)$ that rescale these mappings and are calibrated jointly to the Ly$\alpha$ luminosity function and angular clustering. Note that our model for Ly$\alpha$ RT through the ISM adopts the widely-used expanding shell geometry. Idealized geometries like this have an established record of successfully
reproducing observed integrated Ly$\alpha$ spectra (e.g. \cite{Verhamme2006, Dijkstra_2006, Yang_2016, 2017A&A...608A.139G, Song:2020ApJ}), making them a practical choice for cosmological simulations. Moreover, the effectiveness of homogeneous models is supported by recent studies showing that spectra from more realistic, inhomogeneous gas distributions can be well approximated by weighted sums of spectra from homogeneous models (see \cite{2025arXiv250919184A}). In the same spirit, \cite{10.1093/mnras/stac1207} showed that spectra from multiphase media can be reproduced with the (homogeneous) expanding-shell model using effective parameters. This justifies our strategy of using a fast, homogeneous RT basis, calibrated against observations, to capture the statistical behavior of Ly$\alpha$ emission from large galaxy populations. In practice the primary effect of replacing our thin shell with a more complex, clumpy or anisotropic ISM/CGM is to shift the effective values of the Ly$\alpha$ RT inputs $(V_{\rm exp},N_{\rm HI},\tau_a)$. The main consequences are an increased object-to-object scatter in the ISM escape fraction, $f_{\rm esc}^{\rm ISM}$ and velocity offsets and subtle changes the Ly$\alpha$ luminosity function and the effective host-mass distribution that sets clustering \citep{10.1093/mnras/stac1207, 2021A&A...653A..83V, Byrohl:2020AA}. The free parameters, $(\alpha,\beta,\gamma)$, in our mappings are central to managing this complexity. By calibrating these parameters to the observed Ly$\alpha$ luminosity function and angular clustering, they act as effective nuisance parameters.

The intrinsic line width $W_{\rm in}$ is determined by 
\begin{align}\label{eq:W_i}
    W_{\rm in} = \frac{V_{\rm max}}{c}\lambda_{\rm Ly\alpha}, 
\end{align}
since the maximum circular velocity of a halo ($V_{\rm max}$, see sec.~\ref{sec:halos} for the definition), traces the depth of halo gravitational potential and thus the kinetic energy of the halo.
We assign the intrinsic EW by the intrinsic continuum level as

\begin{align}\label{eq:EW_i}
    \mathrm{EW}_{\rm in} = \frac{L_{\rm Ly\alpha}^{\rm int}}{L_{\lambda,{\rm UV}}^{\rm int}}, 
\end{align}
where $L_{\lambda,{\rm UV}}$ is the rest-frame luminosity density of the continuum at $\lambda=1500\,$\AA, extrapolated from the intrinsic UV magnitude $M^{\rm int}_{\rm UV}$ assuming the UV slope of -2 \citep{Meurer_1999, Bouwens_2014}.\\ 

In summary, for a given set of the three dimensionless free parameters  $(\alpha,\beta,\gamma)$, we specify $\theta_{g}$ in Eq.~(\ref{eq:theta_g}) which further depends on the parameters available in \texttt{UniverseMachine}, $(M_{*},{\rm sSFR}, V_{\rm max})$ as well as the unavailable parameters, $(r_{1/2}, M_{\rm HI}, Z_{\rm cold})$.

\subsection{Scaling Relations}
\label{subsec:scale}

We outline below how we determine the parameters unavailable in \texttt{UniverseMachine}, $(r_{1/2}, M_{\rm HI}, Z_{\rm cold})$, for each galaxy.
For simplicity, we rely on empirical relations motivated by previous observational results. 
This is distinct from \cite{Gurung-Lopez:2019mn_a} who determined $r_{1/2}$, $M_{\rm HI}$, and $Z_{\rm cold}$ within their semi-analytic model.\\ 

\begin{figure*}
\centering\includegraphics[scale=0.42]{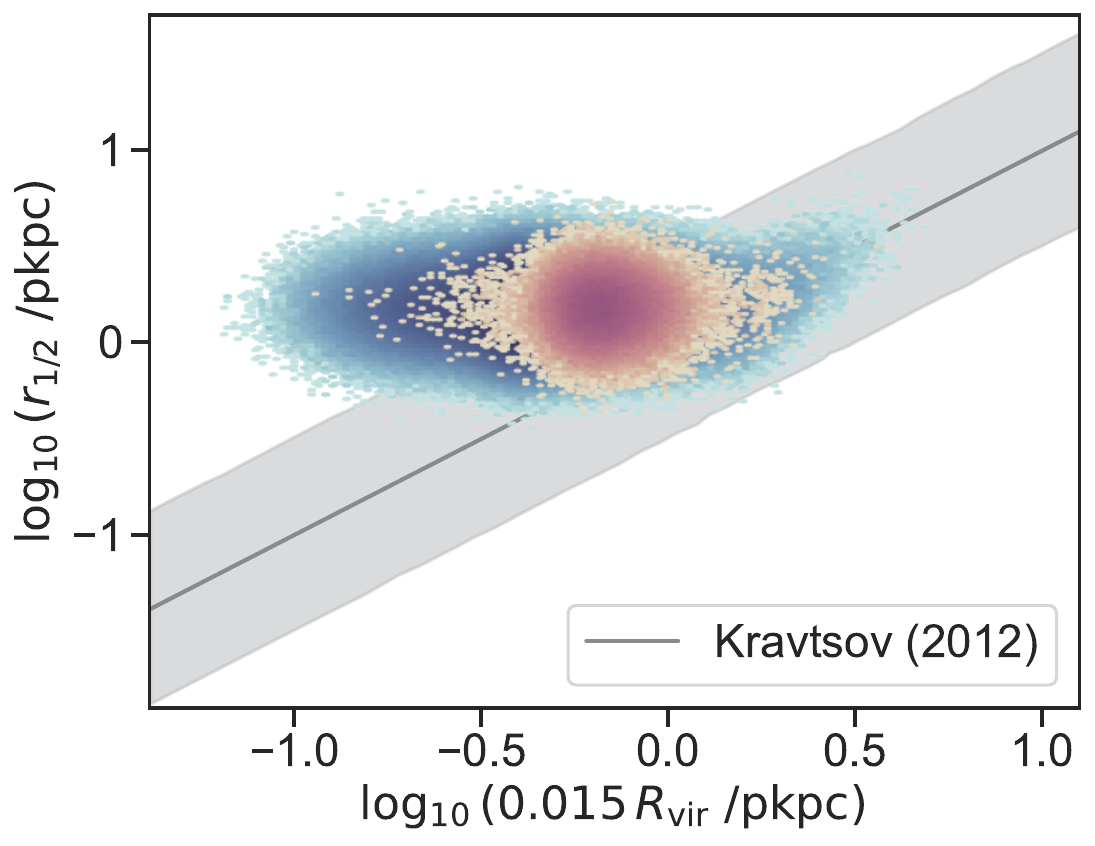}
\includegraphics[scale=0.42]{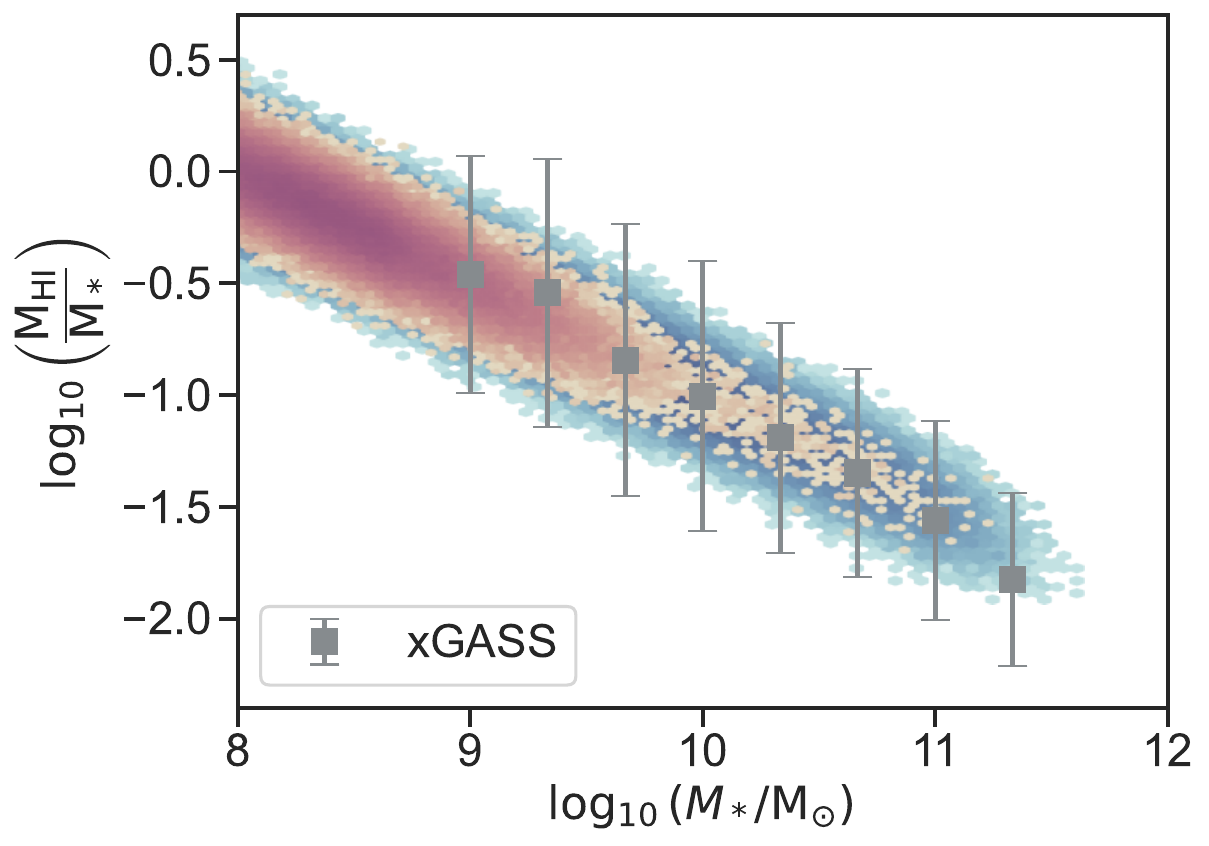}
\includegraphics[scale=0.42]{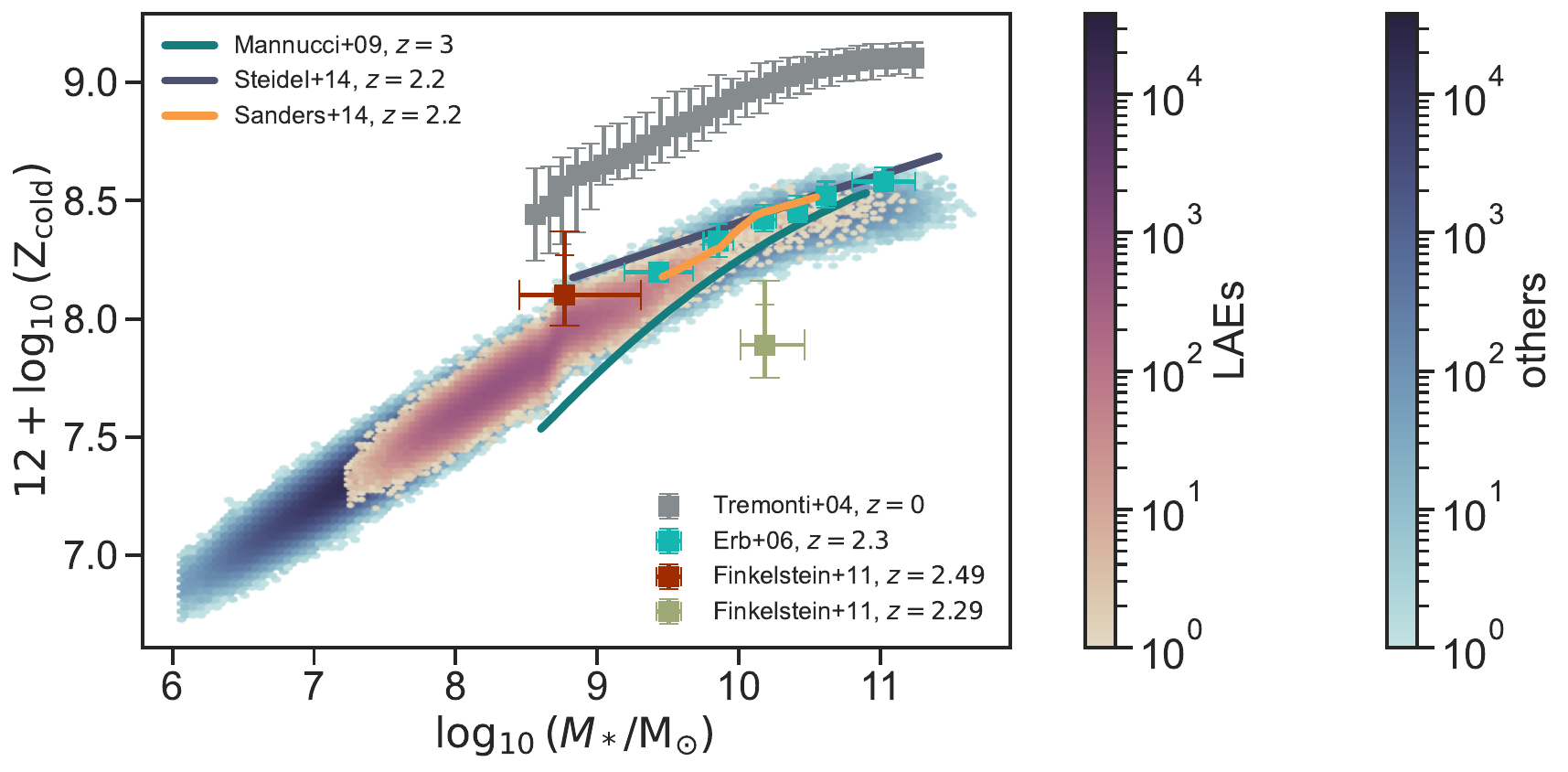}
\caption{Scaling relations we adopt in this work. 
We show our galaxy sample at $z\sim 3$ where LAEs \big($ \mathrm{log}_{10}(L_{Lya}/\,\rm erg\,s^{-1}) \geq 41.7$ and $\mathrm{EW}_0 > 40 \,\, \text{\AA}$\big) are shown in pink while others are represented in blue.
{\it Top left}: Stellar half-mass radius, $r_{1/2}$, versus the halo virial radius, $R_{\rm vir}$. The gray solid line represents the empirical relation $r = 0.015 \, \rm R_{\rm vir}$ from \protect\cite{Kravtsov_2013} with the shaded region indicating the $2\sigma$ scatter. 
{\it Top right}: Neutral atomic hydrogen gas mass fraction, $\rm log_{10} (M_{\rm HI}/M_*)$, as a function of stellar mass, serving as a scaling relation for inferring the neutral hydrogen column density. Measurements from the XGASS survey \citep{10.1093/mnras/sty089} are shown in gray. 
{\it Bottom}: Cold gas metallicity (gas-phase oxygen abundance, $12+\rm log_{10}(Z_{\rm cold})$) as a function of stellar mass compared with several observational mass-metallicity relations including \protect\cite{2004ApJ...613..898T, Erb_2006, mannucci2009lsd, 2011ApJ...729..140F,  2014ApJ...795..165S, sanders2015mosdef} as depicted.}
\label{fig:scaling_rels}
\end{figure*}

To model the galaxy size, we adopt the stellar half-mass radius, $r_{1/2}$, scaled with the halo virial radius, $R_{\rm vir}$, with an associated scatter of $0.12 \, \rm dex$ at fixed virial radius. \cite{Kravtsov_2013} found that $r_{1/2} \approx 0.015 \, R_{\rm vir}$ at $z\sim 0$, a relation widely adopted in theoretical and observational studies \citep[see, e.g.,][]{2017ApJ...838....6H,2018MNRAS.473.2714S, 10.1093/mnras/stz2251, 2020MNRAS.492.1671Z}. Observations particularly on high redshifts show that LAEs are systematically more compact than other star-forming galaxies and that low-mass galaxies have an approximately constant half-mass radius \citep{2015ApJS..219...15S}, suggesting a mass-dependent size relation \citep[see also][for other high-redshift studies.]{2004ApJ...604L...9R,2004ApJ...600L.107F, 2014ApJ...788...28V, Shibuya_2019}. 
Recent simulations also confirm this trend \citep{2018MNRAS.473.4077P, 10.1093/mnras/stu1654, 10.1093/mnras/stu1536, vogelsberger2014properties, 2025arXiv250203679S}. To incorporate this effect, we adopt a modified relation from \cite{Kravtsov_2013} that flattens the galaxy size-halo relation for smaller halos to account for the deviations seen in both simulations and observations. 
In practice, our prescription for $r_{1/2}$ is given by
\begin{equation}
\log_{10}(r_{1/2}/{\rm pkpc}) = 
\begin{cases}
\log_{10}(0.015 R_{\rm vir}/{\rm pkpc}) + \epsilon, & \text{if } R_{\rm vir} \leq R_{\rm piv} \\
\log_{10}(0.015 R_{\rm vir}/{\rm pkpc}) + \epsilon, & \text{if } R_{\rm vir} > R_{\rm piv}
\end{cases}
\end{equation}
where $R_{\rm piv}=113\,{\rm pkpc}$ is the pivot radius, and the scatter term, $\epsilon$, is drawn from a normal (Gaussian) distribution with mean zero and standard deviation $\sigma = 0.12$, corresponding to a 0.12 dex scatter in $\rm log_{10} (r_{1/2}/\rm pkpc)$. We match \cite{Kravtsov_2013} for $R_{\rm vir}>113$ pkpc and adopt a constant $r_{1/2}=0.015 \,R_{\rm vir}$ below that pivot. \cite{Kravtsov_2013} reported a scatter from 0.1 to 0.2 dex, while \cite{2022MNRAS.510.3967R}, using the FIRE simulations, reported a slightly smaller scatter of 0.084 dex. \cite{2022MNRAS.510.3967R} further demonstrated that the galaxy-halo size relation exhibits very weak evolution from $z=0$ to $z=5$, with a scatter remaining nearly constant throughout this redshift range. 
This justifies our choice not to incorporate explicit redshift evolution in our size relation. 
Another technical reason to flatten the relation at a small radius is that $N_{\rm HI}$ would become unphysically larger as $r_{1/2}$ decreases (see Eq.~(\ref{eq:NHI})).
In the top left panel of Fig.~\ref{fig:scaling_rels}, we show the assumed size relation of all galaxies (colored scatter) compared to the empirical relation from \cite{Kravtsov_2013} (solid gray line). 
The shaded region around Kravtsov's relation illustrates a broader scatter corresponding to the $2\sigma$ scatter expected from variations in dark matter halo spin. 
We also discuss the comparison with simulations in Appendix~\ref{sec:app-scalerel}.\\

To model the HI mass, $M_{\rm HI}$, we adopt the scaling relation in terms of the stellar mass, $M_{*}$, motivated by results from the XGASS survey \citep{10.1093/mnras/sty089}. 
Specifically, we parametrize the cold gas mass as
\begin{equation}
\log_{10}\left(\frac{M_{\rm HI}}{M_{*}}\right) = -0.5\log_{10}(M_{*}/M_{\odot}) + 4 +\epsilon, 
\end{equation}
where the scatter term, $\epsilon$, is drawn from a Gaussian distribution with mean zero and standard deviation $\sigma=0.1$, corresponding to a 0.1 dex scatter in $\log_{10}\left(M_{\rm HI}/M_{*}\right)$.
This choice of scatter is taken from observational results, where a typical scatter in the gas fraction is around 0.1 dex \citep{2015MNRAS.452.2479B, 2017ApJS..233...22S, 10.1093/mnras/sty089}. 
In addition, recent hydrodynamic simulations, such as \texttt{IllustrisTNG} \citep{10.1093/mnras/sty2206, 10.1093/mnras/sty618, 10.1093/mnras/stx3040, 10.1093/mnras/stx3112, 10.1093/mnras/stx3304}, \texttt{SIMBA}  \citep{10.1093/mnras/stz937} and \texttt{EAGLE} \citep{10.1093/mnras/stu2058, 10.1093/mnras/stv725} also show a comparable scatter of around 0.1 dex \citep{2020MNRAS.497..146D}. Although observational data at relevant high redshifts is not available, several hydrodynamical simulations (\texttt{EAGLE, IllustrisTNG, SIMBA, FIRE}) consistently suggest only mild evolution in the $M_{\rm HI}$-$M_{*}$ relation out to $z \sim 2\!-\!3$ \citep{2017MNRAS.464.4204C, 10.1093/mnras/stz937}. 
This minimal evolution arises because galaxy gas reservoirs at different epochs self-regulate, with increased gas inflows at higher redshift balanced by enhanced star formation rates and stronger feedback-driven outflows. 
Thus, the atomic hydrogen content of galaxies at fixed stellar mass remains relatively stable across cosmic time, justifying our approach. This relation extends previous studies on gas fraction trends, derived from a robust sample to $M_* \sim 10^9 M_{\odot}$. Other observations consistently show that the fraction of atomic gases increases with decreasing stellar mass \citep{Huang_2012, 2015MNRAS.452.2479B, 2017ApJS..233...22S, 10.1093/mnras/sty089}. The top right panel of Fig.~\ref{fig:scaling_rels} shows the assumed relation together with the XGASS results (gray points with errorbars).\\

The cold gas metallicity, $Z_{\rm cold}$, refers to the abundance of elements heavier than helium in a galaxy's cold gas phase, conventionally expressed as the ratio of the gas phase oxygen abundance to hydrogen. We adopt an empirical relation linking stellar mass, $M_*$,  and cold gas metallicity, $Z_{\rm cold}$, modified from the local observations by \citet{2004ApJ...613..898T}:
\begin{equation}
12 + \log_{10}(Z_{\text{cold}}) \;=\; 0.38\log_{10}(M_*/M_{\odot})+7.81+\epsilon, 
\end{equation}
where we retain the slope of the \citet{2004ApJ...613..898T} relation but lower its normalization to achieve better consistency with observational measurement at $z=2\!-\!3$ \citep{Erb_2006, mannucci2009lsd, 2011ApJ...729..140F,  2014ApJ...795..165S, sanders2015mosdef}. The stochastic term, $\epsilon$, is drawn from a Gaussian distribution with mean zero and standard deviation $\sigma=0.05$, corresponding to 0.05 dex scatter in $Z_{\text{cold}}$. In the bottom panel of Fig.~\ref{fig:scaling_rels}, we present our adapted scaling relation alongside observational measurements at various redshifts. 
Our choice of scatter (0.05 dex) is conservative and consistent with observational uncertainties. 
A comparison of our adopted relation with results from hydrodynamical simulations is presented in Appendix \ref{sec:app-scalerel}, further validating our approach. We treat the Fig.~\ref{fig:scaling_rels} relations as effective means that feed Eqs.~\ref{eq:vexp} - \ref{eq:tau_a}. If the true scalings have non-linear slopes or extra scatter, $(\alpha,\beta,\gamma)$ absorb normalization to first order, while residual non-linearities mainly appear as gentle luminosity function tilts and a slight change in effective host mass (and thus large-scale bias).\\

\begin{figure}
\centering
\includegraphics[scale=0.38]{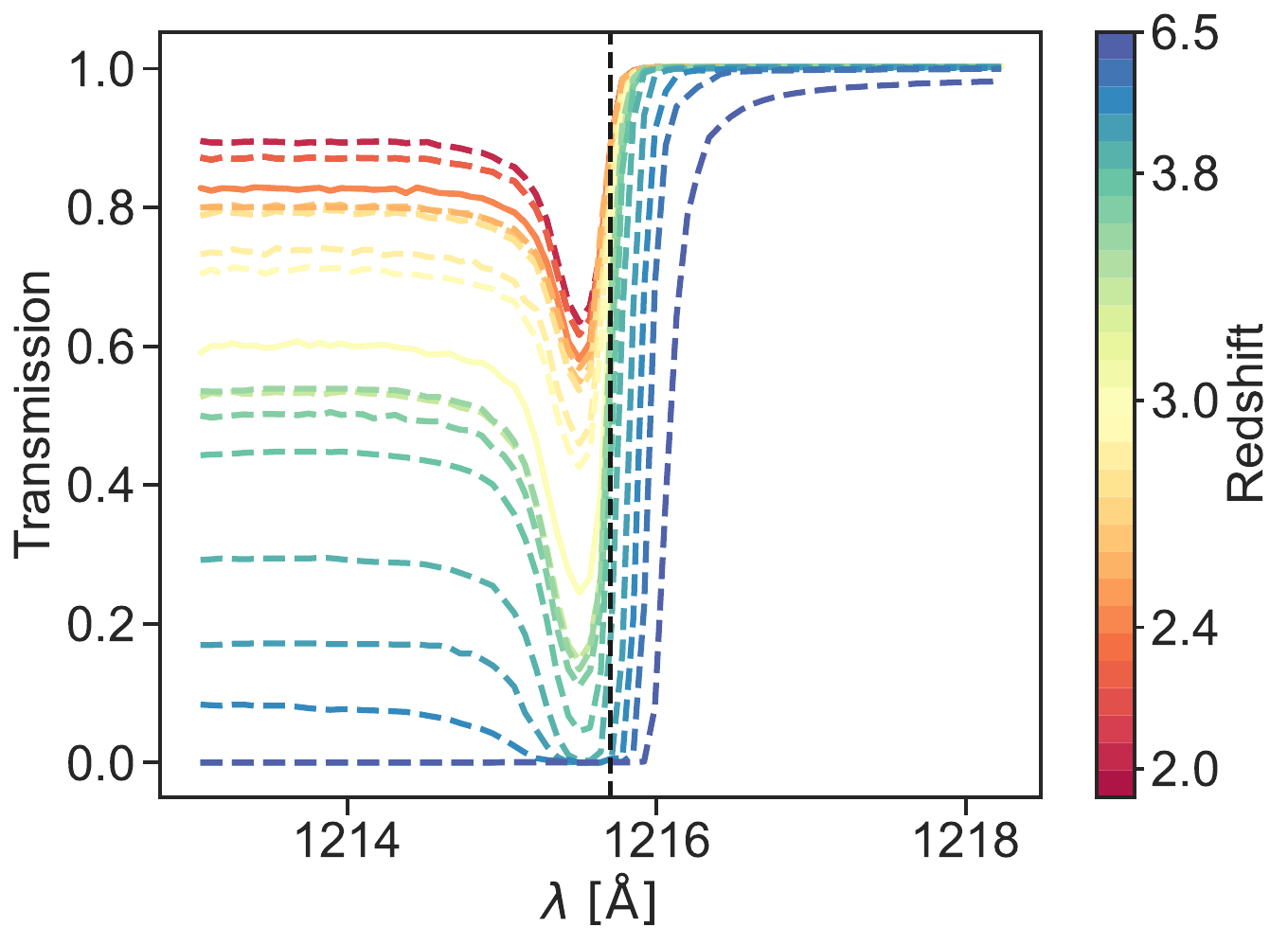}
\includegraphics[scale=0.38]{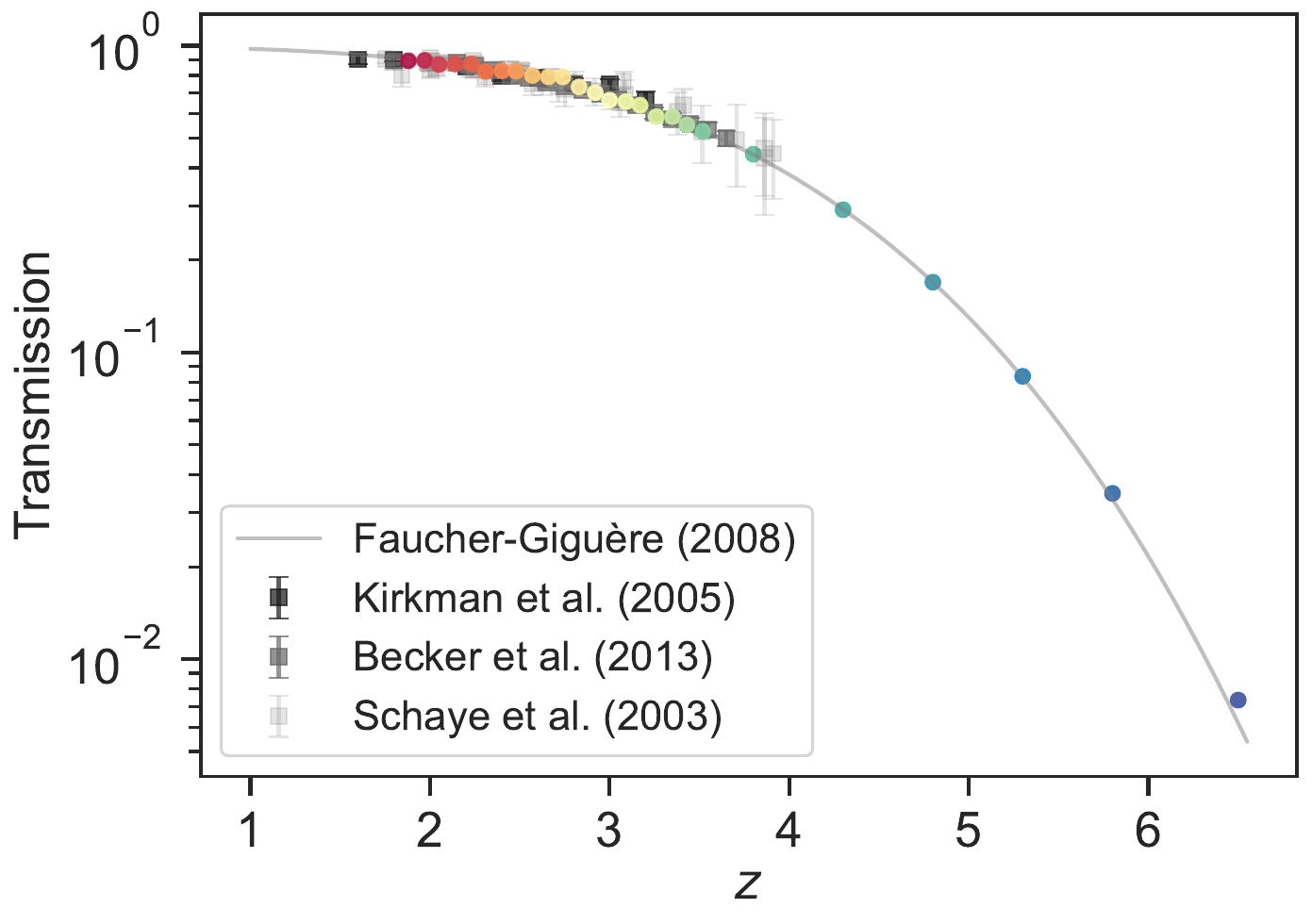}
\caption{
{\it Left}: The derived IGM transmission as a function of wavelength, color-coded for various redshifts. 
The vertical dashed line indicates the rest frame Ly$\alpha$ line center wavelength ($\lambda = 1215.7 \,\text{\AA}$). 
Dashed curves represent the transmission function originally provided by \protect\cite{Laursen:2011ApJ}, while solid curves are the ones adjusted by assuming that the mean transmission function blueward of the Ly$\alpha$ line follows the empirical relation represented in Eq.~(\ref{eq:T_z}).  
{\it Right}: A comparison between observations and the model used in this study, illustrating transmitted flux blueward of the Ly$\alpha$ line with respect to redshift. The solid line corresponds to Eq.~(\ref{eq:T_z}).}
\label{fig:transmission}
\end{figure}

\subsection{Modeling the Radiative Transfer of Lyman-\texorpdfstring{$\alpha$}{alpha} Photons within the Intergalactic Medium}
\label{subsec:igm}

To model the Ly$\alpha$ RT on IGM scales, we adopt a rather simple approach in this work. At $z\!\sim\!2$–3 we adopt a single, redshift-dependent mean IGM transmission. Simulations show that once convolved with typical red-peaked LAE line profiles, the angle-averaged Ly$\alpha$ transmission depends only weakly on halo mass or large-scale environment, while the emergent line’s velocity offset sets the dominant attenuation by the IGM (e.g. \cite{Laursen:2011ApJ,Byrohl:2020AA,10.1093/mnras/stab2762}; see also \citealt{Gurung-Lopez:2019mn_b}, their Fig.~6). In short, we adopt a \textit{fixed} transmission curve for all galaxies at a given redshift. 
Let us briefly review Ly$\alpha$ RT modeling on IGM scales before explaining our procedure.
In an optically thin limit, the impact of Ly$\alpha$ RT on IGM scales is calculated by the transmission curve, $T^{\rm IGM}(\lambda)$, and the optical depth, $\tau_{\rm IGM}(\lambda)$, given by \citep{Miralda-Escude_1993, Madau_1995, Dijkstra:2007mn, Laursen:2011ApJ}
\begin{equation}
\label{eq:tau_igm}
\begin{aligned}
    \tau_{\rm IGM}(\lambda) &= \int_{s_0}^{\infty} ds\, n_{\text{HI}}(s) \, \sigma \left[ \lambda \left( 1+\frac{v(s) + H(z)s}{c} \right), T_{\rm HI} \right], \\
    T^{\rm IGM}(\lambda) &= \exp \left\{-\tau_{\rm IGM}(\lambda)\right\},
\end{aligned}
\end{equation}
where the integration is performed in terms of the distance from the galaxy, $s$. 
$n_{\rm HI}$ is the neutral hydrogen number density, $\sigma(\lambda)$ represents the cross section of neutral hydrogen in the gas rest frame that accounts for the wavelength shift due to the peculiar velocity of a galaxy in the gas rest frame, $v$, and the Hubble flow $H(z)$ at redshift $z$. 
This equation highlights a complexity in modeling transmission due to $n_{\rm HI}$ and $v$.
Namely, we need to model both the neutral hydrogen gas density and velocity profiles at various scales. 
Previous studies used diverse approaches such as analytical models \citep{Zheng_2002, Dijkstra:2007mn} and hydrodynamic simulations \citep{Laursen:2011ApJ, 10.1111/j.1365-2966.2012.20486.x, 10.1093/mnras/stu1600, Gronke_2018, 2019MNRAS.484...39S, Byrohl:2020AA, 10.1093/mnras/stab990, 10.1093/mnras/stab2762, Park_2021, Smith:2022mn}. 
Other approaches include absorption statistics directly from observed quasar spectra \citep{2014MNRAS.442.1805I, 2021ApJ...908...36H} or the Fluctuating Gunn-Peterson Approximation that is commonly adopted in Ly$\alpha$ forest studies \citep[e.g.,][]{2000ApJ...543....1M, Zaldarriaga_2001, 10.1111/j.1365-2966.2004.08224.x, 2005ApJ...635..761M, 2006MNRAS.365..231V,  2006ApJS..163...80M, 10.1111/j.1365-2966.2011.18245.x, 10.1093/mnras/stw3372, 2017PhRvL.119c1302I,  Baur_2017, 2018PhRvD..98h3540M, 2019PhRvL.123g1102M, 2020ApJ...901..153D, Qezlou_2022, 2023JCAP...10..037B,10.1093/mnras/stad1920}.\\  

To model Ly$\alpha$ spectra, we need to model the wavelength dependence of the IGM transmission.
In fact, the transmission has the largest impact around the line center, which is typically due to a large neutral hydrogen density around dark-matter halos \citep{Laursen:2011ApJ}.
HI gas distribution even around the CGM and IGM scales is not still converged among various hydrodynamical simulations \citep[e.g.,][]{Faucher-Giguere_2016,2019MNRAS.482L..85V, Villaescusa-Navarro:2021ApJ}. Due to these complications, we simply adopt a fixed transmission curve motivated by previous simulation work in \citet{Laursen:2011ApJ} but adjust only the blue side of the transmission such that it becomes consistent with the observed mean transmission from \cite{Faucher-Giguère_2008, 10.1111/j.1365-2966.2005.09126.x, 2003ApJ...596..768S}.\\  

The transmission curves in \citet{Laursen:2011ApJ} were computed with $s_{0}=1.5 R_{\rm vir}$ and thus can be separated from our ISM/CGM RT calculation with the shell model. 
However, they were calculated at particular redshift output choices, and the blue side of their transmission curves were not fully consistent with the observed mean transmission whose redshift evolution is well described by the following form \citep{Faucher-Giguère_2008}:
\begin{align}\label{eq:T_z}
    \langle T^{\rm IGM}(\lambda=1213\text{\AA};z) \rangle =  \exp\left\{-1.330 \times 10^{-3} \times (1+z)^{4.094} \right\}.
\end{align}
In practice, we performed a Random Forest regression using a multi-output regressor to predict the IGM attenuation \citep{scikit-learn}. 
As training data, we compiled a grid of transmission curves, $T(\lambda)$, from \cite{Laursen:2011ApJ}, each corresponding to a fixed redshift in the range $z=1$ to $z=6.5$. 
Each input consists of the redshift value and the corresponding output as a vector of transmission sampled across the Ly$\alpha$ wavelength range. 
After formatting the data into a single matrix of shape $(N_{\rm redshift}, N_{\lambda})$, we trained the model to learn the mapping $z \rightarrow T(\lambda)$. Once trained, the regressor was used to predict transmission curves at intermediate redshifts. 
This procedure allows for a smooth interpolation of IGM attenuation across redshifts.
Fig.~\ref{fig:transmission} shows the transmission function at different redshifts as a function of the rest-frame wavelength.
We stress again that $f^{\rm IGM}_{\rm esc}$ depends on each galaxy property through the observed line profile $\phi^{\rm obs}(\lambda)=\phi^{\rm ISM}(\lambda;\vec{\theta}_{g})\times T^{\rm IGM}(\lambda;z)$, even though we fix the mean transmission curve at a given redshift.
One may concern that we do not fully model the transmission in each line-of-sight dependent way \citep[see e.g.,][]{Byrohl:2020AA}.
We take this simple approach because we calibrate our model against the Ly$\alpha$ luminosity function and the angular correlation function  where the impact of line-of-sight dependent transmission is expected to be minimal.
In future work, we plan to model the transmission curve in each line of sight and quantify the impact of the RT on the redshift-space clustering \citep{Zheng:2011ApJ,Behrens:2018AA}.\\ 

\subsection{Model calibration}
\label{subsec:calibration}

In this section, we explain how we calibrate the three dimensionless free parameters in our LAE model, $\alpha$, $\beta$ and $\gamma$, as introduced in Eqs.~(\ref{eq:vexp}), (\ref{eq:NHI}), and (\ref{eq:tau_a}). 
These parameters describe the scaling of the neutral hydrogen expansion velocity, $V_{\rm exp}$, neutral hydrogen column density, $N_{\rm HI}$, and the dust attenuation optical depth, $\tau_a$. 
The calibration process ensures that our model simultaneously reproduces both the observed Ly$\alpha$ luminosity function and the LAE angular correlation function at $z\sim 2$ and $z\sim 3$. These two summary statistics provide complementary information as the amplitude and shape of the luminosity function constrain the Ly$\alpha$ escape fraction as a function of intrinsic galaxy properties, while the LAE angular correlation function is sensitive to the properties of dark-matter halos.
We show in Appendix~\ref{app:free-params} a detailed discussion on parameter dependence of the luminosity function as well as the angular correlation function. As described in Sec.~\ref{sec:obs}, we calibrate our model using the luminosity function in \citet{Konno_2016} and the angular correlation function in \citet{White:2024JC} at $z\sim 2$, while we adopt the luminosity function in \citet{Ouchi_2008} and the angular correlation function in \citet{White:2024JC} at $z\sim 3$. 
We explore the full parameter space within the ranges $\alpha \in (0, 10]$, $\beta \in (0, 0.2]$ and $\gamma \in (0,100]$ that we empirically find sufficiently wide for our analysis.\\  

The luminosity function, denoted as $\Phi[\log_{10}(L_{\rm Ly\alpha})]$, for LAEs is commonly defined as their comoving number density per logarithmic luminosity bin \citep{1968ApJ...151..393S}: 
\begin{equation}\label{eq:lf_def}
   \Phi[\log_{10}(L_{\rm Ly\alpha})] = \frac{N(L_{\rm Ly\alpha})}{V_{\rm sim}\Delta \log_{10}(L_{\rm Ly\alpha}) },
\end{equation}
where $N(L_{\rm Ly\alpha})$ is the number of LAEs in a given logarithmic luminosity bin, $V_{\rm sim}=(400\,{\rm cMpc}\,h^{-1})^{3}$ is the comoving volume of the simulation, and $\Delta \log(L_{\rm Ly\alpha})$ is the logarithmic bin width. We measure the luminosity function using $100$ logarithmic bins spanning from $10^{41}$ to $10^{46}\, {\rm erg\,s^{-1}}$, although we fit only up to $10^{43}\,{\rm erg\,s^{-1}}$ to avoid the AGN contribution as explained in Sec.~\ref{sec:obs}.\\ 

The angular correlation function, $w(\theta)$, quantifies the clustering of galaxies by measuring the excess probability of finding a pair of sources at an angular separation $\theta$ compared to a random distribution \cite{1980lssu.book.....P}, defined by
\begin{align}\label{eq:tpcf_def}
dP_{12} = N(1+w(\theta_{12})) d\Omega_1\,d\Omega_2,
\end{align}
where $dP_{12}$ represents the probability of finding two LAEs within two solid angle regions, $d\Omega_1$ and $d\Omega_2$ with an angular separation of $\theta_{12}$ given the sample's number density $N$. To estimate $w(\theta)$, we employ \textsc{halotools}\footnote{\href{https://halotools.readthedocs.io/en/latest/}{https://halotools.readthedocs.io/en/latest/}} \citep{Hearin_2017}, using the \texttt{Landy-Szalay} estimator \citep{1993ApJ...412...64L}, expressed as:
\begin{align}\label{eq:LS}
w(\theta) = 1+ \left( \frac{N_R}{N_D} \right)^2 \frac{DD(\theta)}{RR(\theta)} -2 \left( \frac{N_R}{N_D} \right) \frac{DR(\theta)}{RR(\theta)},
\end{align}
where $DD$, $DR$, and $RR$ represent the number of data–data, data–random, and random–random LAE pairs, respectively with data and random galaxies having the exact same geometry in the sky. 
Furthermore, $N_R$ and $N_D$ denote the number of random and data (LAE), respectively. 
Note that we prepare random particles because the angular corelation function is somewhat sensitive to the radial selection function (i.e., the width of the redshift slice). 
We first divided the cubic simulation box into five separate subboxes, and then transformed the 3D Cartesian coordinates into the sky coordinates.
This transformation places all the galaxies in a common observer frame, ensuring a fair comparison with observational surveys. 
To further align our simulation with observational data, we apply different radial selection function. 
When comparing with \citet{White:2024JC}, at $z\sim2$ (for the N419 filter) and $z\sim3$ (for N501), we match their redshift distribution by adapting their redshift dependent selection function $dn/dz$ (see Fig.~3 in \citet{White:2024JC}) and randomly down-sample our LAEs and randoms accordingly. We also apply selection cuts to refine our sample, including a luminosity cut. To compute $w(\theta)$, we used $16$ logarithmic bins spanning from $1.5$ to $1505.6$ arcseconds that roughly correspond to the comoving separation of 0.02 and 20 cMpc$/h$ in these redshifts. 
Note that we adopt the same angular binning size as the measurements in \citet{White:2024JC}. 
We repeat these processes for each subbox and average the measurements $w(\theta)$ in the five subboxes.\\ 

\begin{figure}
\includegraphics[scale=0.36]{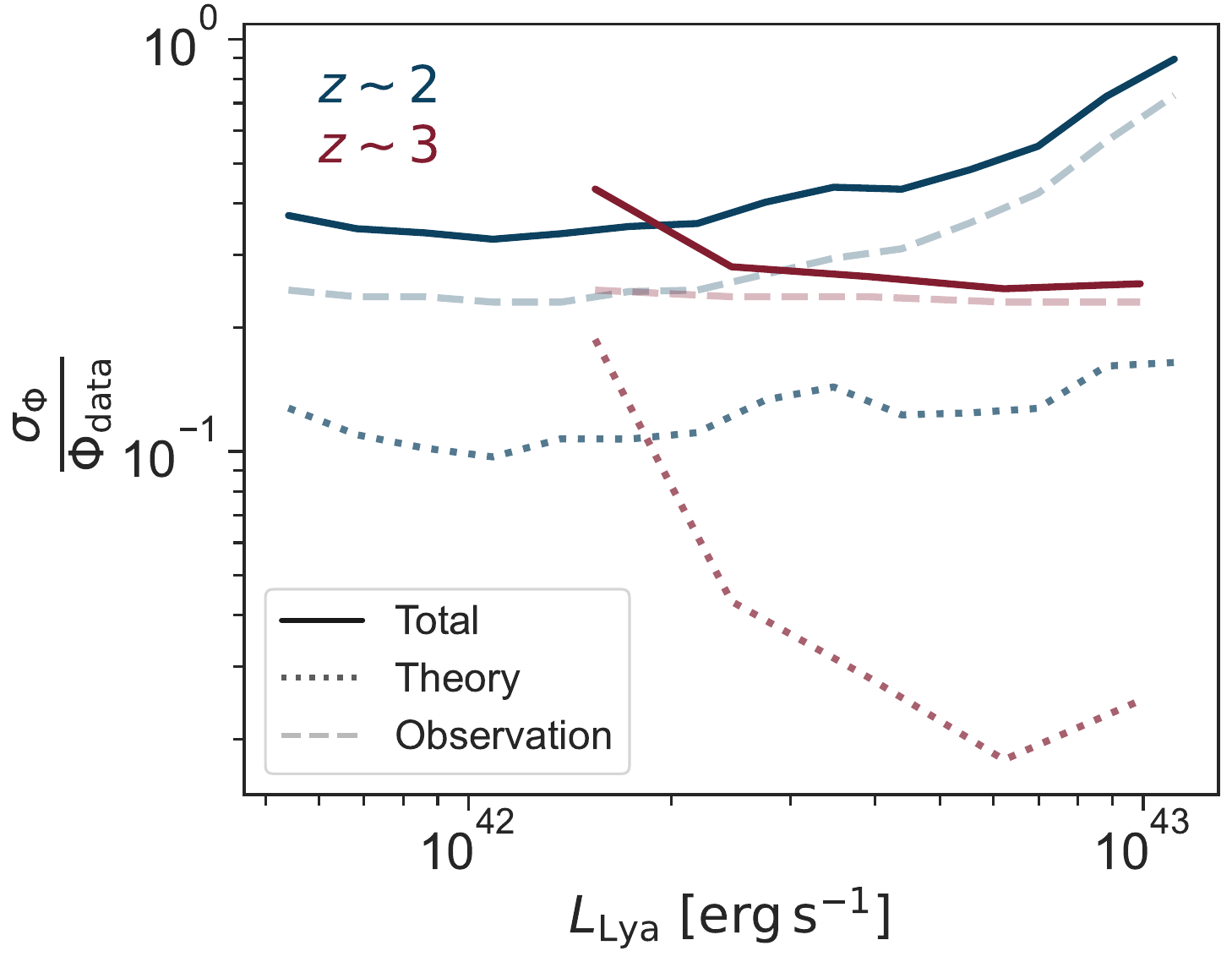}
\includegraphics[scale=0.36]{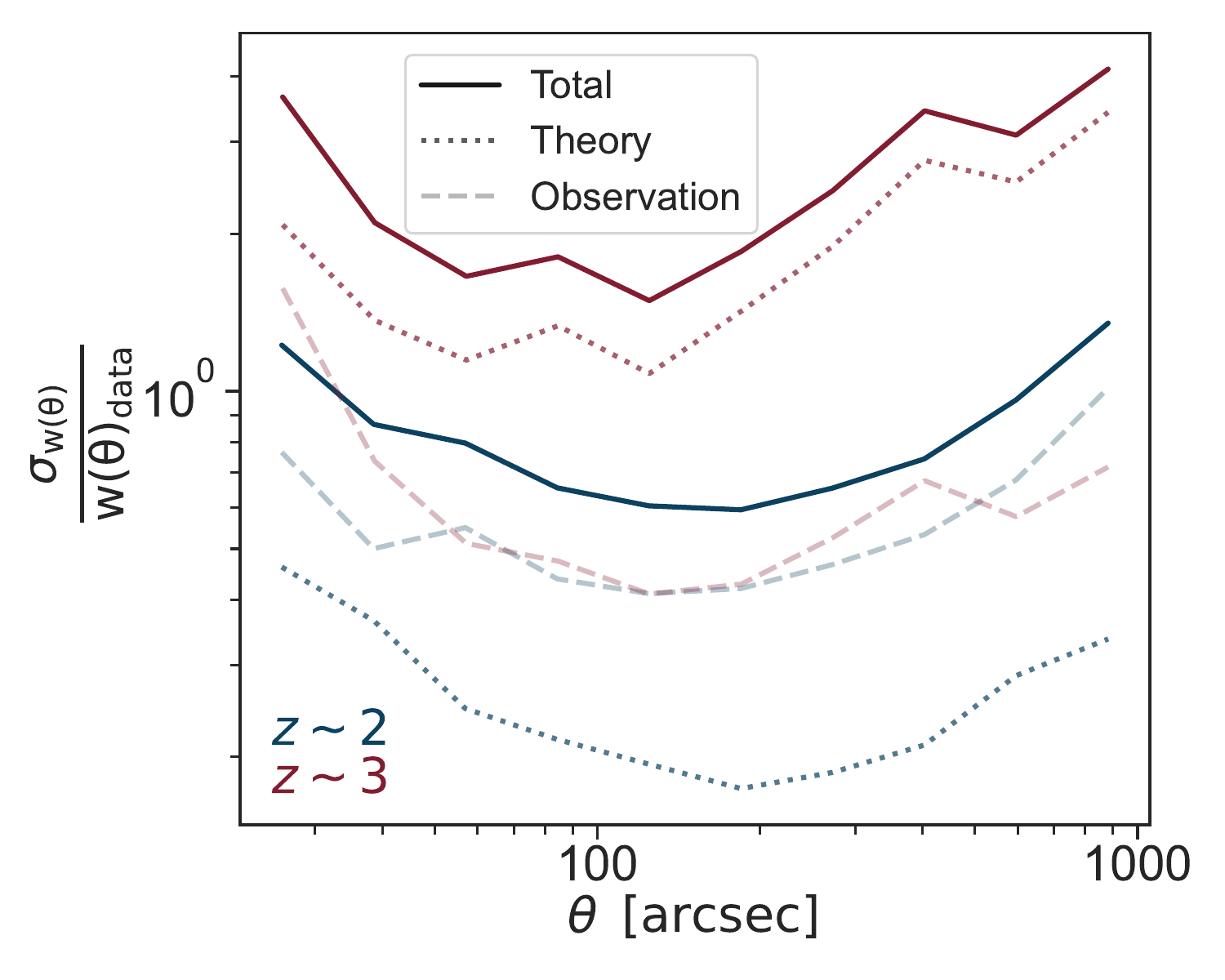}
\caption{
Relative error budgets for Ly$\alpha$ luminosity function and the angular correlation function at both redshifts. 
\textit{Left:} Relative uncertainty in the luminosity function as a function of Ly$\alpha$ luminosity for $z\sim2$ in blue and $z\sim3$ in red. \textit{Right:} Relative uncertainty in the angular correlation function as a function of angular scales with the same color scheme. 
In both panels, dotted curves show the theory errors from the jackknife covariance matrices, dashed curves show the observational errors from the fiducial luminosity function (left) and angular correlation function (right) measurements, and solid curves give the total error.}
\label{fig:covmat}
\end{figure}

Due to the limited simulation box size ($L_{\rm box}=400\,{\rm cMpc}\,h^{-1}$), we estimate the theoretical covariance matrix of the luminosity function and the angular correlation function from our simulation.
We adopt the jackknife (JK) resampling method \citep{2009MNRAS.396...19N} given by 
\begin{align}\label{eq:covmat_JK}
    C^{\rm th}_{ij} = \dfrac{(N_{\rm JK} - 1)}{N_{\rm JK}} \sum_{k=1}^{N_{\rm JK}} (x_i^k - \bar{x}_i ) (x_j^k - \bar{x}_j ).
\end{align}
The total sample is divided into $N_{\rm JK} = 50$ jackknife subregions only in the RA-Dec positions with the \textsc{k-means}\footnote{\href{https://github.com/esheldon/kmeans_radec?tab=readme-ov-file}{https://github.com/esheldon/kmeans-radec}} algorithm \citep{2017MNRAS.464.4045K}.  
$x_i^k$ represents the measurements of the summary statistics in the $i$-th bin, computed using all jackknife regions except the $k$-th one, while $\bar{x}_i$ is the mean of all resamplings. 
We use the best-fitting model that we will show in the next section to evaluate the jackknife theoretical errors. In Fig.~\ref{fig:covmat}, we show the estimated theoretical errors (dotted lines) for the luminosity function (left) and the angular correlation function (right), and compare them with the observation measurement (dashed lines) and the total errors (solid lines) at $z\sim 2$ (blue) and $z\sim 3$ (red). 
For the luminosity function, the theoretical errors are subdominant, since our simulation volume is much larger than the observed ones.
Meanwhile, for the angular correlation function, the theoretical error is not negligible, and larger than the measurement error at $z\sim 3$. 
Our simulation volume is larger than that of \citet{White:2024JC} by a factor of $\sim 2.6$ at both $z\sim 2$ and $z\sim 3$. 
In addition, we downsample our LAEs to match the selection function in \citet{White:2024JC} to ensure the same number density. 
Therefore, the larger theoretical error at $z\sim 3$ could be explained by the overestimated cosmic variance of the jackknife method. 
We leave the investigation of this issue with a larger box simulation for future work.
We also find the off-diagonal elements of the covariance matrix too noisy to stably conduct the parameter estimation, and thus ignore them for simplicity.\\ 

We determine the best-fitting free parameters and their uncertainties by calculating $\chi^2$ (or the likelihood function $\exp(-\chi^2)/2$) given by
\begin{align}\label{eq:chi2}
    \chi^2 = \sum_i^n \dfrac{(x_i^{\rm obs} - x_i^{\rm th})^2}{(\Delta x_i)^2},
\end{align}
where $x_{i}$ denotes either the luminosity function or the angular correlation function at $i$-th bin, and $(\Delta x)^2 = (\Delta x^{\rm obs})^2 + C^{\rm th}_{ii}$ combines uncertainties from the observations and theoretical predictions.
As noted above, we ignore the correlation among different bins, and thus the total $\chi^2$ is simply the sum of two contributions from the luminosity function and the angular correlation function, $\chi^{2}_{\rm tot}=\chi^{2}_{\rm LF}+\chi^{2}_{\rm ACF}$.

\begin{figure}[t!]
\includegraphics[scale=0.42]{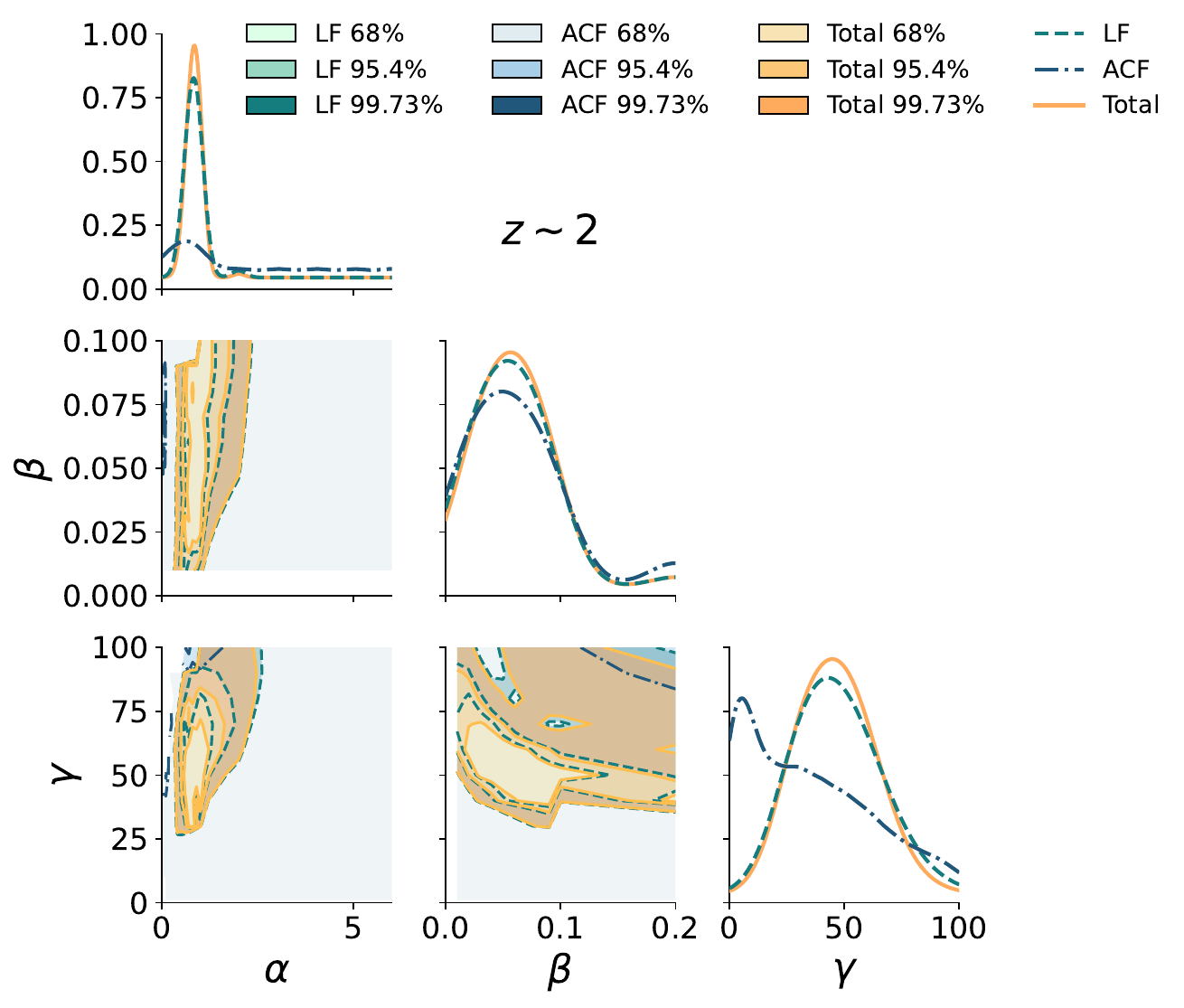}
\includegraphics[scale=0.42]{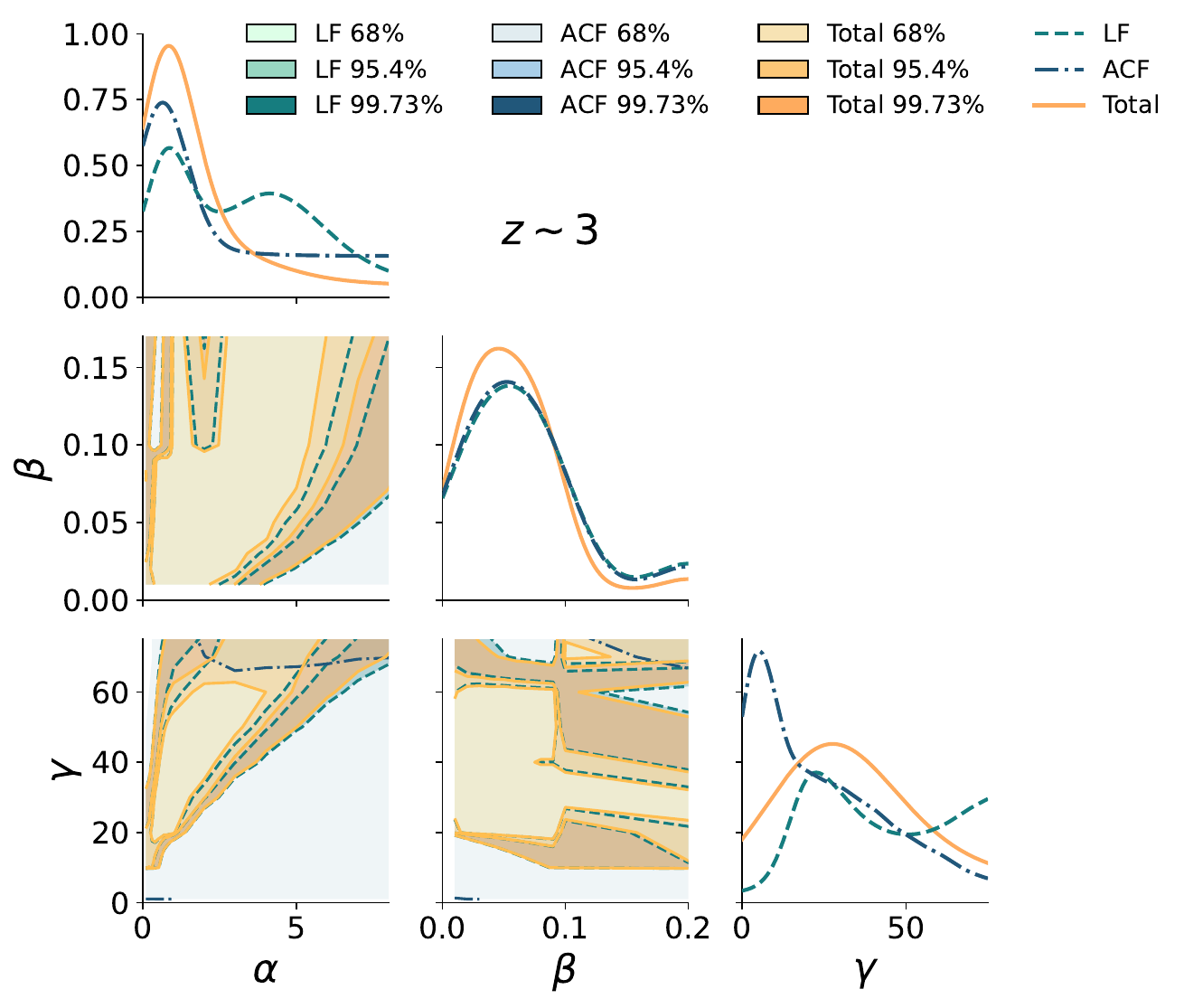}
\caption{
Corner plots showing the parameter constraints for $\alpha$, $\beta$ and $\gamma$ at $z\sim2$ (\textit{left}) and $z\sim3$ (\textit{right}).
The contours represent the 1$\sigma$ ($68\%$), 2$\sigma$ ($95.4\%$), and 3$\sigma$ ($99.73\%$) confidence intervals, with darker colors corresponding to the higher-confidence regions. 
The parameter constraints are derived from fitting the Lyman-$\alpha$ luminosity function (LF, green), the angular correlation function (ACF, blue), and the combined total fit (yellow). 
We also show the one-dimensional likelihood function, $\exp(-\chi^2/2)$, marginalized over other two parameters and normalized to one at those best-fitting values.}
\label{fig:contours}
\end{figure}

\begin{table}[t!]
\refstepcounter{table}\label{Table:data-fit}

  {\small\textsc{Table}~\thetable.\textemdash\ }%
  {\footnotesize
    Summary of best‐fit parameters at two redshifts.  The first column lists each quantity, the
    second column shows its value at \(z\sim2\), and the third column shows the corresponding value at \(z\sim3\).
  }\medskip

  \centering
  \small
  \begin{tabular}{c|c|c}
    Redshift 
      & \(z \sim 2\) 
      & \(z \sim 3\)\\
    \hline\hline
    Luminosity Function ($N_{\rm bin}$)
      & \citet{Konno_2016} ($N_{\rm bin}=13$)
      & \citet{Ouchi_2008} ($N_{\rm bin}=5$) \\
    \hline
    Angular Correlation Function \(\bigl[f_{\rm cont}\bigr]\) ($N_{\rm bin}$)
      & \citet{White:2024JC}\,[8\%]  ($N_{\rm bin}=10$)
      & \citet{White:2024JC}\,[11\%] ($N_{\rm bin}=10$)\\
    \hline
    $\chi^{2}_{\rm tot} = \chi^{2}_{\rm LF}+\chi^{2}_{\rm ACF}$ 
      & $4.71 = 2.00 + 2.71$
      & $1.90 = 0.06 + 1.84$ \\
    \hline
    \(\bigl(\alpha,\ \beta,\ \gamma\bigr)\) 
      & $\bigl(0.90^{+0.21}_{-0.16},\;0.07\pm 0.03,\;50.00^{+16.33}_{-17.94}\bigr)$  
      & $\bigl(0.80^{+0.61}_{-0.52},\;0.04 \pm 0.03,\;30.00^{+23.21}_{-10.01}\bigr)$ \\
    \hline
    \(\overline{M}_{h}\) \(\Bigl(\log (L/\mathrm{erg\,s^{-1}}) \geq 41.7, \,\, \mathrm{EW}_0 > 40 \,\, \text{\AA} \,\Bigr)\,\,\mathrm{M_{\odot}}/h\) 
      & \(1.12 \times 10^{11}\)
      & \(1.41 \times 10^{11}\)  \\
    \hline
    \(f_{\rm sat}\) \(\Bigl(\log (L/\mathrm{erg\,s^{-1}}) \geq 41.7, \,\, \mathrm{EW}_0 > 40 \,\, \text{\AA} \,\Bigr)\) 
      & \(12.37\%\)
      & \(8.86 \%\) \\
    \hline
    \(\overline{M}_{h}\) \(\Bigl(\log (L/\mathrm{erg\,s^{-1}}) \geq 42.22, \,\, \mathrm{EW}_0 > 40 \,\, \text{\AA} \,\Bigr)\,\, \mathrm{M_{\odot}}/h\) 
      & \(2.06 \times 10^{11}\)  
      & \(1.55 \times 10^{11}\) \\
    \hline
    \(f_{\rm sat}\) \(\Bigl(\log (L/\mathrm{erg\,s^{-1}}) \geq 42.22, \,\, \mathrm{EW}_0 > 40 \, \, \text{\AA} \,\Bigr)\) 
      & \(10.06 \%\)
      & \(8.67 \%\) \\
  \end{tabular}
\end{table}

\section{Results}
\label{sec:result}

In this section, we present the results of our LAE model calibration and compare our simulated LAEs with observational measurements at $z\sim2$ and $z\sim3$. 
This section is organized as follows. 
We first describe the best-fit parameters obtained from our modeling approach and discuss their impact on the inferred Lyman-$\alpha$ properties. 
We then compare the observational LAE properties, such as Lyman-$\alpha$ escape fractions, equivalent width distributions, and observed Lyman-$\alpha$ velocity offsets, with our model predictions. 
Lastly, we present the implications and predictions of our LAE model, including the escape fraction and the LAE-halo connection. 
We will investigate the relationship between LAEs and their host dark-matter halos, specifically predicting the Halo Occupation Distribution (HOD) and potential assembly bias signatures.

\begin{figure}
\centering
\includegraphics[scale=0.35]{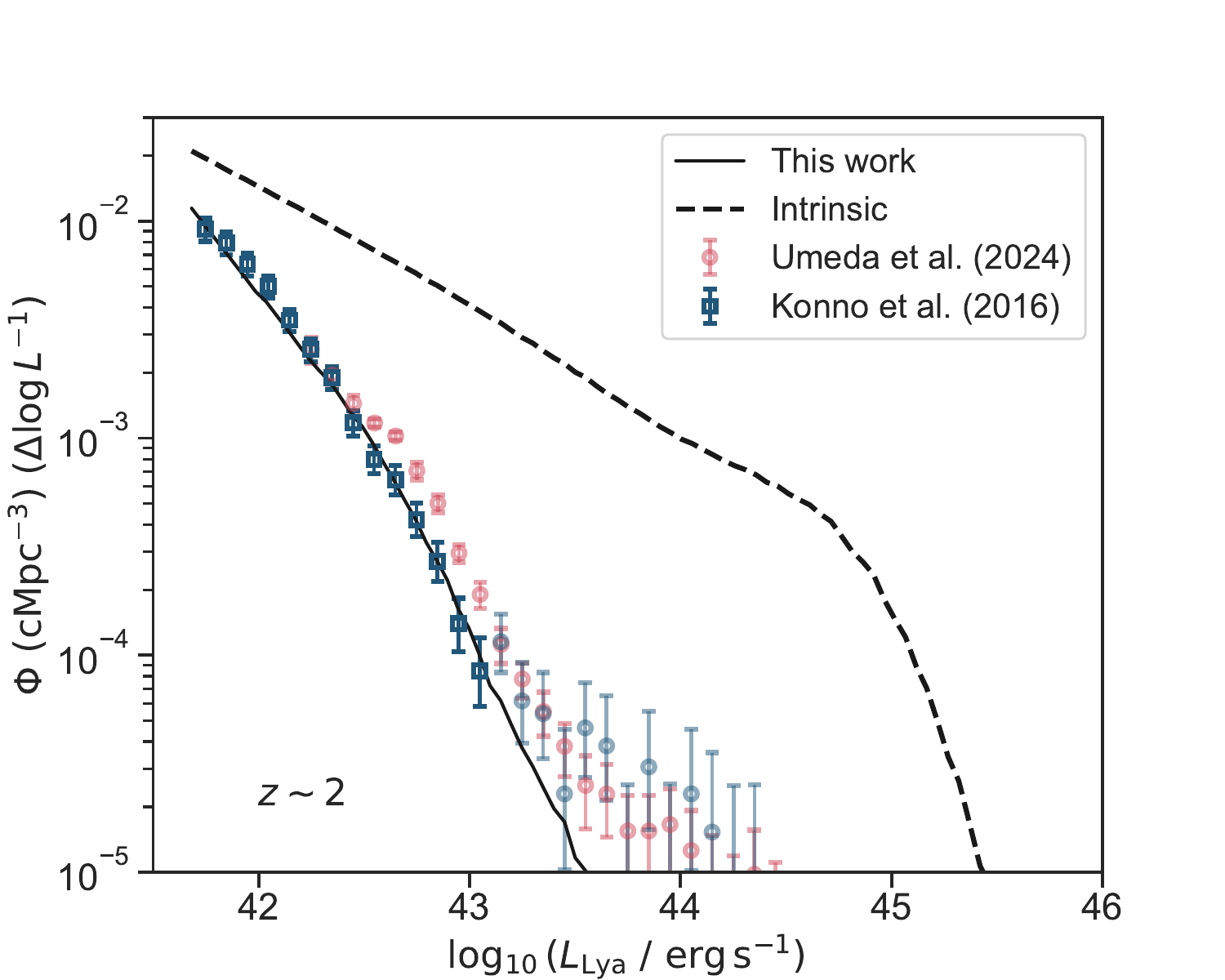}
\includegraphics[scale=0.35]{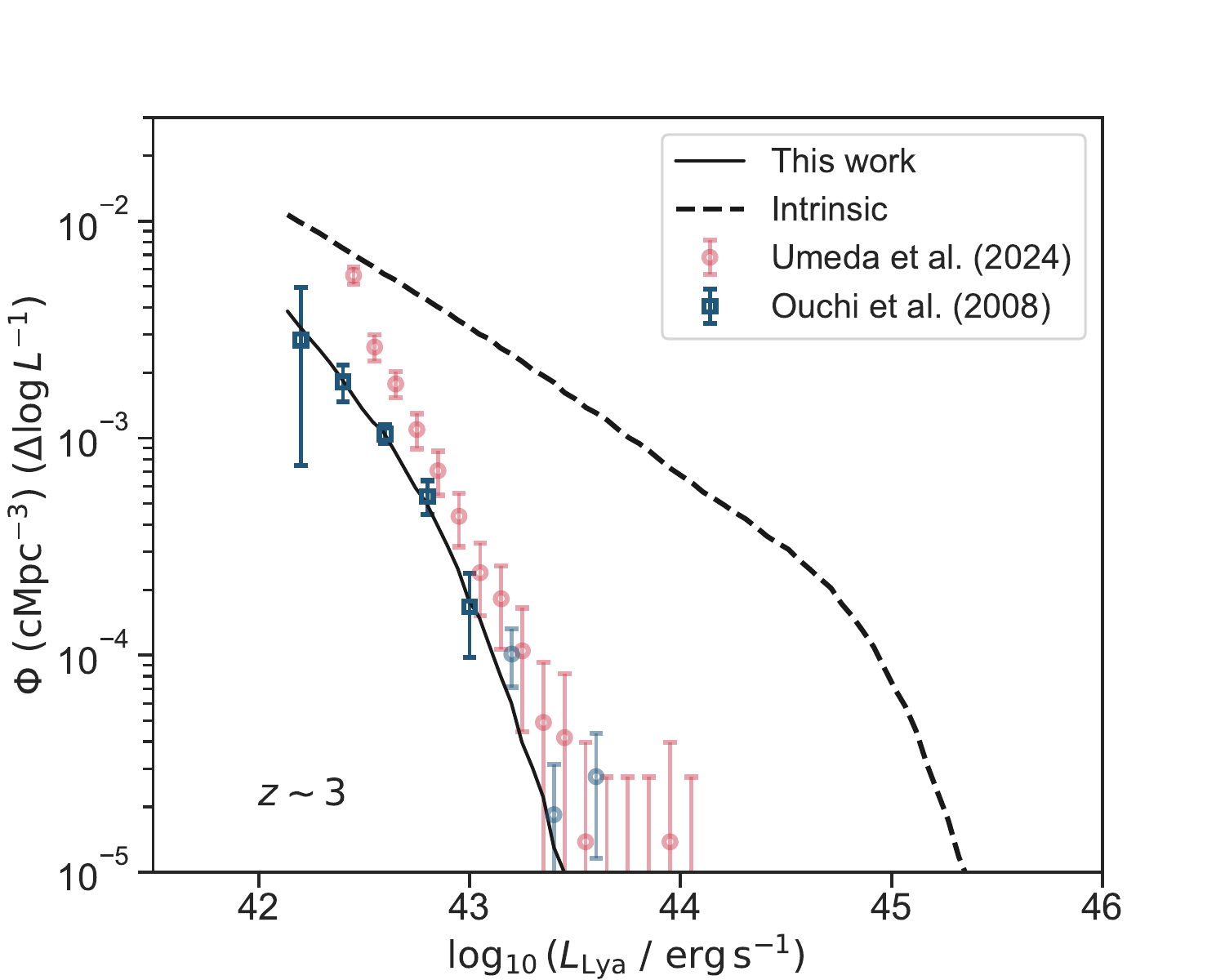}
\caption{
Lyman-$\alpha$ luminosity function for our best-fit model at $z\sim2$ (left panel) and $z\sim3$ (right panel). 
The solid black line shows our best-fit model, while the black dashed line indicates the intrinsic luminosity function as computed from Eq.~(\ref{eq:L_int}). 
The blue data points represent the observational data at each redshift, while 
the fainter blue data points are not used to avoid contamination from AGNs (see Sec.~\ref{sec:obs}).
For comparison, we also show the measurements in \cite{Umeda:2025ApJS} (faint red points). 
}
\label{fig:LF}
\end{figure}

\begin{figure}
\centering
\includegraphics[scale=0.35]{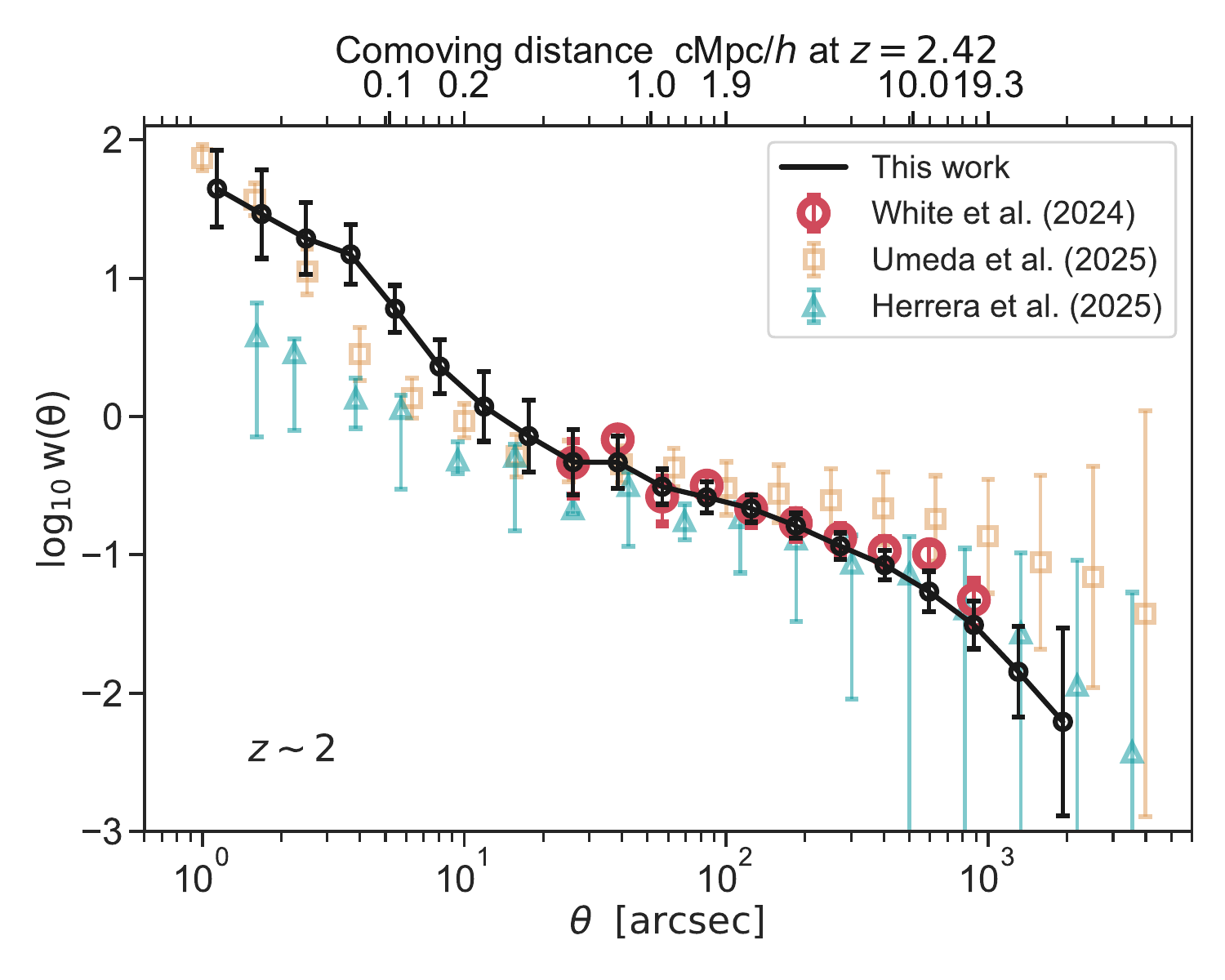}
\includegraphics[scale=0.35]{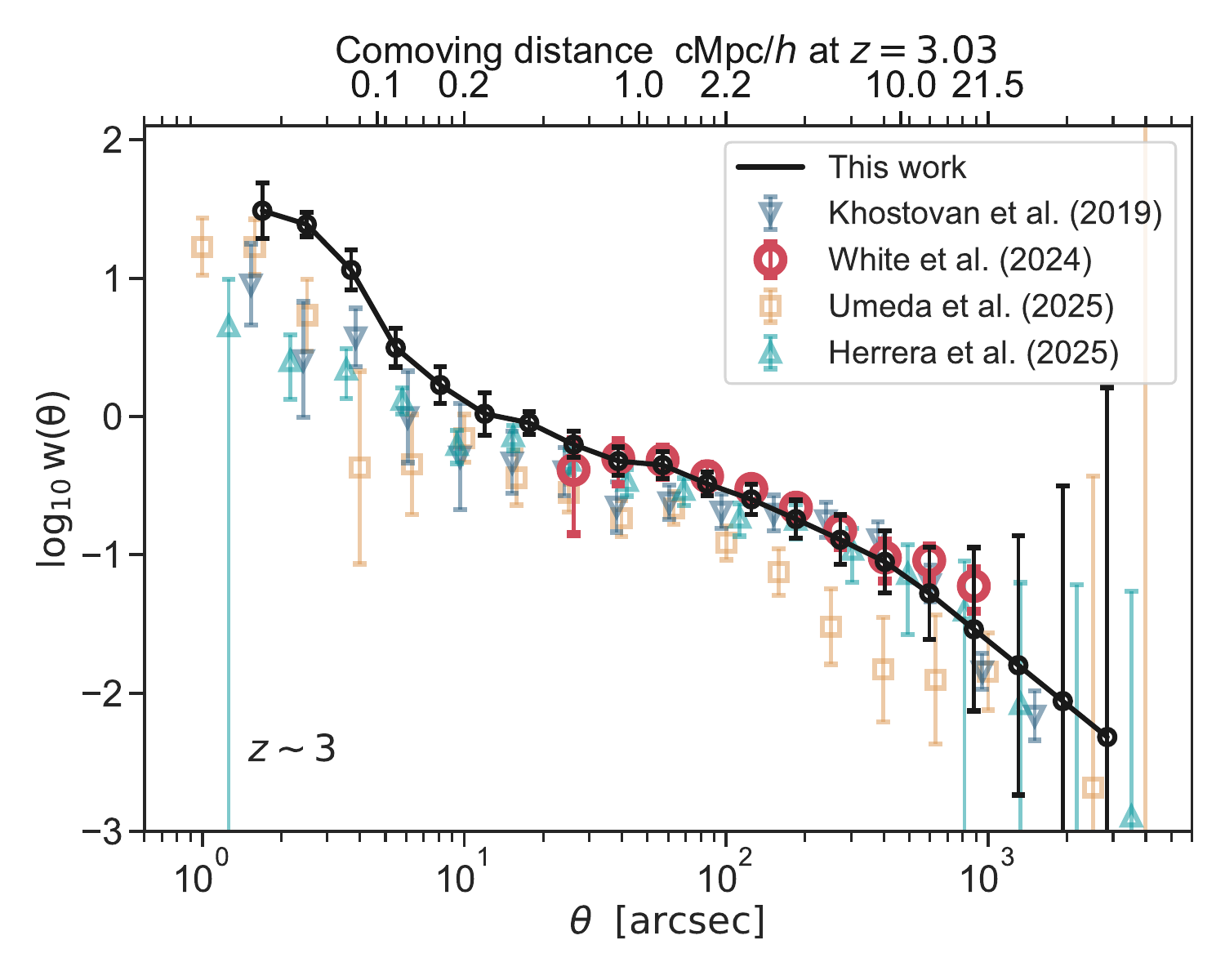}
\caption{
Angular correlation function, $w(\theta)$, predictions for our fiducial model compared with observational data at $z\sim 2$ ({\it left}) and at $z\sim 3$ ({\it right}). 
In both panels the black curve shows the best‐fit model with the \cite{White:2024JC} dataset, which incorporates the redshift distribution $(dn/dz)$, luminosity threshold, and contamination fraction. For reference, we compare other observational measurements; \cite{Khostovan:2019mn} (blue triangle) at $z\sim 3$, \cite{Umeda:2025ApJS} (orange square) and \cite{Herrera:2025ar} (green triangle) at both $z\sim 2$ and $z\sim 3$.
Note that $w(\theta)$ is highly sensitive to the redshift interval since it measures the projected clustering in RA–Dec; a broader redshift range would result in a different projection and clustering amplitude. 
Error bars in each bin were estimated from the diagonal elements of the covariance matrix computed via jackknife resampling.
}
\label{fig:CF}
\end{figure}

\subsection{Parameter constraints and model fitting}
\label{sec:fit}

Fig.~\ref{fig:contours} displays the resulting parameter constraints and the likelihood (normalized to one at the best-fit values) for $\alpha, \beta$ and $\gamma$. The contours correspond to the $1 \sigma$ ($68\%$), $2 \sigma$ ($95.4\%$) and $3 \sigma$ ($99.73\%$) confidence intervals. 
Darker colors indicate higher confidence intervals. We notice that the angular correlation function weakly constrains the parameter space. This is expected since the angular correlation function has larger errors than the luminosity function in our setup (see Fig.~\ref{fig:covmat}). 
In addition, the angular correlation function is primarily sensitive to the halo bias, and thus our results show that the LAE clustering only does not constrain a detailed model of the Lyman-$\alpha$ escape fraction. 
Meanwhile, the Ly$\alpha$ escape fraction can be well constrained by the amplitude and shape of the luminosity function. 
We confirm expected degeneracies among the three free parameters constrained by the luminosity function. 
A larger escape fraction can be achieved by either a larger $\alpha$, a smaller $\beta$, or a smaller $\gamma$, as discussed in Sec.~\ref{sec:lae_model}. 
However, $\alpha$ and $\gamma$ are relatively well constrained by the luminosity function, since $\alpha$ mainly changes the slope, while $\gamma$ changes the overall amplitude as demonstrated in Fig.~\ref{fig:app_LF_evol} in Appendix~\ref{sec:app-LF-abc}. 
$\beta$ is the parameter most weakly constrained by both the luminosity function and the angular correlation function, while it is essential to have $\beta$ in this range to recover the observed velocity offset measurement, as we discuss in the next subsection.\\ 

We present the best-fitting values of the model parameters and the goodness of fit, $\chi^{2}$, in Table.~\ref{Table:data-fit}. 
Although we find that our best-fitting model is consistent with the observed measurements, the $\chi^{2}$ values tend to be small; we obtain $\chi^{2}_{\rm tot}=4.7$ for $20=13+10-3$ degrees of freedom at $z\sim 2$. 
This is most likely because our theory errors are overestimated and/or due to the fact that we ignored the correlation among different bins for both theoretical and observational measurements.  
Nevertheless, it is highly encouraging that the best-fit results of $(\alpha,\beta,\gamma)$ of two redshifts are consistent within these uncertainties. 
The best-fitting $(\alpha,\beta,\gamma)$ at $z=2.4$ and $z=3$ are statistically consistent within their credible intervals, which is encouraging for the stability of our effective mapping across $z\!\sim\!2$--3. However, given the proximity of these redshifts and current uncertainties, this agreement should be viewed as suggestive rather than conclusive; firmer validation of the Sec.~3.3 scalings will require a wider redshift baseline and additional observables (e.g., EW/line-profile) to break degeneracies. This fact also justifies our various assumptions in the scaling relations (see Sec.~\ref{subsec:scale}), since it does not require an additional redshift evolution.\\

We compare our best-fit predictions of the luminosity function in Fig.~\ref{fig:LF}.
In both left $(z\sim2)$ and right $(z\sim3)$ panels of Fig.~\ref{fig:LF}, our best-fit model consistently traces the fitted observational data points (thick blue points). 
Our prediction is clearly underestimated at $\log_{10}(L_{\rm Ly\alpha}\, /{\rm erg\,s^{-1}}) > 43.0$ compared to the thick blue points, which could be explained by the contribution of AGNs. 
We also show the intrinsic Ly$\alpha$ luminosity function (dashed line), whose difference from the best-fit model is the consequence of our escape fraction modeling. 
For comparison, we overplot another observational measurement from \cite{Umeda:2025ApJS}.
Interestingly, there are non-negligible discrepancies between the two measurements. 
In Appendix.~\ref{sec:app-umeda}, we also show the results fitted to the luminosity function in \cite{Umeda:2025ApJS}.\\ 

The comparison of the angular correlation function is also presented in Fig.~\ref{fig:CF}. 
At both $z\sim 2$ (left) and $z\sim 3$ (right), our best-fitting model is consistent with the observational data in \cite{White:2024JC} (large red circle points). 
Notice that the observational data in \cite{White:2024JC} only cover the two-halo regime, since a typical virial radius of LAEs is given by $\sim 200\,{\rm pkpc}/h$ (see Fig.~\ref{fig:scaling_rels}), which roughly corresponds to 30 and 40 arcseconds at $z\sim 2$ and $z\sim 3$, respectively. 
In this sense, the angular correlation function in the 1-halo regime is purely a prediction of our best-fitting model. 
For reference, we compare other observational measurements; \cite{Khostovan:2019mn} (blue triangle) at $z\sim 3$, \cite{Umeda:2025ApJS} (orange square) and \cite{Herrera:2025ar} (green triangle) at both $z\sim 2$ and $z\sim 3$.
It may seem that our model overestimates the 1-halo term in comparison to these measurements, but we stress that this is not a fair comparison, since the different measurements have different selection criteria in terms of luminosity threshold and radial selection function. 
As an example, we show in Appendix.~\ref{sec:app-khostovan} that we can reproduce the \cite{Khostovan:2019mn} measurements (including the 1-halo term regime) with a different set of free parameters.\\

\subsection{Other observational supports}
\label{sec:other_obs_supports}

\begin{figure}
\centering
\includegraphics[scale=0.46]{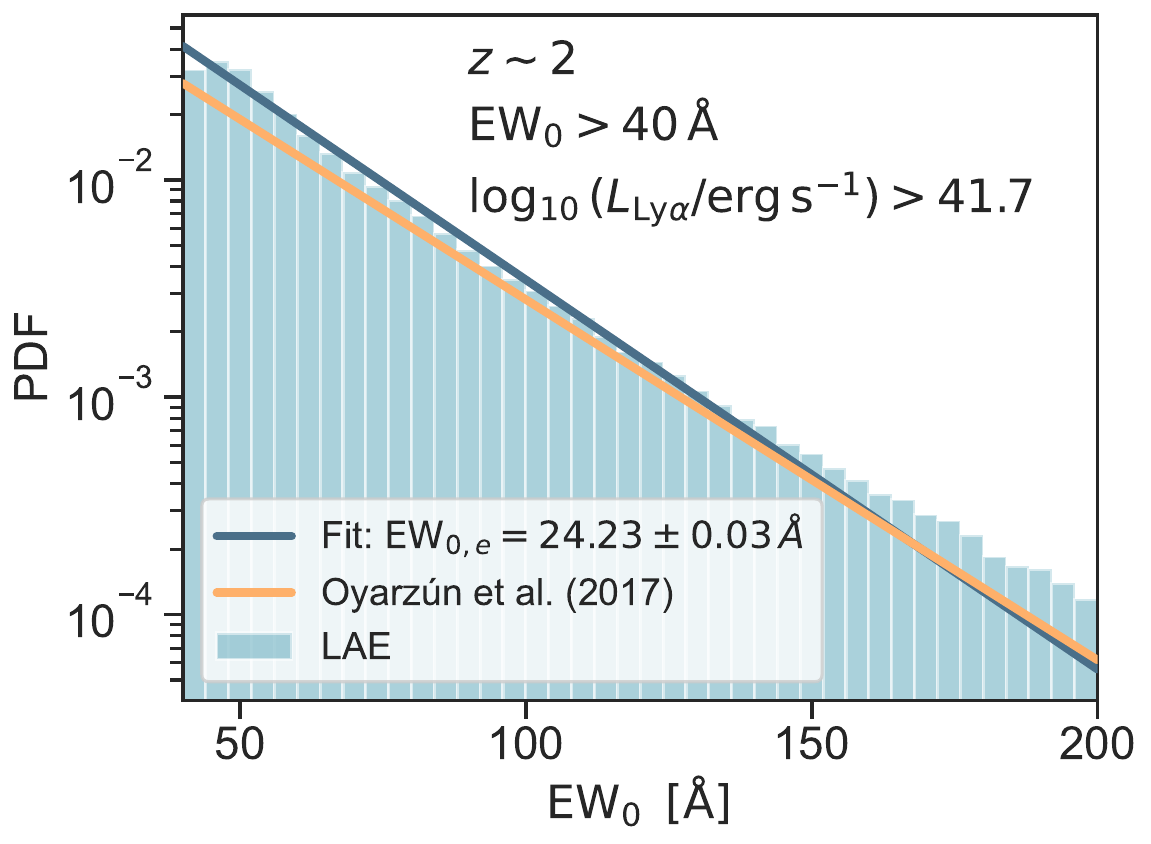}
\includegraphics[scale=0.46]{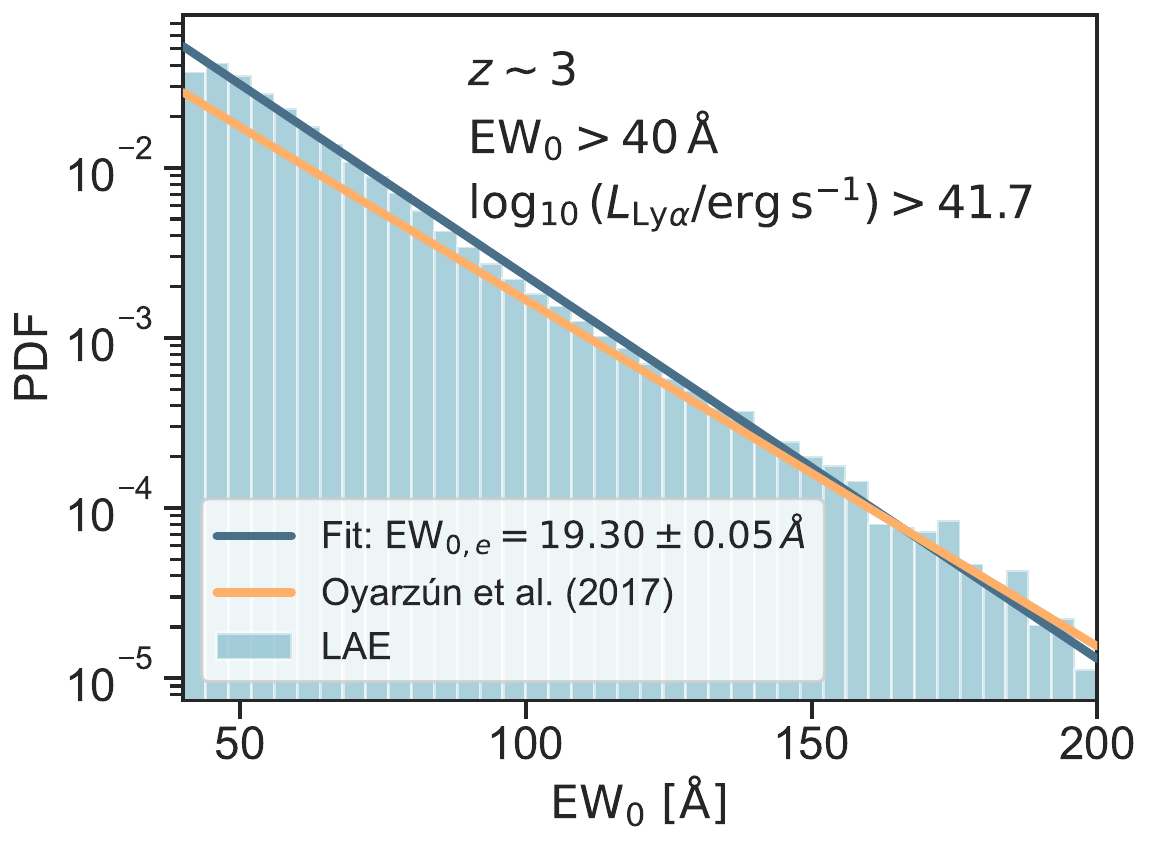}
\caption{
The probability density function (PDFs) of the rest-frame Ly$\alpha$ equivalent widths ($\mathrm{EW}_0$) for our fiducial LAE samples at $z\sim2$ ({\it left}) and $z\sim3$ ({\it right}).  
Histograms show the $\mathrm{EW}_0$ distributions for LAEs with $40 \mathrm{\AA} < \mathrm{EW}_{0} < 200,\mathrm{\AA}$ and a luminosity cut of $\log_{10}(L_{\mathrm{Ly}\alpha}/\rm erg\,s^{-1})>41.7$, all normalized to unit area.
The dark blue solid line represents the best-fit exponential function, $P(\mathrm{EW}_{0}) = (A/\mathrm{EW}_{0,e})\exp(-\mathrm{EW}/\mathrm{EW}_{0,e})$, to our LAE model.  
The orange solid line is the \cite{Oyarzún_2017} exponential fit, re-normalized over the same $\mathrm{EW}_0$ range.
}
\label{fig:EW_dist}
\end{figure}

We have shown that our LAE model successfully reproduces the Ly$\alpha$ luminosity function and the angular correlation function at $z\sim 2$ and $3$. 
However, a natural question is if our best-fit parameters of $(\alpha,\beta,\gamma)$ physically make sense. 
Since they are all parameters that are supposed to only effectively capture Ly$\alpha$ RT physics, it is not straightforward to interpret the direct physical meaning of the best-fit values obtained.
Instead, we will check the ability of our LAE model to match additional observables, thereby providing further validation. Note that it is not straightforward to make a fair comparison for most of the following observables, since a different data set has different LAE selection criteria. 
We stress that we did not fit the observables in this subsection to derive our best-fit LAE model.\\  

A relation between Ly$\alpha$ emissions and UV luminosity can be used as a diagnostic to check how LAEs are populated in the entire young star-forming galaxy population. 
Fig.~\ref{fig:EW_dist} presents the rest-frame Ly$\alpha$ equivalent width (EW$_0$) distribution for our simulated LAEs at redshifts $z\sim2$ and $z\sim3$. 
In both panels, the histograms indicate a steep exponential decline, well described by a normalized probability density function of the form $P(\mathrm{EW}_{0}) = (1/\mathrm{EW}_{0,e})\exp(-\mathrm{EW_0}/\mathrm{EW}_{0,e})$ (dark blue line). We find a characteristic scale $\mathrm{EW}_{0,e} = 24.23 \pm 0.03 \text{\AA}$ at $z\sim2$ and $\mathrm{EW}_{0,e} = 19.30 \pm 0.05 \text{\AA}$ at $z\sim3$, indicating most LAEs have moderate equivalent widths. For comparison the orange solid line show the exponential distribution from \cite{Oyarzún_2017} ($\mathrm{EW_{0,e}}\sim 38\,\, \text{\AA}$ for their $M_*$-selected sample at $z\sim 3-4.6$), re-normalized over $40 < \mathrm{EW_0} < 200\,\, \text{\AA}$ to match our histogram's range. 
The exponential form and scale of our prediction are in good agreement with observations, though minor discrepancies (slight over-estimation at low $\mathrm{EW_0}$ and underestimation at high $\mathrm{EW_0}$) are expected given sample differences (Ly$\alpha$-luminosity selected sample in this work vs. mass-selected sample in \cite{Oyarzún_2017} as well as the redshift range differences) and observational limitations (relatively small size of \cite{Oyarzún_2017} sample) increasing fit uncertainties. 
Too many small EW objects could be explained by the fact that we include LAEs down to relatively small Ly$\alpha$ luminosity, $\log_{10}(L_{\mathrm{Ly}\alpha}/\rm erg\,s^{-1})>41.7$. 
Too few large EW at $z\sim 3$ could be explained by the lack of AGNs or bright Ly$\alpha$ sources at $\log_{10}(L_{\mathrm{Ly}\alpha}/\rm erg\,s^{-1})>43$ in our model.\\

\begin{figure}
\centering
\includegraphics[scale=0.39]{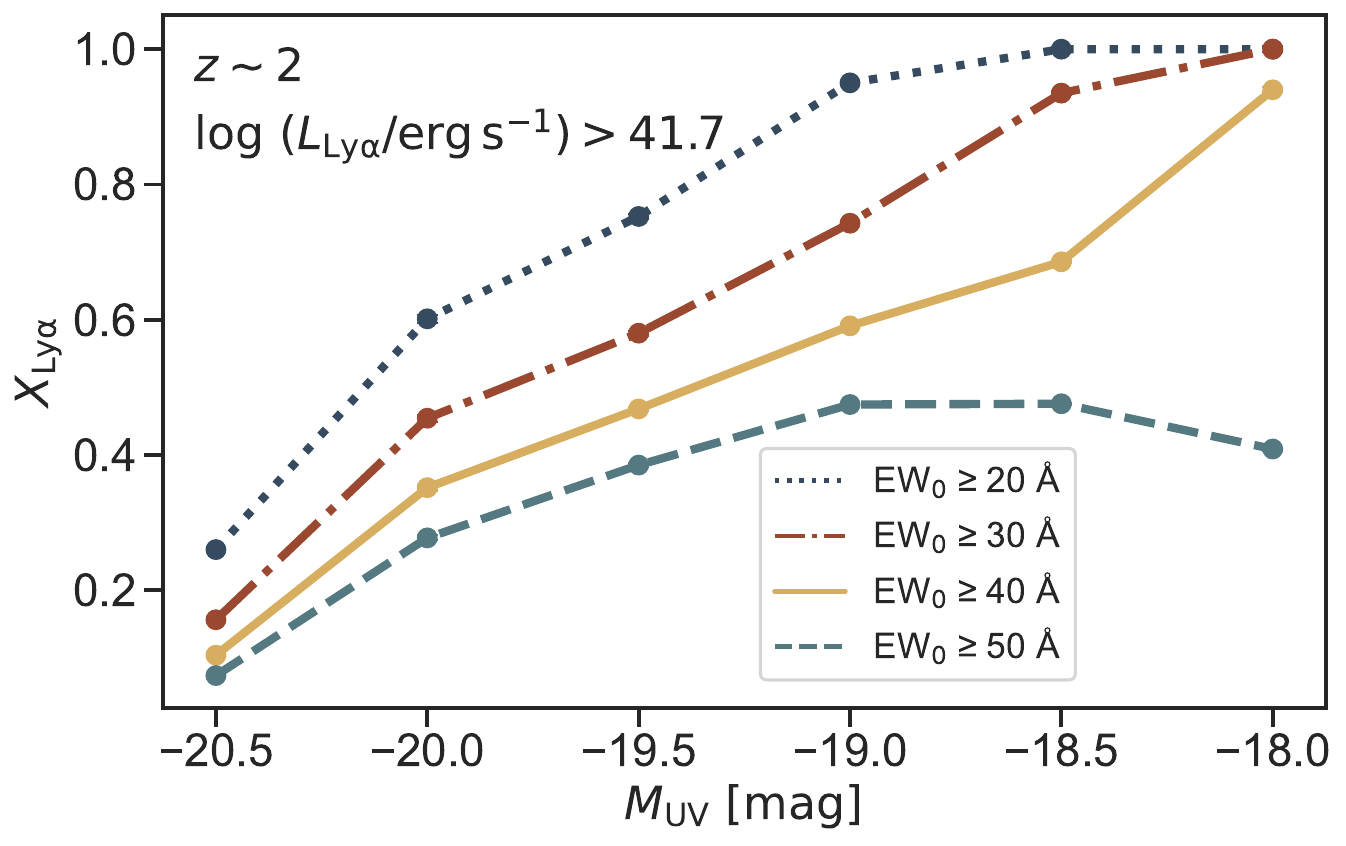}
\includegraphics[scale=0.39]{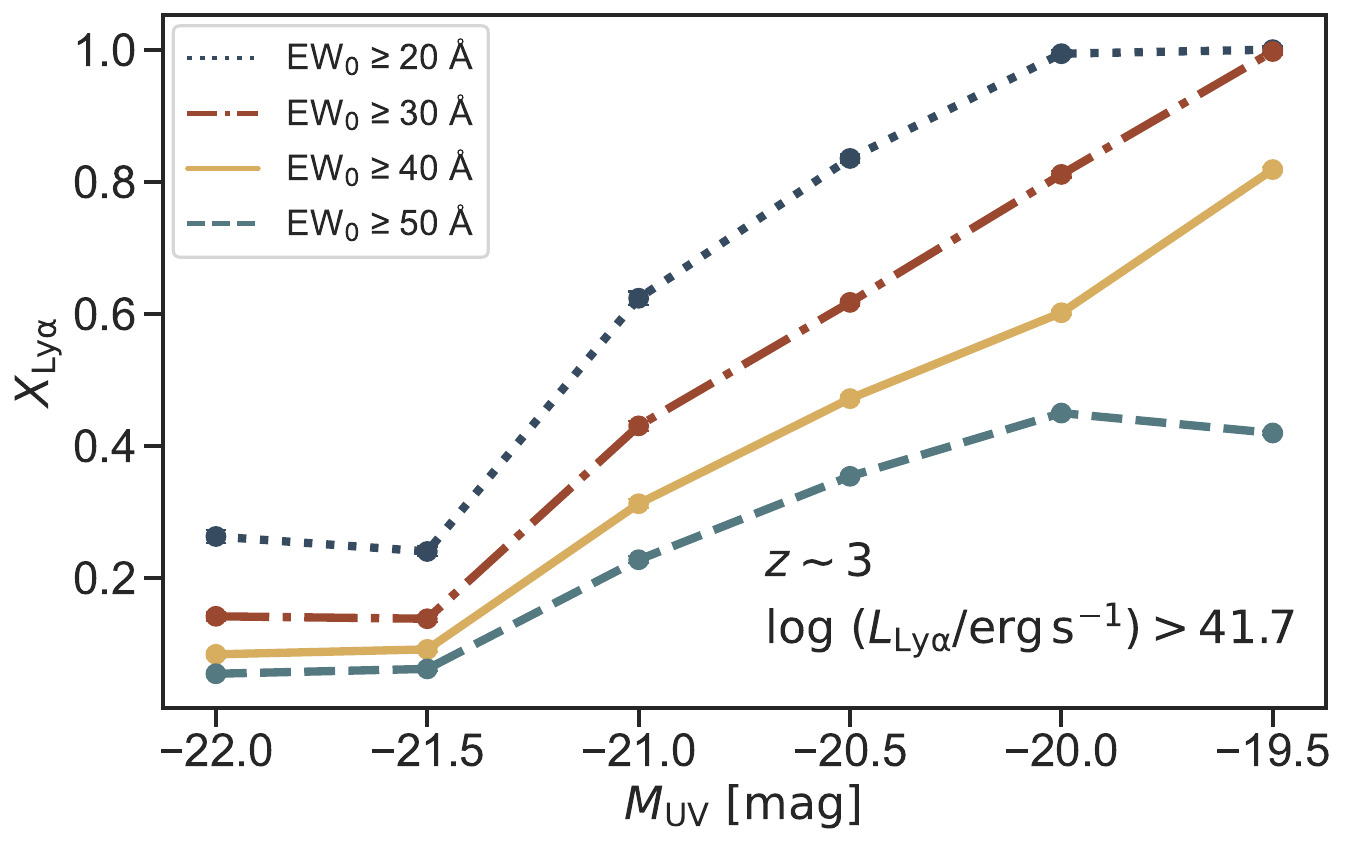}
\caption{
Fraction of galaxies displaying observable Lyman-$\alpha$ emission $(X_{\rm Ly\alpha})$ as a function of rest-frame UV absolute magnitude $(M_{\rm UV})$ for different Ly$\alpha$ equivalent width (EW) thresholds. 
The left and right panels present our model predictions at $z\sim2$ and $z\sim3$, respectively. 
Line colors and styles correspond to different minimum EW cuts. 
In both panels, a fixed Ly$\alpha$ luminosity threshold of $\log_{10}(L_{\rm Ly\alpha}/ \rm erg\,s^{-1}) > 41.7$ is applied. 
The error bars are derived from Poisson error propagation: $\sigma_{X_{\rm Ly\alpha}} = ({N_{\rm LAE}}/{N_{\rm UV}^{2}} + {N_{\rm LAE}^{2}}/{N_{\rm UV}^{3}})^{1/2}$, where $N_{\rm LAE}$ is the number of Ly$\alpha$ emitters and $N_{\rm UV}$ is the total number of continuum-detected galaxies in each $M_{\rm UV}$ bin.
}
\label{fig:XLya}
\end{figure}

\begin{figure}
\centering
\includegraphics[scale=0.36]{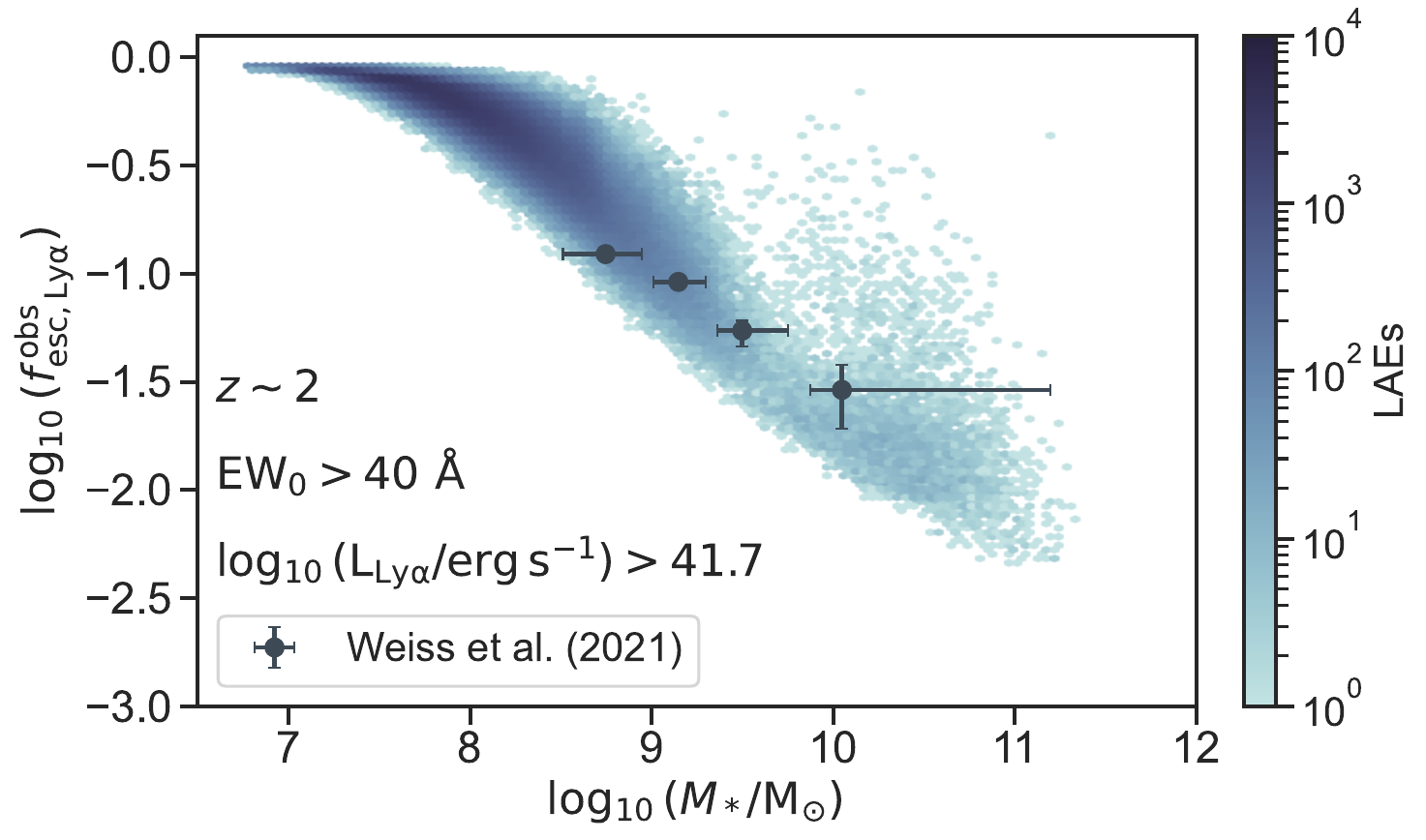}
\includegraphics[scale=0.36]{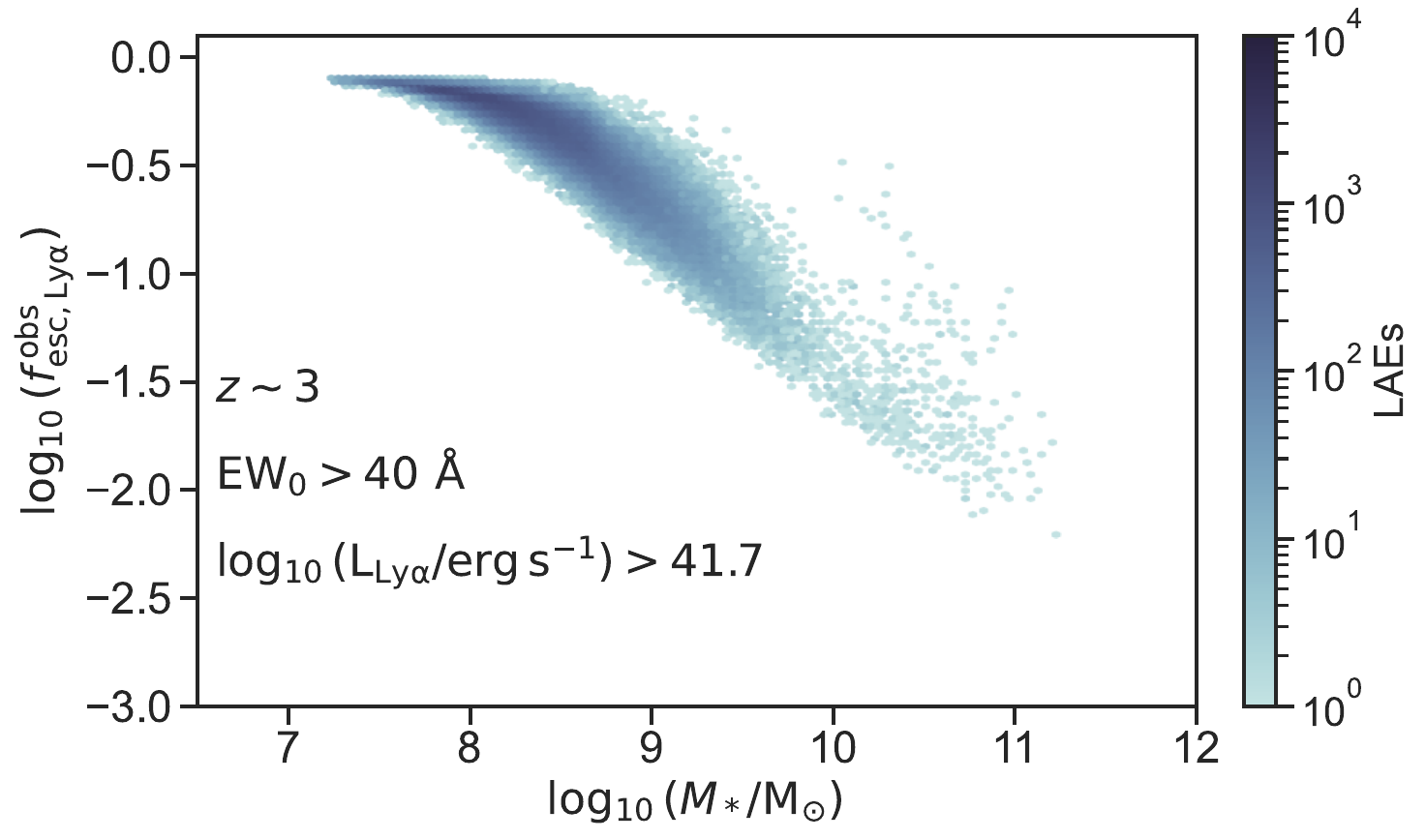}
\caption{
Observed Ly$\alpha$ escape fraction ($f_{\mathrm{esc,Ly\alpha}}^{\mathrm{obs}}$) as a function of stellar mass ($M_*$) for individual LAEs in our fiducial model at $z\sim2$ (\textit{left}) and $z\sim3$ (\textit{right}). 
The colored scatter points show simulated LAEs with observed Ly$\alpha$ luminosities above $\log_{10} (L_{\mathrm{Ly\alpha}}/ \rm erg\,s^{-1}) > 41.7$ and rest-frame equivalent width $\mathrm{EW_{0}} > 40 \, [\text{\AA}]$. 
Dark data points with error bars represent observational measurements from \citet{Weiss_2021}, who inferred $f_{\mathrm{esc,Ly\alpha}}^{\mathrm{obs}}$ using stacked Ly$\alpha$ and H$\beta$ emission from a large sample of [OIII]-selected galaxies at $1.9 \leq z \leq 2.35$. 
Our model qualitatively reproduces the observed trend of decreasing escape fraction with increasing stellar mass.
}
\label{fig:fesc_obs}
\end{figure}

\begin{figure}
\centering
\includegraphics[scale=0.38]{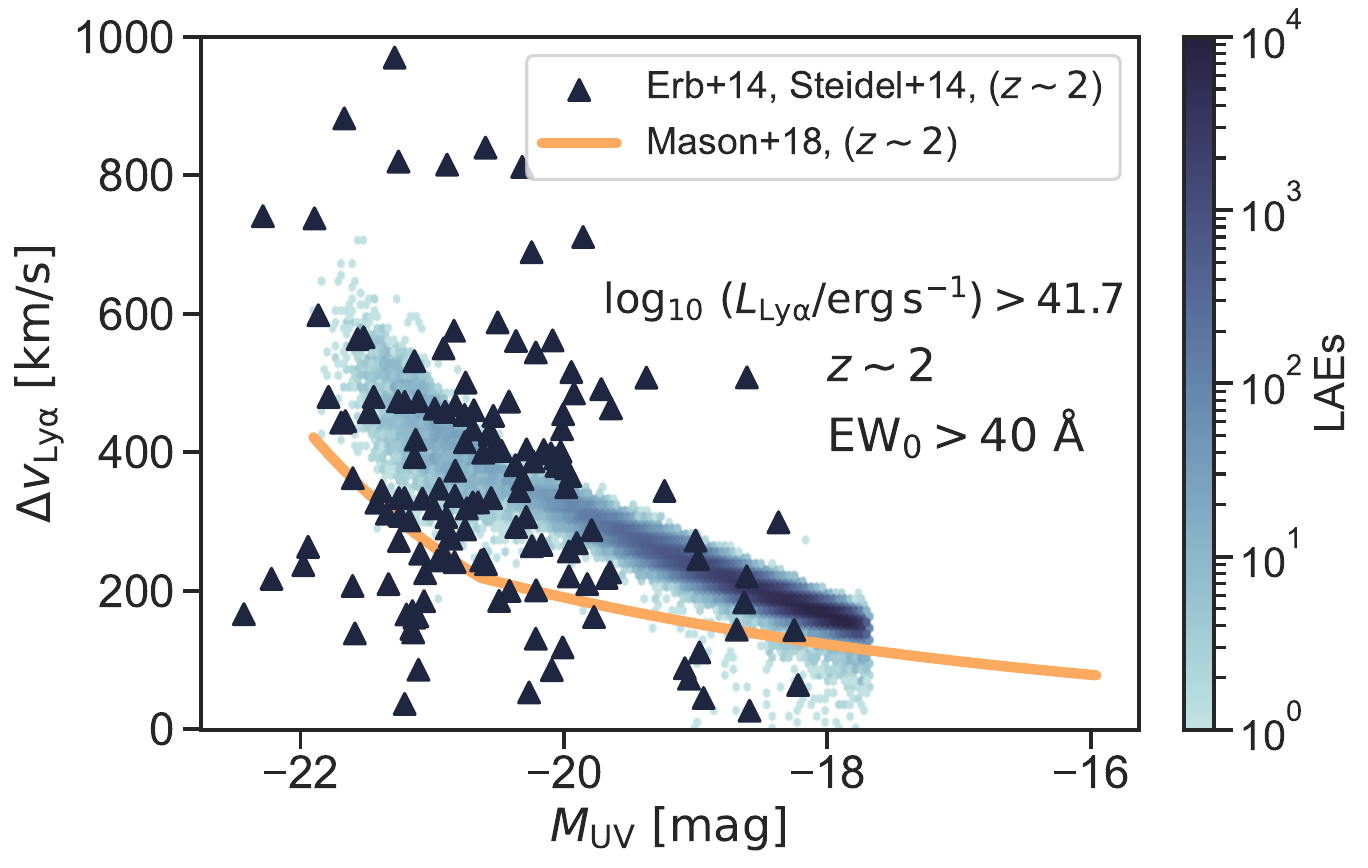}
\includegraphics[scale=0.38]{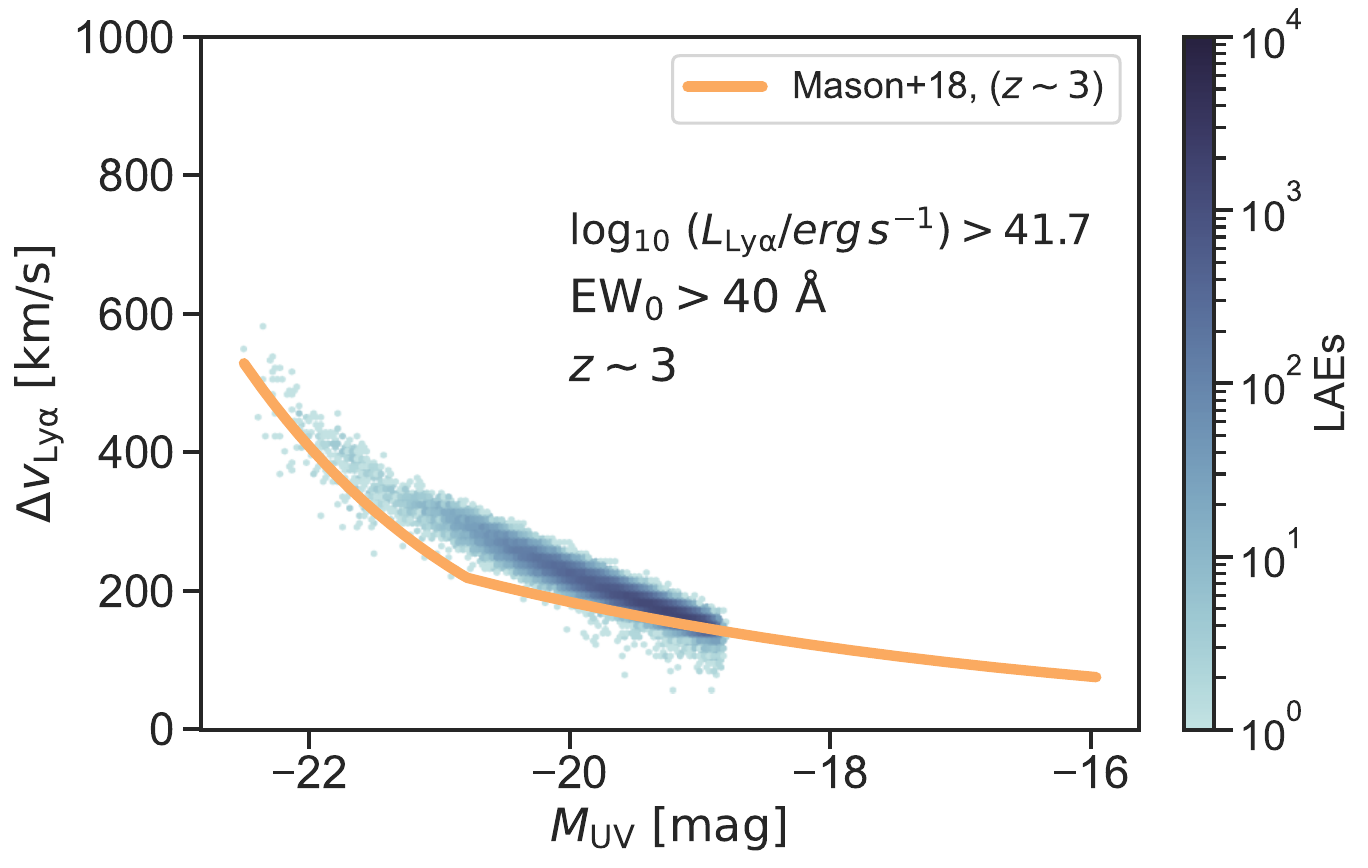}
\caption{Lyman-$\alpha$ velocity offset ($\Delta v_{\mathrm{Ly}\alpha}$) as a function of rest-frame UV absolute magnitude ($M_{\mathrm{UV}}$) for simulated LAEs (shaded area with the color-bar) at $z\sim2$ ({\it left}) and $z\sim3$ ({\it right}). 
The simulated points correspond to galaxies with $\log_{10} (L_{\mathrm{Ly\alpha}}/\rm erg\,s^{-1}) > 41.7$ and $\mathrm{EW}{_0} > 40\,$\AA. 
Observational data from \citet{Erb_2014} and \citet{2014ApJ...795..165S} are represented by dark blue triangles at $z\sim2$. 
The orange solid line depicts the prediction from \citet{2018ApJ...856....2M} who empirically modeled the median velocity offset as a function of UV magnitude and redshift via halo abundance matching as $\log_{10}\Delta v_{\mathrm{Ly}\alpha} = 0.32\,\gamma\left(M_{\mathrm{UV}} + 20.0 + 0.26z\right) + 2.34$, where $\gamma = -0.3$ for $M_{\mathrm{UV}}\geq -20 - 0.26z$ and $\gamma = -0.7$ otherwise.
}
\label{fig:deltav-MUV}
\end{figure}

\begin{figure}[t!]
\centering
\includegraphics[scale=0.39]{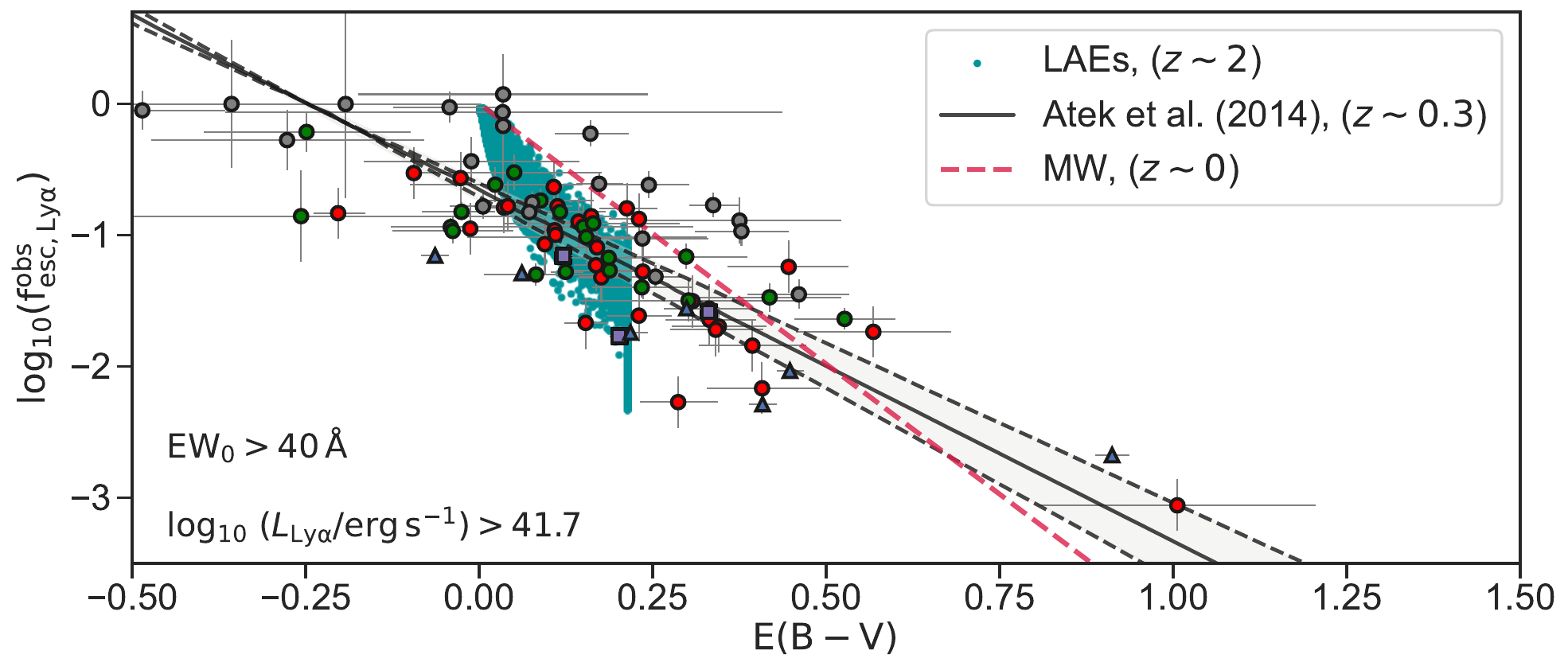}
\caption{
Observed Ly$\alpha$ escape fraction ($f_{\mathrm{esc,Ly\alpha}}^{\mathrm{obs}}$) as a function of dust reddening $E(B-V)$ for our LAEs at $z\sim2$. 
Green points show individual LAEs with $\log(L_{\mathrm{Ly\alpha}}/ \rm erg\,s^{-1}) > 41.7$ and $\mathrm{EW}_{0} > 40,\mathrm{\AA}$. 
The $E(B-V)$ values are derived from the model dust attenuation (see Eq.~(\ref{eq:tau_a})) via the \citet{Calzetti_2000} attenuation law using $E(B-V) = \tau_a \times 1.086 / R_V$ with $R_V = 4.05$. 
For comparison, we include empirical fits from \citet{2014A&A...561A..89A} (black line), who studied local star-forming galaxies, and the Milky Way extinction law (red dashed line). 
Our results show a broad agreement with observed trends, particularly the anti-correlation between $f_{\mathrm{esc,Ly\alpha}}^{\mathrm{obs}}$ and $E(B-V)$, which is a hallmark of dust-regulated Ly$\alpha$ escape. The observational data points shows selected LAEs at $z\sim0.3$ as well as local measurements ($z\sim0$) from the following studies: Gray circles show GALEX-selected LAEs at $z\sim0.3$ \citep{Martin_2005} from \citet{2014A&A...561A..89A}; green circles are from \citet{2009ApJ...704L..98S} and red circles from \citet{Cowie_2011}, both also at $z\sim0.3$. Blue triangles compile local ($z\sim0$), ``net'' Ly$\alpha$ emitters observed with IUE \citet{Boggess_1978}, i.e. galaxies whose IUE spectra retain net Ly$\alpha$ emission after accounting for absorptions. Purple squares show nearby starbursts imaged in  HST/ACS  \citep{2008A&A...488..491A}.}
\label{fig:EBV-fesc}
\end{figure}

Another way to check the relation with the UV SFR is the fraction of LAEs as a function of UV magnitude. 
Fig.~\ref{fig:XLya} quantifies the fraction of LAEs, defined as $X_{\rm Ly\alpha}$, as a function of rest-frame UV magnitude $(M_{\rm UV})$ for different equivalent width (EW) thresholds. 
Specifically, we calculate:
\begin{align}
    X_{\rm Ly\alpha}(M_{\rm UV}, \mathrm{EW_{min}}) = \frac{N_{\rm LAE}(M_{\rm UV}, \mathrm{EW}\geq \mathrm{EW_{min}}, L_{\rm Ly\alpha}\geq L_{\rm min})}{N_{\rm UV}(M_{\rm UV}, L_{\rm Ly\alpha}\geq L_{\rm min})},
\end{align}
where $N_{\rm LAE}$ is the number of LAEs in a given $M_{\rm UV}$ bin that meet both the EW and Ly$\alpha$ luminosity thresholds, and $N_{\rm UV}$ is the total number of continuum-selected sources above the same luminosity threshold. For this analysis, we adopt $\log_{10}(L_{\rm Ly\alpha}/\rm erg\,s^{-1}) > 41.7$. Statistical uncertainties on $X_{\rm Ly\alpha}$ are simply estimated by the Poisson error. 
As expected, at both redshifts, at a given UV magnitude, the fraction of galaxies displaying observable Ly$\alpha$ emission naturally decreases as the EW threshold increases, simply because higher EW threshold exclude galaxies with weaker Ly$\alpha$ emission. This trend indicates that galaxies with higher Ly$\alpha$ EW represent a smaller fraction of the overall galaxy population at these UV magnitudes. 
For instance at $M_{\mathrm{UV}}\sim -20$, for $\mathrm{EW}_0 \geq 20\, \text{\AA}$, more than $\sim 60\%$ of the galaxies are LAEs, while for $\mathrm{EW}_0 \geq 50\, \text{\AA}$ this fraction lowers to $\sim 20\%$ at $z\sim2$.
At both $z\sim2$ and $z\sim3$, our results indicate a clear trend: the fraction of galaxies showing strong Ly$\alpha$ emission increases toward fainter UV magnitudes. 
This trend becomes especially prominent at higher EW thresholds $(\geq 40\, \text{\AA}, \geq 50\,\text{\AA})$, suggesting that galaxies with lower UV luminosities are more likely to host physical conditions that facilitate greater Ly$\alpha$ escape. 
These trends predicted by our model, are qualitatively consistent with observations \citep[e.g.,][]{Stark_2011, 10.1093/mnras/stx2307,Kusakabe:2020AA}.\\ 

We next investigate how the Ly$\alpha$ escape fraction varies with stellar mass in our simulated LAE population. 
Fig.~\ref{fig:fesc_obs} shows $f_{\mathrm{esc,Ly\alpha}}^{\mathrm{obs}}$ as a function of $M_*$ for individual LAEs at $z\sim2$ and $z\sim3$, selected with $\mathrm{EW_{0}} > 40 \,\,\text{\AA}$ and $\log (L_{\mathrm{Ly\alpha}}/ \rm erg\,s^{-1}) > 41.7$. 
In both redshift bins, our model exhibits a clear declining trend in $f_{\mathrm{esc,Ly\alpha}}^{\mathrm{obs}}$ with increasing stellar mass. 
This behavior emerges naturally from our radiative transfer prescriptions, where more massive galaxies typically host denser HI and dustier interstellar media, resulting in more significant attenuation of Ly$\alpha$ photons. 
We compare our model predictions with the binned observational measurements from \citet{Weiss_2021}, who analyzed the Ly$\alpha$ escape fraction of [OIII]-emitting galaxies between $1.9 \leq z \leq 2.35$. 
Their escape fractions were estimated by stacking UV grism spectra from HST to measure H$\beta$ and combined with Ly$\alpha$ fluxes from HETDEX. 
By comparing the stacked H$\beta$/Ly$\alpha$ ratios, they inferred how $f_{\mathrm{esc,Ly\alpha}}^{\mathrm{obs}}$ varies with stellar mass and other galaxy properties. 
Their analysis identified stellar mass and internal extinction as the dominant factors that regulate Ly$\alpha$ escape, as naturally predicted by our LAE model.\\ 

A systemic velocity offset is an indicator of the Ly$\alpha$ radiative transfer. 
Fig.~\ref{fig:deltav-MUV} illustrates the relationship between the Ly$\alpha$ velocity offset, defined as the offset between the rest-frame intrinsic Ly$\alpha$ wavelength ($\lambda_{\rm Ly\alpha}=1215.67$ \AA) and the position of the red peak, given by $\Delta v_{\rm Ly\alpha}=c(\lambda_{\rm peak}/\lambda_{\rm Ly\alpha}-1)$. 
The simulated galaxies in both panels (shaded area color-coded by density) represent LAEs selected with rest-frame Ly$\alpha$ equivalent width $\mathrm{EW}_{\mathrm{0}}>40\,\,\text{\AA} \,$  and a luminosity threshold of $\log_{10}(L_{\mathrm{Ly\alpha}}/\mathrm{erg\,s^{-1}}) > 41.7$. 
We compare our results with observational measurements obtained by \citet{Erb_2014} and \citet{2014ApJ...795..165S} for galaxies at similar redshifts ($z \sim 2$). 
Furthermore, we include the analytic relation derived from \citet{2018ApJ...856....2M}, which describes the median velocity offset as a function of UV luminosity and redshift. 
This relation was constructed by combining semi-analytic modeling and abundance matching techniques, adopting a redshift-dependent halo velocity offset relation and translating it into an observable $M_{\mathrm{UV}}$ dependence $\log_{10}\Delta v_{\mathrm{Ly}\alpha} = 0.32\,\gamma\left(M_{\mathrm{UV}} + 20.0 + 0.26z\right) + 2.34$, where the slope parameter $\gamma$ transitions between two regimes: $\gamma=-0.3$ at fainter luminosities ($M_{\mathrm{UV}}\geq -20-0.26z$) and $\gamma=-0.7$ at brighter luminosities. 
Our LAEs show a clear decline in Ly$\alpha$ velocity offset toward fainter UV magnitudes, consistent with both observational data and the analytical predictions of \citet{2018ApJ...856....2M}. 
We also find that the amplitude of the velocity offset is highly correlated with $\beta$ or $N_{\rm HI}$. 
For example, we could have a model with much larger $\beta$ (say $\beta\sim 50$) but much smaller $\gamma$ (say $\gamma\sim 1$) that fits only to the LAE luminosity function. 
However, we find that such a model gives too large a velocity offset, inconsistently with the range shown here. Although we include observationally motivated scatters in the Sec.~\ref{subsec:scale} scaling relations, the mappings in Eqs.~\ref{eq:vexp} -- \ref{eq:tau_a} are currently deterministic; as a result, the model underestimates the intrinsic scatter of $\Delta v_{\rm Ly\alpha}$ at fixed $M_{\rm UV}$ relative to the data. We plan to incorporate an explicit $\sim$0.2--0.25\,dex intrinsic scatter or an equivalent stochasticity in $V_{\rm exp}$/$N_{\rm HI}$ and/or a sightline-dependent IGM to recover the observed width.\\ 

Finally, let us check the strength of the dust reddening. 
Fig.~\ref{fig:EBV-fesc} explores the relationship between the observed Ly$\alpha$ escape fraction ($f_{\mathrm{esc,Ly\alpha}}^{\mathrm{obs}}$) as a function of the dust reddening $E(B-V)$ for our LAEs at $z\sim2$ and $z\sim3$ (green points). 
To directly compare with observational studies, we convert the model-derived dust optical depth $\tau_a$ (computed via Eq.~(\ref{eq:tau_a})) into $E(B-V)$ using the \citet{Calzetti_2000} attenuation law, commonly adopted for actively star-forming galaxies. 
Specifically, we adopt $E(B-V) = 1.086\tau_a/R_{V}$ with $R_V = 4.05$. 
Interestingly, the range of $f^{\rm obs}_{\rm esc,Ly\alpha}$ and $E(B-V)$ for our LAEs is consistent with the observational results in \citet{2014A&A...561A..89A}. 
In addition, we confirm a clear negative correlation between $f_{\mathrm{esc,Ly\alpha}}^{\mathrm{obs}}$ and $E(B-V)$: galaxies with higher dust content tend to have lower observed Ly$\alpha$ escape fractions. 
This trend is consistent with observational results from \citet{2014A&A...561A..89A}, who derived similar relations for local star-forming galaxies based on GALEX and H$\alpha$ data. 
We overlay their best-fit relation (black line) and the theoretical Milky Way extinction curve (red dashed) for comparison.
We note that $E(B-V)$ is observationally derived from the Balmer decrement that generally shows that the dust attenuation for emission lines tend to be greater than that for the stellar continuum \citep[see e.g.,][]{Calzetti_2000,Kashino:2013ApJ,Saito:2020mn}.
This fact would be consistent with the high value of $\gamma \sim \mathcal{O}(10)\gg 1$.
A potential caveat of this comparison is that all the observational data points are derived at $z\sim 0$.
We do not see a strong evolution in our LAE model prediction from $z\sim 3$ to $z\sim 2$.\\

\subsection{Implications of our LAE model}
\label{sec:lae_properties}
So far we have discussed observational results that support our LAE modeling. 
In this section, we focus instead on \textit{predictions} and \textit{implications} of our LAE model.\\

\begin{figure}[t]
\includegraphics[scale=0.45]{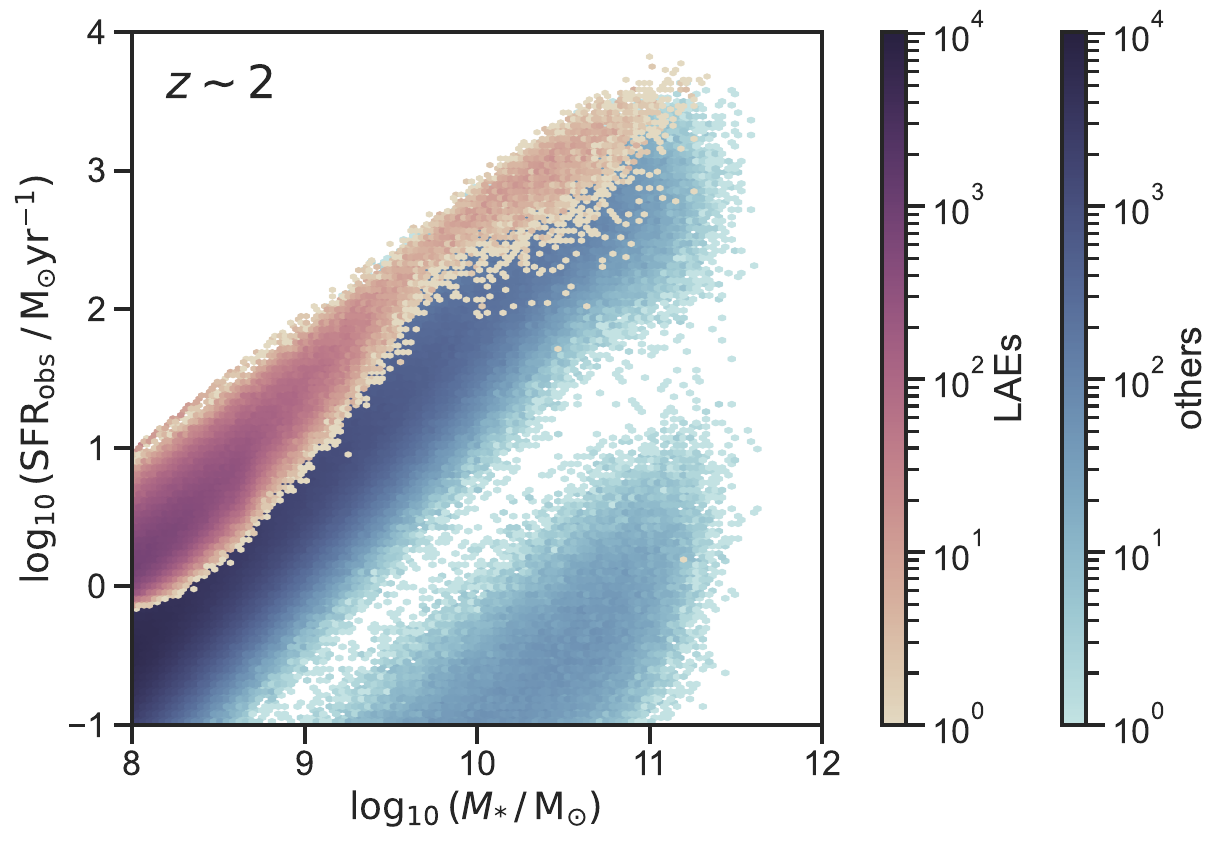}
\includegraphics[scale=0.45]{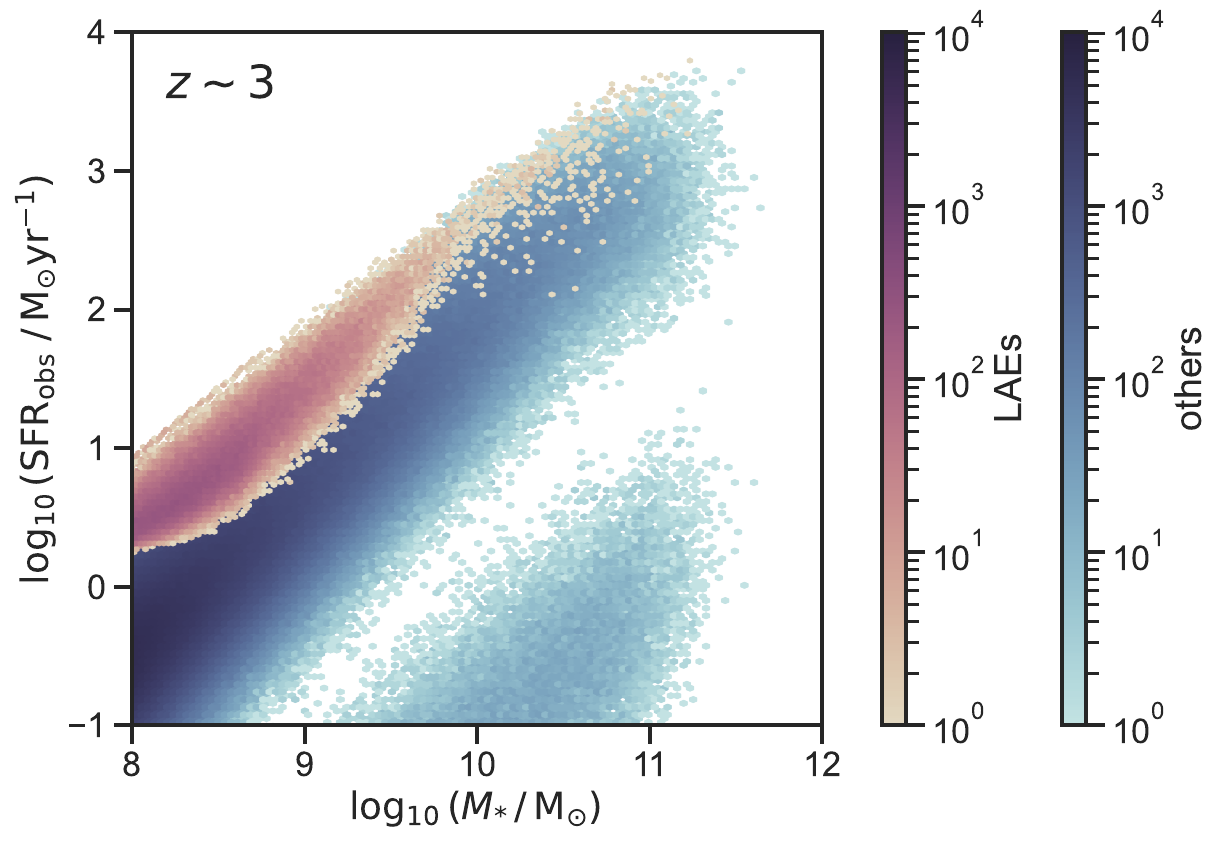}
\caption{
The relation between stellar mass $M_{*}$ and SFR for LAEs \big( ($\mathrm{log}_{10}(L_{Lya}/\,\rm erg\,s^{-1}) > 41.7$ and $\mathrm{EW}_0 > 40 \,\, \text{\AA}$) \big)  (in pink) and others (in blue) at $z \sim 2$ ({\it left}) and $z \sim 3$ ({\it right}). 
}
\label{fig:sfr-sm}
\end{figure}

\begin{figure}
\centering
\includegraphics[scale=0.38]{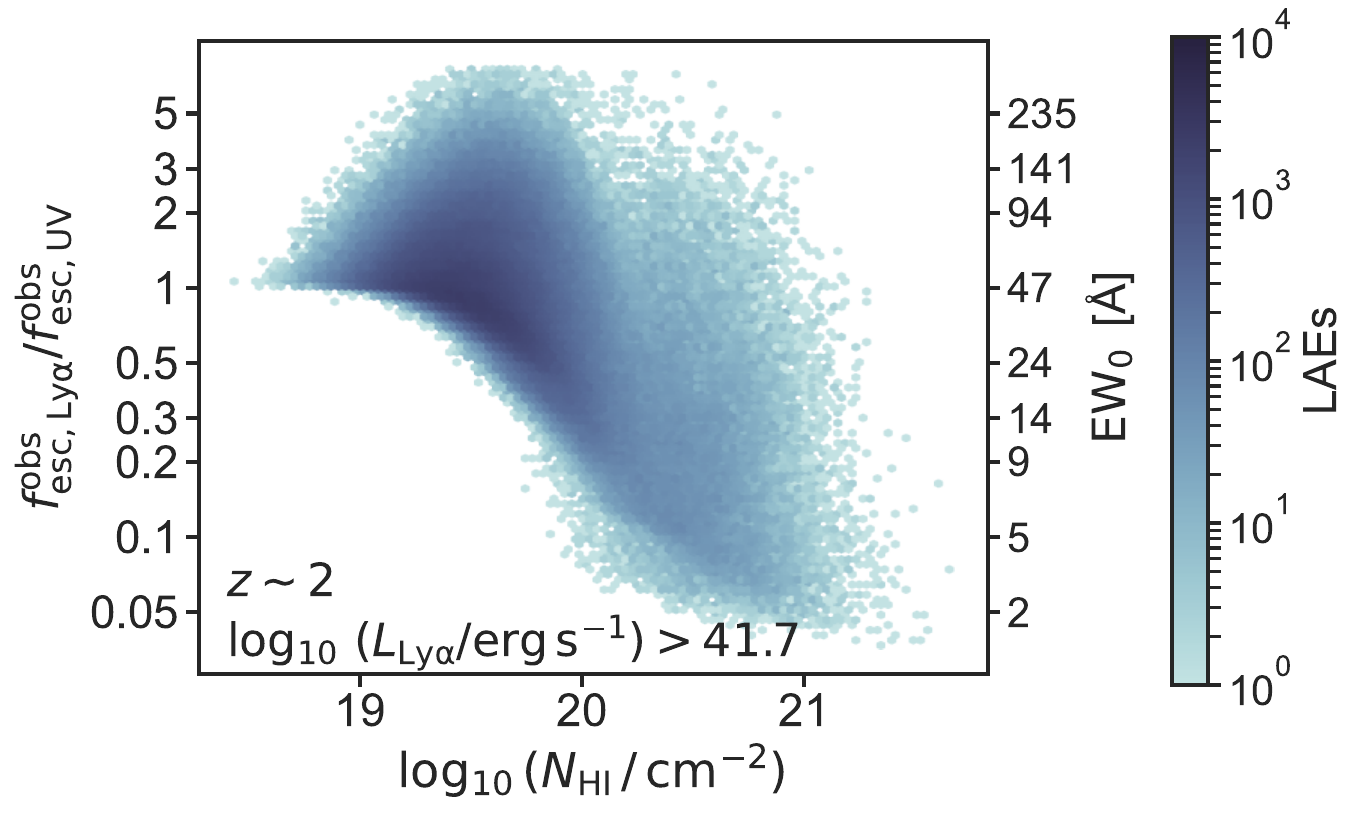}
\includegraphics[scale=0.38]{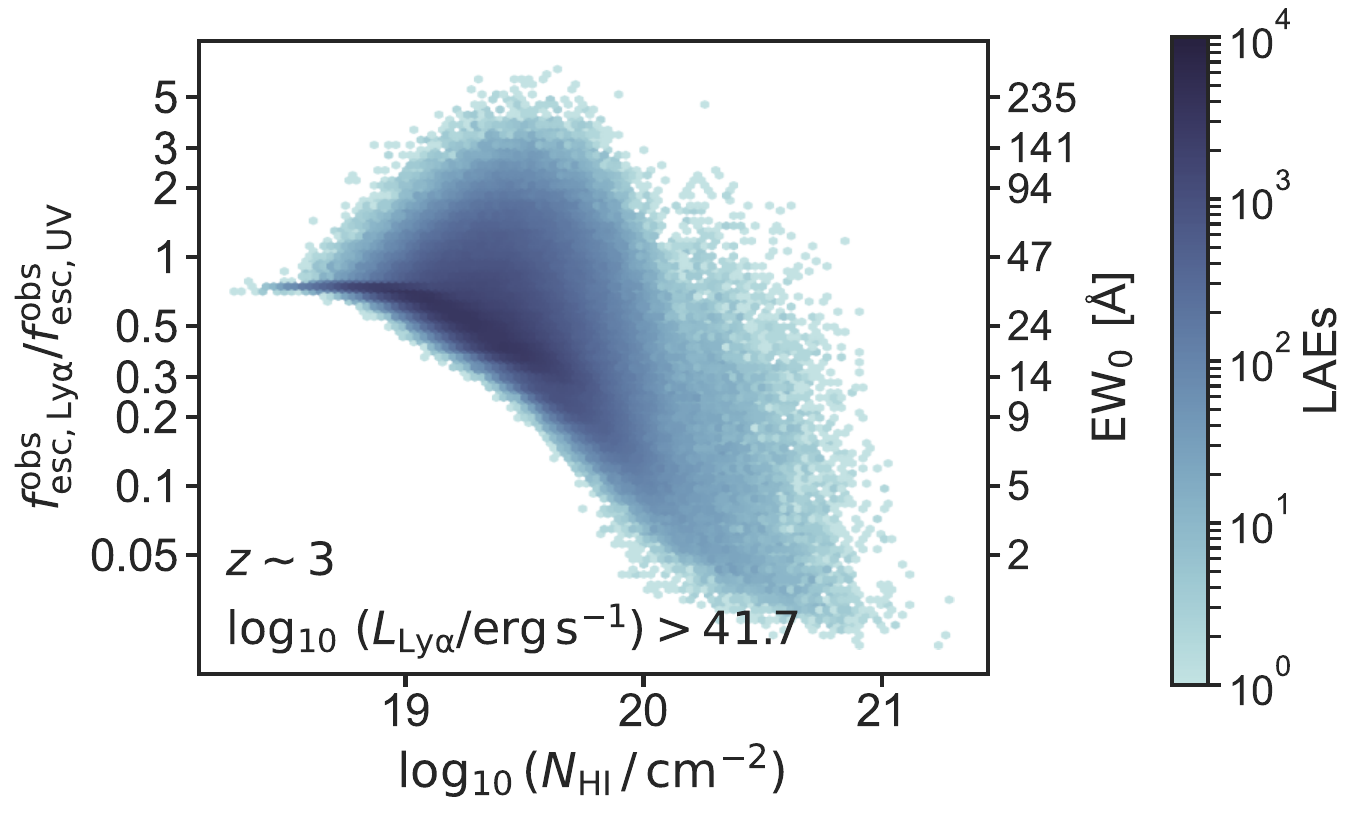}
\caption{
Ratio of Lyman-$\alpha$ to UV photon escape fractions $(f_{\rm esc,Ly\alpha}^{\rm obs}/f_{\rm esc,UV}^{\rm obs})$ as a function of neutral hydrogen column density $(N_{\rm HI})$ (x-axis) and the rest-frame Ly$\alpha$ equivalent width (twin y-axis). Each point represents an individual LAE from our fiducial best-fit model, with color coding by their number counts. The left and right panels display results for redshifts $z\sim2$ and $z\sim3$, respectively. Only galaxies exceeding a luminosity threshold of $\log(L_{\rm Ly\alpha}/ \rm erg\,s^{-1}) > 41.7$ are included. 
}
\label{fig:fesc_ratio}
\end{figure}

\subsubsection{Ly$\alpha$ escape fraction}

A simple question to ask is why massive galaxies are not considered LAEs at $z=2\!-\!3$ in our model. 
Is it because star formation in massive galaxies is quenched and thus intrinsic Ly$\alpha$ luminosity is small, or because massive galaxies tend to have smaller Ly$\alpha$ escape fraction? 
Fig.~\ref{fig:sfr-sm} shows the relation between stellar mass and SFR. 
Our LAEs tend to be in an upper region of the star formation main sequence, and the number of LAEs generally decreases as a function of $M_{*}$ (see also Fig.~\ref{fig:fesc_obs}).  
This is expected because, at fixed $M_{*}$, galaxies with higher SFR have larger intrinsic Ly$\alpha$ luminosity and $V_{\rm exp}$ and thus larger $f^{\rm ISM}_{\rm esc}$. 
Interestingly, the \texttt{UniverseMachine} model predicts a clear bimodal population in ($M_{*}$,SFR), and our model has no LAE in the quenched population of galaxies. 
This is not because the intrinsic SFR is too small, but because the escape fraction is too small at high $M_{*}$, since galaxies with similar SFR (${\rm SFR}<10\,{\rm M_{\odot}/yr}$) are considered LAEs at $\log_{10}(M_{*}/{\rm M_{\odot}})<9$.\\ 

In addition, Fig.~\ref{fig:fesc_ratio} shows the ratio of Ly$\alpha$ to UV escape fractions, $f^{\mathrm{obs}}_{\mathrm{esc,Ly}\alpha}/f^{\mathrm{obs}}_{\mathrm{esc,UV}}$, as a function of the neutral hydrogen column density, $N_{\mathrm{HI}}$, for individual LAEs in our model at $z\sim2$ (left) and $z\sim3$ (right). 
The ratio of the Ly$\alpha$ and the UV escape fraction is an indicator of the efficiency with which Ly$\alpha$ photons can escape compared to UV photons due to radiative transfer effects.
In our model, this ratio can be expressed as follows:
\begin{equation}\label{eq:Lya_to_UV_ratio}
\begin{split}
  \frac{f^{\rm obs}_{\rm esc,Ly\alpha}}{f^{\rm obs}_{\rm esc,UV}}
    &\equiv 
     \biggl(\frac{L^{\rm obs}_{\rm Ly\alpha}}{L^{\rm int}_{\rm Ly\alpha}}\biggr)
     \biggl(\frac{L^{\rm int}_{\lambda,UV}}{L^{\rm obs}_{\lambda,UV}}\biggr)
     = \frac{f^{\rm obs}_{\rm esc,Ly\alpha}}{e^{-\tau_{\rm UV}}}
  \\
    &= \biggl(\frac{3\times1.4\times10^{40}\,\text{erg\,s}^{-1}\,\text{\AA}^{-1}
                  \times \mathrm{SFR_{obs}}}
                 {1\times10^{42}\,\text{erg\,s}^{-1}\times \mathrm{SFR_{obs}}}\biggr)
       \biggl(\frac{\lambda_{\rm Ly\alpha}}{\lambda_{\rm UV}}\biggr)^{-2}
       \mathrm{EW}_{0}
       = \frac{\mathrm{EW}_{0}}{47\,\text{\AA}}\,. 
\end{split}
\end{equation}
where the observed Ly$\alpha$ escape fraction, $f^{\rm obs}_{\rm esc,Ly\alpha}$, was introduced in Sec.~\ref{subsec:overview}, and the observed UV escape fraction, $f^{\rm obs}_{\rm esc,UV}=e^{-\tau_{\rm UV}}$, was computed in Sec.~\ref{sec:halos}. 
The second line of Eq.~(\ref{eq:Lya_to_UV_ratio}) is derived with the Kennicutt relations we adopted for intrinsic Ly$\alpha$ luminosity and UV continuum luminosity density \citep{Kennicutt:1998ARAA}. 
In Fig.~\ref{fig:fesc_ratio}, we find a strong anti-correlation between $f_{\mathrm{esc,Ly\alpha}}^{\mathrm{obs}}/f_{\mathrm{esc,UV}}^{\mathrm{obs}}$ and $N_{\mathrm{HI}}$, indicating that denser hydrogen environments significantly suppress Ly$\alpha$ escape compared to the UV continuum.
This is expected, since higher $N_{\rm HI}$ gives smaller observed Ly$\alpha$ escape fraction, $f_{\rm esc, Ly\alpha}^{\rm obs}$ as seen in Fig.~\ref{fig:fISM}.
At column densities $N_{\mathrm{HI}} \lesssim 10^{19}\,\mathrm{cm}^{-2}$, Ly$\alpha$ photons escape almost as efficiently as UV continuum photons ($f_{\mathrm{esc, Ly\alpha}}^{\mathrm{obs}}/f_{\mathrm{esc,UV}}^{\mathrm{obs}} \approx 1$). 
Note also that the ratio can exceed unity, that is, $f_{\rm esc,Ly\alpha}^{\rm obs}>e^{-\tau_{\rm UV}}$. 
As expected from Eq.~(\ref{eq:Lya_to_UV_ratio}), the ratio is defined in terms of the luminosity density for UV but the luminosity for Ly$\alpha$, leading to the dependence of $f_{\mathrm{esc,Ly\alpha}}^{\mathrm{obs}}/f_{\mathrm{esc,UV}}^{\mathrm{obs}} \propto \mathrm{EW}_{0}$ and the ratio is greater than unity for LAEs with $\mathrm{EW}_{0}>47\,$\AA.
$f_{\mathrm{esc, Ly\alpha}}^{\mathrm{obs}}/f_{\mathrm{esc,UV}}^{\mathrm{obs}}\gg 1$ has been invoked in the past to theoretically explain objects with unusually large Ly$\alpha$ EWs \citep{Neufeld:1990ApJ,HansenOh2006,10.1093/mnras/stu1513}. 
In reality, this would be possible in geometries in which the HI can `shield' the dust from Ly$\alpha$ radiation. 
Such configurations require optically tick clumps embedded in a low-dust medium \cite{2014A&A...562A..52D}.
However, more recently it has been argued that these conditions are not realistic and such `equivalent width boosting' are quiet rare and unlikely to be common in realistic galaxies \citep{LaursenNonEnhancement,Gronke2017_AA}. It is reassuring that our modeling approach, relying on physically motivated but less extreme assumptions, does not require ratios significantly greater than unity at $z \sim 2\!-\!3$. 
We also notice that some of our LAEs have a small ratio, $f_{\mathrm{esc,Ly\alpha}}^{\mathrm{obs}}/f_{\mathrm{esc,UV}}^{\mathrm{obs}}<  0.1$,  particularly for LAEs with high HI column density.
These systems could be regarded as Damped Lyman-$\alpha$ Absorbers (DLAs) with $\log_{10}(N_{\rm HI}/{\rm cm}^{-2}) \gtrsim 20.3$, which is consistent with a scenario where DLAs are hosted by massive halos \citep[e.g.,][]{Bird:2014MN}.\\

\begin{figure}[t]
\centering
\includegraphics[scale=0.32]{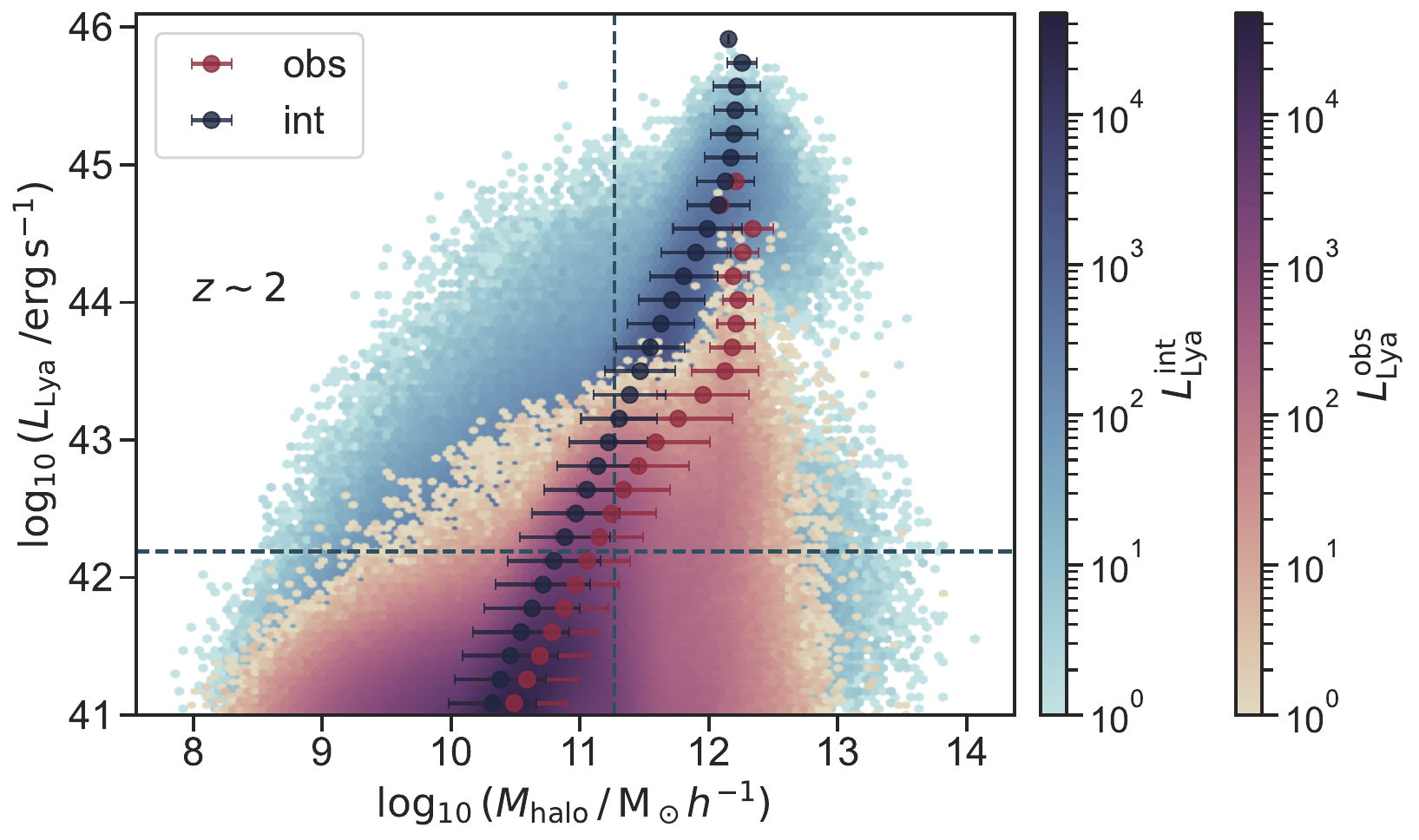}
\includegraphics[scale=0.32]{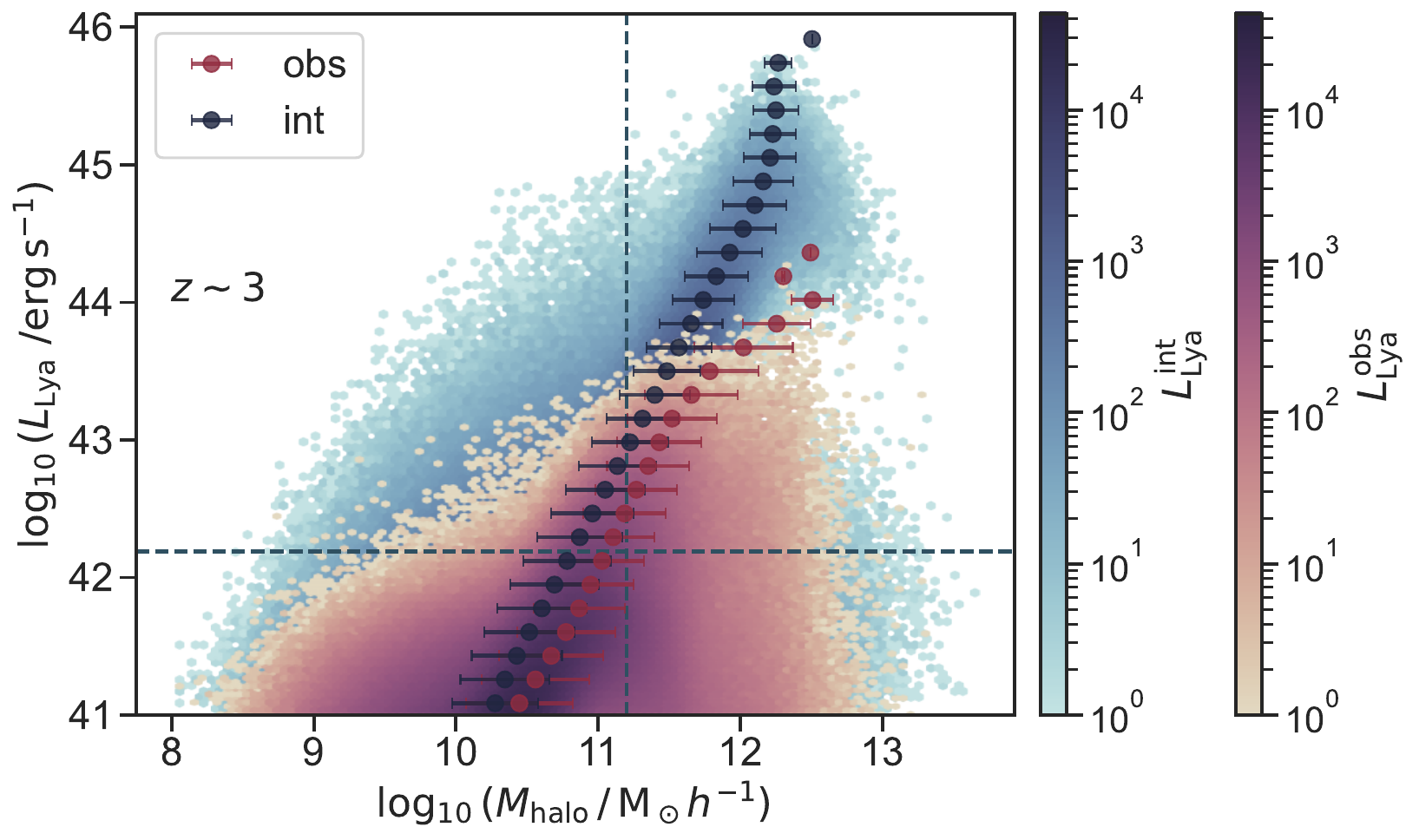}
\caption{
Ly$\alpha$ intrinsic (blue) and observed (pink) luminosities as a function of halo mass. The left panel presents our fiducial model results at $z\sim2$, and the right panel shows the corresponding results at $z\sim3$. The vertical dashed line marks the mean halo mass of the observed LAE sample, while the horizontal dashed line indicates the minimum luminosity threshold adopted for the angular correlation function analysis.
}
\label{fig:lum_mhalo}
\end{figure}

\begin{figure}
\centering
\includegraphics[scale=0.38]{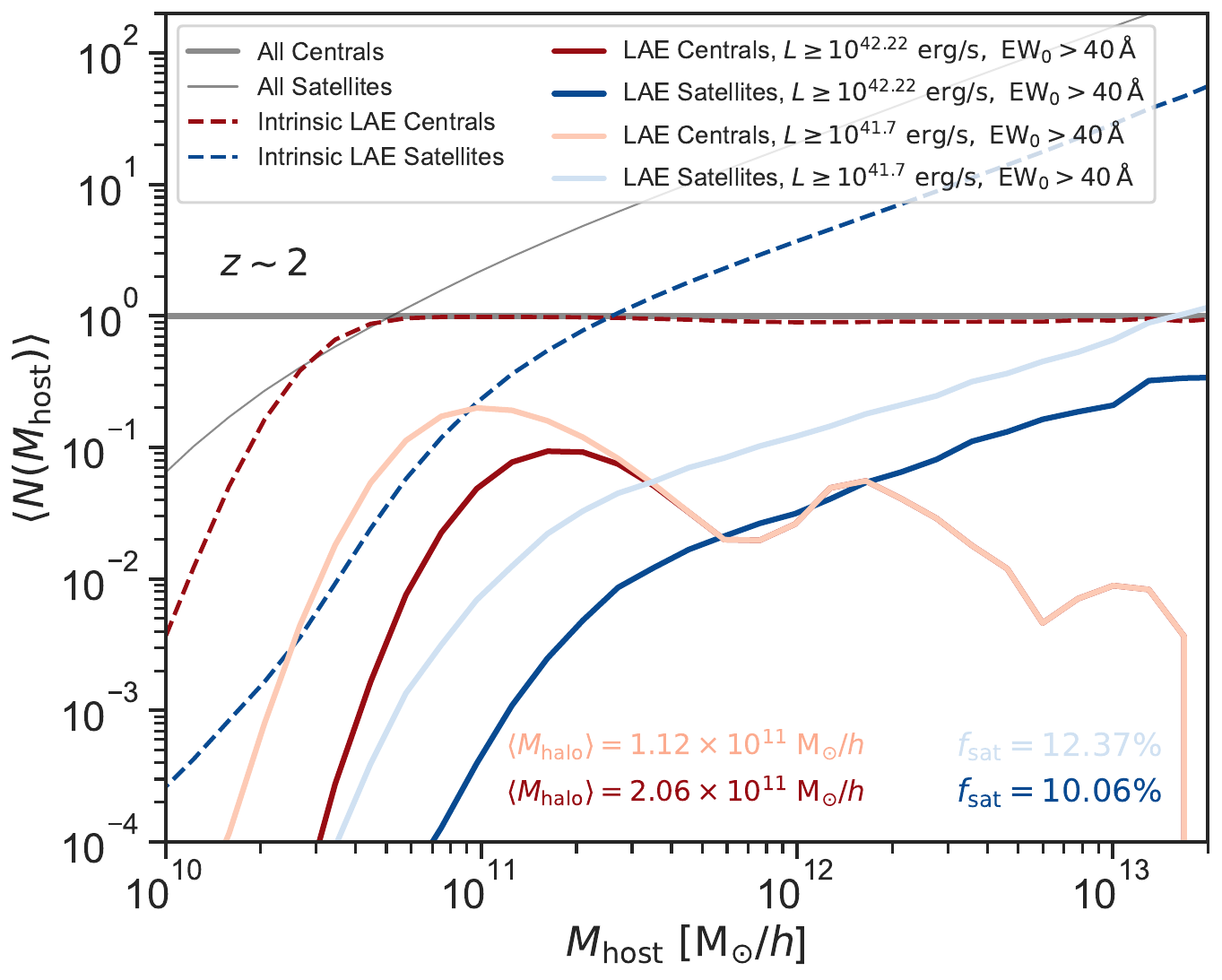}
\includegraphics[scale=0.38]{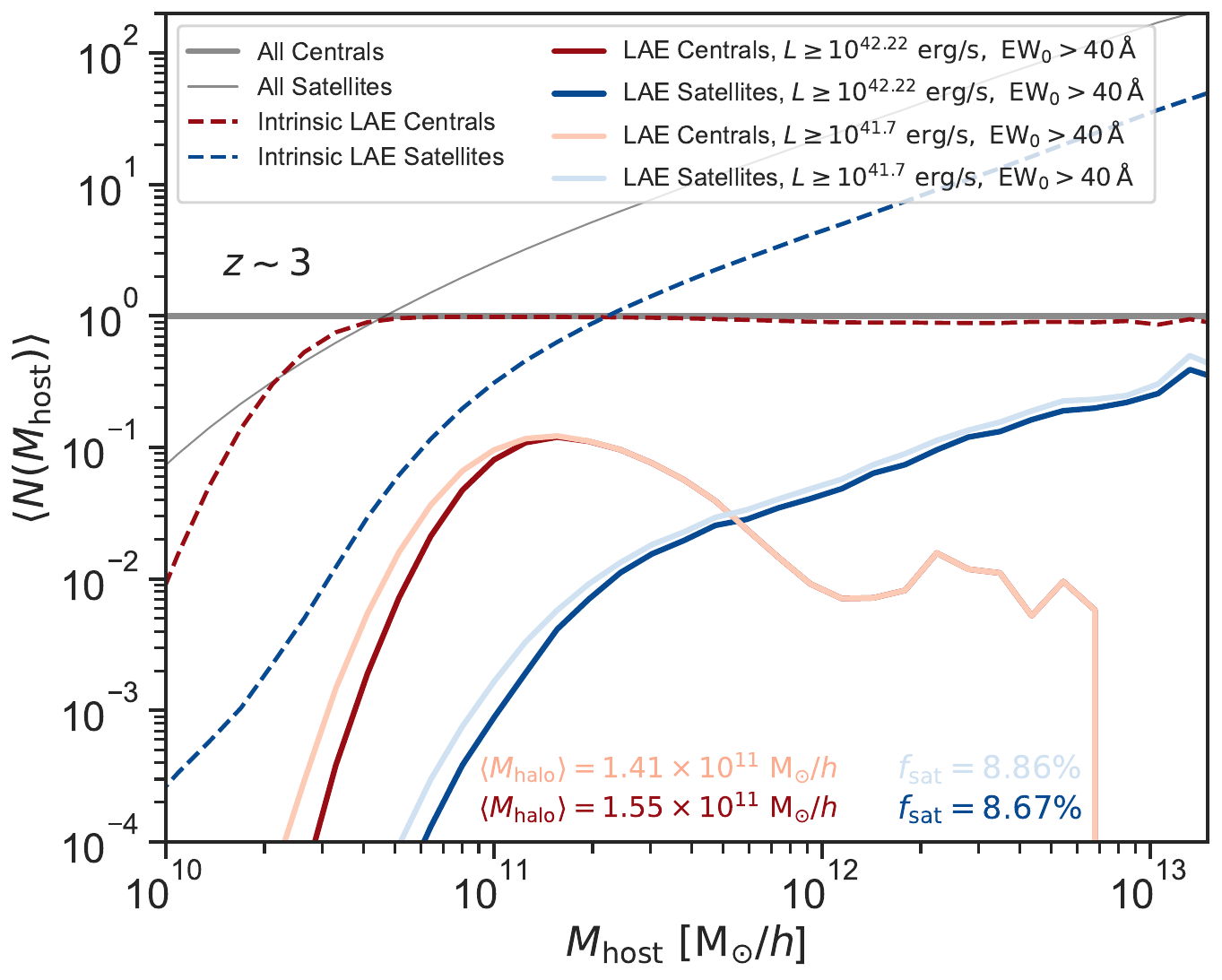}
\caption{
Halo Occupation Distribution (HOD) of LAEs at $z\sim2$ (left) and $z\sim3$ (right). Gray curves show the total central and satellite galaxy occupation functions in \texttt{UniverseMachine}. 
Dashed red and blue lines represent intrinsic LAE central and satellite populations selected based solely on SFR (see Eq.~\ref{eq:L_int}). 
Solid red and blue lines show the final observed LAE populations after accounting for ISM and IGM radiative transfer effects for two Ly$\alpha$ luminosity thresholds. 
Mean halo masses and satellite fractions are quoted for each sample.
}
\label{fig:hod}
\end{figure}

\begin{figure}
\centering
\includegraphics[scale=0.42]{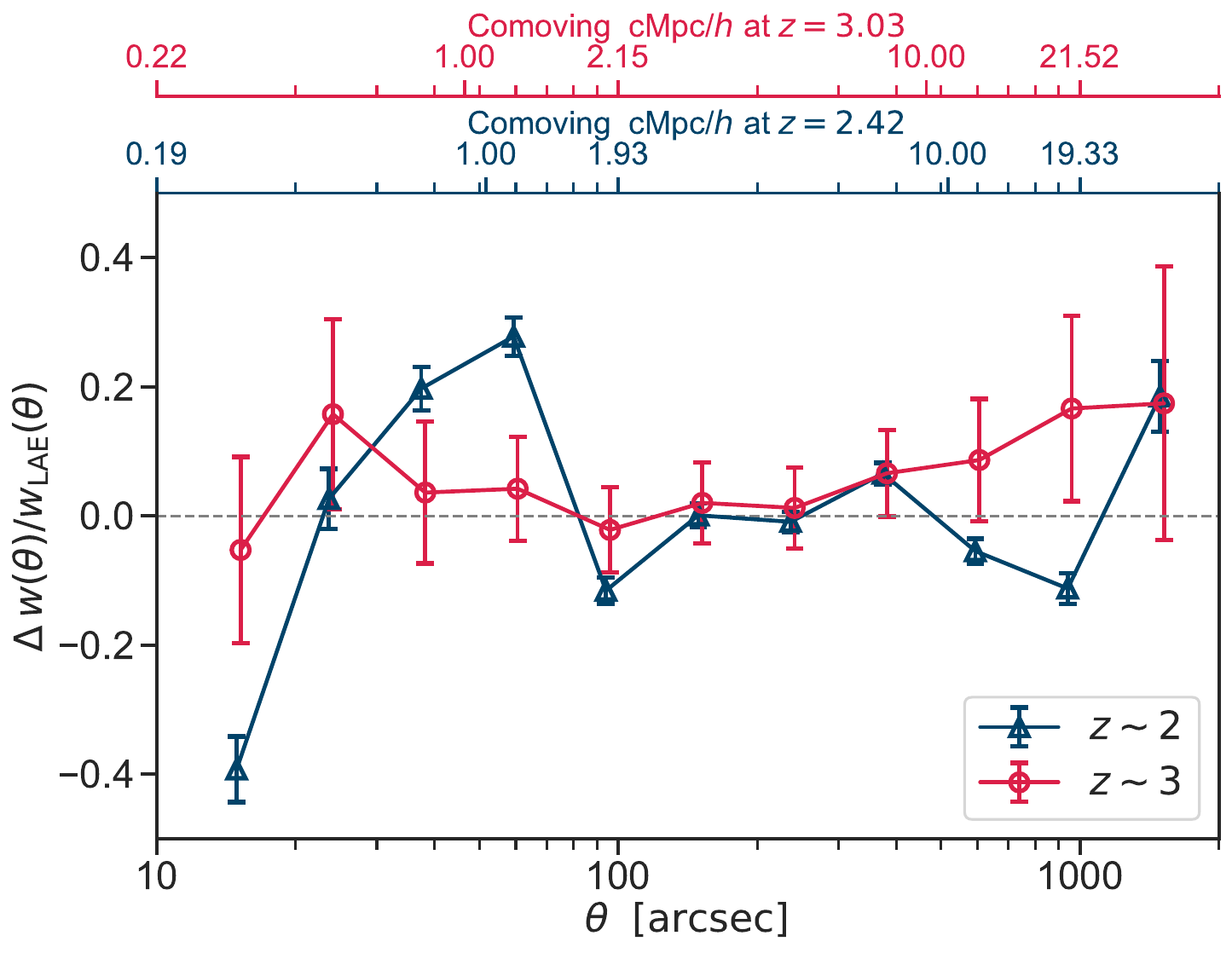}
\caption{
The measured strength of the assembly bias effect in our LAE model. 
Fractional difference between the angular correlation function of shuffled \texttt{UniverseMachine} galaxies and observed LAEs at $z \sim 3$ (red circles) and $z \sim 2$ (blue triangles). 
Shuffling preserves the mass-dependent halo occupation distribution but removes dependencies on secondary halo properties. The error bars are per-bin $1\sigma$ from the jackknife on $\Delta w/w$. Neighboring bins are correlated, and the differing bar sizes reflect the different $w(\theta)$ error budgets at large $\theta$ (see Fig.\ref{fig:covmat}).}
\label{fig:assembly_bias}
\end{figure}

\subsubsection{LAE-halo Connection}
Understanding how observed galaxies are connected with underlying dark-matter halos is important for both cosmology and galaxy evolution studies. 
This is because the clustering amplitude and its scale dependence on scales greater than the typical virial radius of halos depend on characteristic quantities of dark-matter halos, such as halo mass and age \citep[see e.g.,][for a review]{Wechsler:2018AR}. 
In addition, the clustering on the scales smaller than the halo virial radius tells us how the evolution of galaxies are affected by the dense environment. 
Some earlier observational work interpreted the large-scale amplitude of the LAE angular correlation function on the basis of a simple framework such as the correlation length or Halo Occupation Distribution (HOD), concluding that LAEs typically occupy halos with masses ranging from $\approx 10^{11}$-$10^{12}\,\rm{M}_{\odot}$ \citep[e.g.,][]{Hamana:2004mn,Ouchi:2018PA,Hong:2019mn,Herrero-Alons2021AA}. 
However, it is particularly unclear if such a simple framework works well for LAEs as HOD is supposed to work for a mass-limited galaxy sample, which is not clearly the case for LAEs. 
We remind here that our LAE model is designed to reproduce the luminosity function and the large-scale angular correlation function and thus \textit{predicts} the LAE-halo connection as a consequence.\\  

In Fig.~\ref{fig:lum_mhalo}, we show the LAE distribution of the Ly$\alpha$ luminosity versus the dark matter halo mass. 
We show intrinsic (blue) and observed (pink) Ly$\alpha$ luminosities as a function of the halo mass. 
Since the halo mass is the same for an intrinsic and observed luminosity for each LAE, blue points moved vertically downwards from the corresponding pink points. The blue and red error bars indicate the mean and standard deviation of the halo mass within each intrinsic and observed luminosity bin, respectively.
At both $z\sim2$ (left) and $z\sim 3$ (right), there is a clear trend of increasing luminosity with increasing halo mass, indicating that more massive halos typically host brighter LAEs.
However, the halo mass scatter at fixed Ly$\alpha$ luminosity is as large as 0.5 dex, contributed by two effects; about 0.3 dex from the intrinsic scatter between SFR and halo mass as well as about 0.4 dex from the additional scatter from the Ly$\alpha$ RT model.
This fact is qualitatively consistent with a little evolution of the LAE bias as a function of Ly$\alpha$ luminosity in \citet{Kusakabe:2018PASJ}. \\ 

We also compute the HOD of our simulated LAEs. 
Fig.~\ref{fig:hod} presents the resulting HODs at $z \sim 2$ (left) and $z\sim 3$ (right). 
The solid gray lines show the mean occupation of all central and satellite galaxies in the \texttt{UniverseMachine} sample, while dashed lines represent the intrinsic LAEs (galaxies that satisfy our SFR-based LAE selection before accounting for Ly$\alpha$ radiative transfer effects with $\rm SFR \geq 0.1 \,\,\rm M_{\odot}\,yr^{-1}$).
These curves serve as a reference to isolate the impact of radiative transfer and selection criteria. 
The solid red and blue curves show the observed LAE populations, split into central and satellite components for different Ly$\alpha$ luminosity thresholds. 
We find that the HOD of the observed LAE centrals generally follows the shape of the intrinsic LAE sample, but exhibits a suppression at high halo masses due to the lower Ly$\alpha$ escape fraction, as discussed earlier. 
The LAE occupation peaks around the halo masses of $10^{11} - 10^{12} \,\rm{M}_{\odot}$ and declines toward higher halo masses, in qualitative agreement with other earlier work \citep[e.g.,][]{Ravi:2024Ph,Sullivan:2025ar}. 
We also quote the mean halo mass and satellite for each Ly$\alpha$ luminosity threshold in Fig.~\ref{fig:hod} and Table.~\ref{Table:data-fit}. 
Since we do not force our model to match the angular correlation function in the one-halo regime, the 8.7-12.4\% satellite fraction is merely a prediction. 
In addition, we find that, out of these LAE satellites, about 25\% of them are hosted by LAE centrals, and the rest of about 75\% are hosted by other centrals.\\

Finally, we quantify the impact of the halo assembly bias in our LAE model. 
The halo assembly bias refers to the fact that the clustering properties of galaxies/halos depend on secondary halo characteristics beyond halo mass alone \citep[see e.g.,][]{Wechsler_2006, 2007MNRAS.377L...5G,2018ApJ...853...84Z, 2018MNRAS.480.3978A}. 
To isolate the impact of occupancy variations in our LAE population, we performed a shuffling test, following \citet{10.1093/mnras/stw840}.
Within each halo mass bin, we randomly select \texttt{UniverseMachine} galaxies to match the exact number of LAEs, computing the angular correlation function, $w(\theta)$, for each randomized realization.
Fig.~\ref{fig:assembly_bias} shows the fractional difference between the shuffled and observed LAE angular correlation functions $(\Delta w(\theta)/w_{\rm LAE}(\theta))$ at both redshifts. 
The deviations from zero would indicate the presence of assembly bias, but in our test the residuals hover around zero at all scales and at both redshifts therefore there is no systematic deviation beyond the 1-$\sigma$ uncertainties. 
Thus, the level of assembly bias is statistically insignificant in our limited size of the simulation box (400 ${\rm cMpc}/h$) and a more solid conclusion would require a bigger simulation.\\

\section{Discussion}
\label{sec:discussion}

\subsection{Comparison with previous works}
\label{subsec:comparison}

We now outline the differences and complementarities among various LAE simulation works in the literature. 
In Table \ref{table:LAE_sims}, we try to provide a brief summary of the comparison among different cosmological simulations of LAEs, focusing on the relevant redshift range, $z=2\!-\!3$.\\  

\begin{table}[t]
  
  \refstepcounter{table}\label{table:LAE_sims}

  {\small\textsc{Table}~\thetable.\textemdash\ }%
  {\footnotesize
    A comparison with other example LAE simulation works at $z\sim 2$–$3$.
    “\Checkmark” symbol means the model is fitted to or compared at the level of
    $\chi^{2}$ with the corresponding observable.
    “\Checkmark$_{\rm cc}$” symbol means the model prediction is just compared
    with the corresponding observable as a consistency check.
    If a comparison with observational data is not available, it is indicated by “–”.
    LF/$\,\overline{n}_{\rm g}$ stands for the luminosity function or the integrated
    number density of LAEs, while $w(\theta)/b_{\rm g}$ stands for the angular
    correlation function or just the clustering amplitude in terms of the LAE galaxy bias.
    The last column “Misc.” refers to miscellaneous observables related to LAEs.
  }\medskip

  \centering
  \begin{tabular}{c|cccccc}
    Reference & 
    \begin{tabular}{c}
      $L_{\rm box}$\\ ${\rm cMpc}/h$
    \end{tabular}
    & Galaxy model                                                       
    & Ly$\alpha$ RT model                        
    & LF/$\,\overline{n}_{\rm g}$ 
    & $w(\theta)/b_{\rm g}$ 
    & Misc. \\
    \hline\hline
    This work & 400                          
    & \begin{tabular}{c}
      empirical \\ \texttt{(UniverseMachine)}
    \end{tabular}
    & 
    \begin{tabular}{c}
      spherical shell RT\\ \& mean IGM transmission
    \end{tabular}  
    & \Checkmark /\Checkmark           
    & \Checkmark /\Checkmark  
    & \Checkmark$_{\rm cc}$               \\
    \hline
    \citet{Byrohl:2023mn}          & 35                           
    & \begin{tabular}{c}
      hydro \\ \texttt{(TNG)}
    \end{tabular}                                                         
    & full RT                        
    & \Checkmark /\Checkmark           
    & –         
    & \Checkmark$_{\rm cc}$                \\
    \hline
    \citet{Nagamine:2010PA}
    & 10–100                          
    & \begin{tabular}{c}
      hydro \\ \texttt{(GADGET2)}
    \end{tabular}
    & constant/empirical $f_{\rm esc}$                               
    & \Checkmark /\Checkmark                    
    & – 
    & –                 \\
    \hline
    \citet{Shimizu:2011mn}
    & 100                          
    & \begin{tabular}{c}
      hydro \\ \texttt{(GADGET2)}
    \end{tabular}
    & empirical $f_{\rm esc}$                               
    & \Checkmark /\Checkmark                    
    & \Checkmark$_{\rm cc}$/\Checkmark$_{\rm cc}$ 
    & \Checkmark$_{\rm cc}$                 \\
    \hline
    \citet{Im:2024ApJ}
    & $90\times 90\times 1000$                          
    & \begin{tabular}{c}
      hydro \\ \texttt{(Horizon Run 5)}
    \end{tabular}
    & empirical $f_{\rm esc}$                               
    & \Checkmark /\Checkmark                    
    & – 
    & \Checkmark$_{\rm cc}$                 \\
    \hline
    \begin{tabular}{c}
      \citet{Ravi:2024Ph}\\ \citet{Sullivan:2025ar}
    \end{tabular}          
    & 500/250                          
    & \begin{tabular}{c}
      hydro \\ \texttt{(MTNG/Astrid)}
    \end{tabular}
    & empirical $f_{\rm esc}$                               
    & –/\Checkmark                    
    & \Checkmark$_{\rm cc}$/\Checkmark  
    & –                 \\
    \hline
    \citet{Garel:2015mn}
    & 100                          
    & \begin{tabular}{c}
      semi analytic \\ \texttt{(GALICS)}
    \end{tabular}      
    & 
    \begin{tabular}{c}
      spherical shell RT\\ \& no IGM transmission
    \end{tabular}
    & \Checkmark /\Checkmark 
    & – 
    & \Checkmark$_{\rm cc}$ \\
    \hline
    \citet{Gurung-Lopez:2019mn_b}
    & 542                          
    & \begin{tabular}{c}
      semi analytic \\ \texttt{(GALFORM)}
    \end{tabular}      
    & 
    \begin{tabular}{c}
      spherical shell RT\\ \& IGM transmission
    \end{tabular}
    & \Checkmark /\Checkmark           
    & \Checkmark$_{\rm cc}$/\Checkmark$_{\rm cc}$         
    & –                 \\
  \end{tabular}
\end{table}

The most physically well-motivated approach would be to run a cosmological hydrodynamical simulation and a Ly$\alpha$ RT code on the fly or as a post process. 
However, neither is it computationally feasible to resolve the dynamic range of relevant spatial scales, nor can such approach explore the large space of unconstrained physical parameters. 
As a result, most of the LAE models based on hydro simulations still relied on empirical modeling of the Ly$\alpha$ escape fraction \citep{Nagamine:2010PA,Shimizu:2010mn,Shimizu:2011mn,Im:2024ApJ,Ravi:2024Ph,Sullivan:2025ar}.
The only exception is the work of \citet{Byrohl:2023mn} where they adopted the high-resolution \texttt{TNG50} simulation and ran a full Monte Carlo Ly$\alpha$ RT calculation on the Voronoi mesh.  
Still, their simulation volume was limited only to $35\,{\rm cMpc}/h$, and they needed to calibrate their dust attenuation by rescaling the Ly$\alpha$ emission from stellar regions to match the observed Ly$\alpha$ luminosity function $z\sim 2$. 
They have also checked that their LAE model is consistent with other observables such as the Ly$\alpha$ escape fraction versus stellar mass and the EW distribution, but do not consider spatial LAE clustering. \\

Semi-analytic or empirical models such as \texttt{UniverseMachine} would be less physical but more flexible than a hydro approach to model the entire galaxy population in a cosmological volume. 
Although earlier work in this category adopted an empirical model of the Ly$\alpha$ escape fraction \citep{LeDelliou:2006mn, Kobayashi:2008as}, more recent work included Ly$\alpha$ RT in a more physically well-motivated way using the spherically expanding shell model \citep{Garel:2015mn,Gurung-Lopez:2019mn_a}. 
Our work could be categorized here.\\

Since none of the work predicts the relevant LAE observables from the first principle, it would be important to know the observables against which each model is calibrated. 
Our work is unique in the sense that it is the only model that simultaneously reproduces the LAE luminosity function and the angular correlation function with the Ly$\alpha$ RT model being included beyond a simple empirical escape fraction model.\\

\subsection{Limitations of our model and future directions}
\label{subsec:limits}

Our framework leaves room for a number of future improvements.
First, the spherical shell Ly$\alpha$ RT model is currently a simplified and effective model, and it does not capture the real complexities of individual systems of gas density and velocity distribution. 
Although the spherical shell model can successfully explain a large number of observed LAE spectra \citep[e.g.,][]{2017A&A...608A.139G,Gurung-Lopez:2021zELDA,GurungLopez2025ar,Nianias:2025ar} and can be mapped onto different, more complex geometries \citep{Gronke:2016ApJ,Li:2022MN}, a single gas shell with fixed outflow velocity cannot explain both the spectral and surface brightness profile at the same time \citep{Song:2020ApJ}.
This means that our model is currently not suitable for simultaneously studying the impact of Ly$\alpha$ RT on spatial diffusion, e.g., Ly$\alpha$ halos \citep[e.g.,][]{Steidel:2011ApJ,Momose:2014MN,Momose:2016MN,Leclercq:2017AA,LujanNiemeyer:2022ApJ} or Lyman-$\alpha$ intensity mapping \citep[e.g.,][]{Schaan:2021JCAP,Croft:2018MN,Lin:2022ApJS,LujanNiemeyer:2023ApJ,LujanNiemeyer2025ApJ}.
Second, the IGM transmission is treated as just an average at each redshift rather than computing the transmission for each galaxy along its particular line of sight. The relevance of any mass-dependent $T_{\rm IGM}$ also depends on the luminosity threshold of the LAE sample. Our calibration targets relatively bright, narrow-band selections that favor red-peaked line profiles; in representative spectra the blue peak typically contributes a small fraction of the red-peak flux within $\pm10 \AA$  (Fig.~\ref{fig:flux}). Excising the blue side would thus shift $L_{\rm Ly\alpha}$ by only $\sim0.1{-}0.2$ dex. Around our selection cut, the $L_{\rm Ly\alpha}-M_h$ relation exhibits large intrinsic+measurement scatter (several-tenths dex; Fig.~\ref{fig:lum_mhalo}), so such a shift moves only a small fraction of sources across the threshold. First-order normalization/slope differences are therefore absorbed by our joint Luminosity function + $w_\theta$ calibration through the nuisance parameters $(\alpha,\beta,\gamma)$, leaving at most weak, second-order, environment-dependent residuals.
This means that we may need to be careful when we compare the clustering in three dimensions, since IGM transmission could be correlated with the underling large-scale stricture that leads to an anisotropic selection effect \citep{Zheng:2010ApJ,Behrens:2018AA,Gurung-Lopez:2019mn_b}.
Third, Lyman-$\alpha$ powering mechanisms other than nebular emission in star-forming regions could be substantial, depending on scales \citep[e.g.,][]{Byrohl:2023mn}. 
Most importantly for cosmological applications, AGNs contribute significantly to the high end of the LAE luminosity function \citep{Zheng:2016ApJS,Spinoso:2020AA,Zhang:2021ApJ,Liu:2022ApJS,Liu:2022ApJ,Torralba-Torregrosa:2023A&A}. 
Contamination from (obscured) AGN systems needs to be treated as an additional emission channel and radiative transfer treatment.
Finally, the volume of our simulation (side length of $400 \, \mathrm{cMpc}/h$) is relatively small by cosmological standards. 
This imposes restrictions on the applicability of our current mock to cosmological studies with large-scale clustering such as BAO and RSD.\\

In the future, this work can be expanded in promising directions to further enhance its fidelity and predictive capabilities. 
We can explore the selection effects by modeling the wavelength-dependent line-of-sight IGM transmission for the given large-scale density and velocity field of the underlying dark matter simulation. 
IGM transmission features near the halo scale, not captured in the $N$-body simulations, can be informed by high-resolution cosmological hydrodynamical simulations \citep{Byrohl:2020AA, byrohl2025} or explore approximate semi-analytical descriptions (e.g., \cite{2025MNRAS.542.2151L}).
Hence, we can compute the observed flux by convolving our LAE spectra with such direction-dependent line-of-sight IGM transmission, incorporating any large-scale correlations affecting the LAE clustering.
In addition, we plan to scale up our LAE catalog by adopting a larger simulation such as \texttt{Uchuu-UM} \citep{Aung:2023MN}.
Addressing these limitations will open up other exciting scientific investigations. 
For example, a simultaneous fit of our model to the LAE luminosity function and clustering in a much wider redshift range would allow us to constrain the HI properties on IGM scales during reionization in a manner consistent with low redshifts \citep[e.g.,][]{Mason_2015,Umeda:2025ApJS}.
In addition, it is critical to explore in more detail what physical relations are key to reproduce the LAE luminosity function, clustering, and other LAE-related observables. 
We showed that the simple and empirical scaling relations with the expanding shell model do an excellent job of reproducing these observables with few parameter choices. 
In our future work, we will explore more systematically how the model is sensitive to the choice of our model assumptions and the implications of the posterior distributions of the model parameters on the observables such as spectral shapes. 
Such a systematic study would help us refine subgrid physics in a more physical approach, such as a hydrodynamical simulation with a full Ly$\alpha$ RT code.\\

\section{Summary and Conclusion}
\label{sec:conclusion}
We have introduced an empirical, yet physically motivated, approach to simulating realistic LAEs at redshifts $z\sim 2\!-\!3$ designed for ongoing and forthcoming large-scale cosmological surveys. Our method integrates the \texttt{UniverseMachine} galaxy-halo model, with spherical expanding shell model for the Ly$\alpha$ radiative transfer capturing essential processes in both the ISM and IGM. By calibrating three free parameters, a single set of their values successfully reproduced several observational diagnostics simultaneously, including  the UV luminosity function (Fig.~\ref{fig:UV_LF}), observed Ly$\alpha$ luminosity function (Fig.~\ref{fig:LF}), angular clustering (Fig.~\ref{fig:CF}), Ly$\alpha$ equivalent width distribution (Fig.~\ref{fig:EW_dist}), observed Ly$\alpha$ escape fraction (Fig.~\ref{fig:fesc_obs}), systemic velocity offsets (Fig.~\ref{fig:deltav-MUV}) and dust reddening (Fig.~\ref{fig:EBV-fesc}) at both redshifts. Beyond matching these key diagnostics, our approach also delivers predictions for LAE-halo connection, relation between Ly$\alpha$ luminosity and the halo mass and the relative escape efficiency of Ly$\alpha$ photons compared to the UV continuum.\\

Below, we summarize the main steps of our implementation and highlight the key findings.\\

\begin{itemize}
  \item We started with \texttt{SMDPL} cosmological $N$-body simulation combined with the \texttt{UniverseMachine} empirical galaxy-halo model which provides a realistic population of galaxies including their positions, velocities, stellar masses and intrinsic star formation rates. The simulation's moderately large volume ($400\,\,\text{cMpc}/h$) and high resolution (dark matter particle mass of $m_{\rm p} = 9.63 \times 10^{7}\, {\rm M_{\odot}}/h$) allowed us to resolve halos hosting LAEs, while \texttt{UniverseMachine} empirically linked these halos to realistic galaxies with observable properties (Sec.~\ref{sec:initial-phase}).

  \item The original \texttt{UniverseMachine} predicted SFRs underestimates the faint end of the observed UV luminosity function at $z\sim 2\!-\!3$. We adjusted the SFRs to match the faint-end UV luminosity function (while staying consistent with observational uncertainties) to ensure sufficient UV photons for Ly$\alpha$ production. To match the bright end, which can be explained by the attenuation of UV photons by dust, we adopt the empirical approach in \citet{2020MNRAS.492.5167V} (Sec.~\ref{sec:halos}).

  \item Using the recalibrated SFR, we computed the intrinsic Ly$\alpha$ luminosity for each galaxy assuming case-B recombination and a Salpeter IMF. This step connects the SFR to the Ly$\alpha$ emission, but the observed Ly$\alpha$ luminosity depends significantly on radiative transfer effects which we modeled next (Sec.~\ref{sec:lae_model}).

  \item We simulated how Ly$\alpha$ photons escape galaxies by modeling RT through expanding spherical gas shells using \citet{Gurung-Lopez:2021zELDA}. The escape fraction depends on three key parameters: outflow velocity, neutral hydrogen column density and dust optical depth. We used \texttt{UniverseMachine} empirical model alongside observationally motivated scaling relations to determine the key radiative transfer properties (Sec.~\ref{subsec:ism}).

  \item After escaping the ISM/CGM, Ly$\alpha$ photons face further absorption by the IGM. We adopted a redshift dependent mean transmission curve (based on \citet{Laursen:2011ApJ}). We modified only the blue side so that it matches the mean IGM transmission from \citet{Faucher-Giguère_2008} (Sec.~\ref{subsec:igm}).

  \item Combining ISM/CGM and IGM effects, we produced observed Ly$\alpha$ spectra for each LAE. The final observed escape fraction determined whether a galaxy was detected as an LAE. The same intrinsic Ly$\alpha$ luminosity could yield dramatically different observed Ly$\alpha$ luminosities due to RT effects, explaining the scatter in the $L_{Ly\alpha}\!-\!M_{\text{halo}}$ relation (Fig~\ref{fig:lum_mhalo}).

  \item We calibrated three free parameters $(\alpha, \beta, \gamma)$, scaling $V_{\rm exp}$, $N_{\rm HI}$ and $\tau_{a}$, to simultaneously match the Ly$\alpha$ luminosity function and the angular correlation function. A single parameter-set sufficed at both $z\sim2$ and $z\sim3$, validating model's robustness (Sec.~\ref{subsec:calibration}). 

  \item Without further tuning, the model is in agreement with equivalent width distribution, observed Ly$\alpha$  escape fraction, dust reddening and the Ly$\alpha$ velocity offset (Sec.~\ref{sec:other_obs_supports}).

  \item Beyond matching observational data, our simulation offers insights into the underlying galaxy-halo connection for LAEs. Specifically, we found that LAEs predominantly occupy halos within the mass range of approximately $\sim  10^{10} \!-\! 10^{13}\,\, \textrm{M}_{\odot}/h$. Our halo-occupation analysis (Fig.~\ref{fig:hod}) shows that the central occupation of observed LAEs closely follows the intrinsic LAE sample at low to intermediate halo masses but shows a pronounced suppression at the high-mass end due to declining Ly$\alpha$ escape fractions. Our satellite fraction prediction changes from $8.7\%$ to $12.4\%$ for our different LAE samples at different redshifts. The central LAE occupation peaks around $10^{11} - 10^{12}\,\,\text{M}_{\odot}$ before falling off towards larger halo masses reflecting the combined effects of the dust attenuation and metallicity. We demonstrate that lower-mass halos typically lack sufficient Ly$\alpha$ luminosity or escape fraction to be detectable, whereas higher-mass halos may suppress observable Ly$\alpha$ emission due to enhanced metallicity or dust (Sec.~\ref{sec:lae_properties}).

  \item We also examined the potential impact of assembly bias but found their influence to be minor, suggesting that halo mass remains primary driver of LAE clustering. Nevertheless, detecting subtler effects from assembly bias might require future studies with larger samples. Our model also provides predictions for the relationship between Ly$\alpha$ luminosity and halo mass, further demonstrating its versatility (Fig.~\ref{fig:assembly_bias}).

  \item Moreover, we predict the ratio of Ly$\alpha$ to UV escape fractions as an indicator of how efficiently Ly$\alpha$ photons escape relative to the UV continuum. We find a strong anti-correlation of this ratio with the neutral hydrogen column density, indicating that denser hydrogen environments more severely suppress Ly$\alpha$ escape compared to UV photons. (Fig.~\ref{fig:fesc_ratio})
\end{itemize}

Our work bridges the gap between galaxy-scale physics and cosmological surveys by demonstrating that Ly$\alpha$ visibility is not just about star formation. It is a complex interplay of gas dynamics, dust and IGM absorption. The success of our model lies in its simplicity: with only three physically interpretable parameters, we reproduce a wide range of observables. This makes the mock catalog particularly valuable for ongoing and upcoming cosmological LAE surveys, where systematic uncertainties in LAE selection could bias cosmological measurements. We make our LAE catalog and spectra publicly available upon publication.\\

In the future, we will incorporate additional physical processes and observational constraints to further improve the fidelity of our LAE model. 
Importantly for precision cosmology, our framework will enable us to explore LAE selection and distortion effects due to the complex radiative transfer across spatial scales.
Nevertheless, our framework offers a crucial first step in linking dark-matter halos to LAE observables, turning the LAEs from cosmological nuisances into powerful probes of both galaxy evolution and the large-scale structure.\\

\section*{Acknowledgments}
The authors are grateful to the HETDEX collaboration for their helpful comments. 
The authors thank Siddhartha Gurung-L\'{o}pez for the development of this work in the early stages. 
HK and SS acknowledge the support for this work from by National Science Foundation grant, NSF-2219212. 
SS is also supported by the U.S. Department of Energy, Office of Science, Office of High Energy Physics under DE-SC0024694. 
SS is supported in part by World Premier International Research Center Initiative, MEXT, Japan.
Part of this work was carried out by CB as JSPS International Research Fellow.
MG thanks the Max Planck Society for support through the Max Planck Research Group, and the European Union for support through ERC-2024-STG 101165038 (ReMMU).
This work used the high-performance computing resources of \textit{The Mill} at Missouri University of Science and Technology (DOI: 10.71674/PH64-N397).
This work was performed in part at the Aspen Center for Physics, which is supported by a National Science Foundation grant, PHY-2210452.\\

The CosmoSim database used in this paper is a service by the Leibniz-Institute for Astrophysics Potsdam (AIP).
The MultiDark database was developed in cooperation with the Spanish MultiDark Consolider Project CSD2009-00064.
The authors gratefully acknowledge the Gauss Centre for Supercomputing e.V. (www.gauss-centre.eu) and the Partnership for Advanced Supercomputing in Europe (PRACE, www.prace-ri.eu) for funding the MultiDark simulation project by providing computing time on the GCS Supercomputer SuperMUC at Leibniz Supercomputing Centre (LRZ, www.lrz.de).

\appendix

\section{A. Simulation Methodology and Technical Details}
\label{app:technical}

In this appendix, we present additional details and tests to support the numerical setup used in our LAE modeling framework. First in Appendix~\ref{sec:app-nbody}, we assess how varying the mass resolution in different $N$-body simulations affects the resulting halo mass function. From this analysis, we identify the mass limit below which the halos become unreliable. Next, Appendix~\ref{sec:app-UVLF-zevol} examines how the UV luminosity function evolves with redshift in our recalibrated \texttt{UniverseMachine} catalog, showing that our model remains consistent within the redshift range $1.9 \leq z \leq 3.2$. Finally, Appendix~\ref{sec:app-orphans} discusses orphan galaxies, which are indeed subhalos that are no longer tracked by the halo finder. We quantify how these orphans influence LAE observables, particularly their impact on the faint-end of the luminosity function and small-scale clustering.\\

\subsection{A.1. The impact of $N$-body simulation mass resolution}
\label{sec:app-nbody}
\begin{figure}
\centering
\includegraphics[scale=0.42]{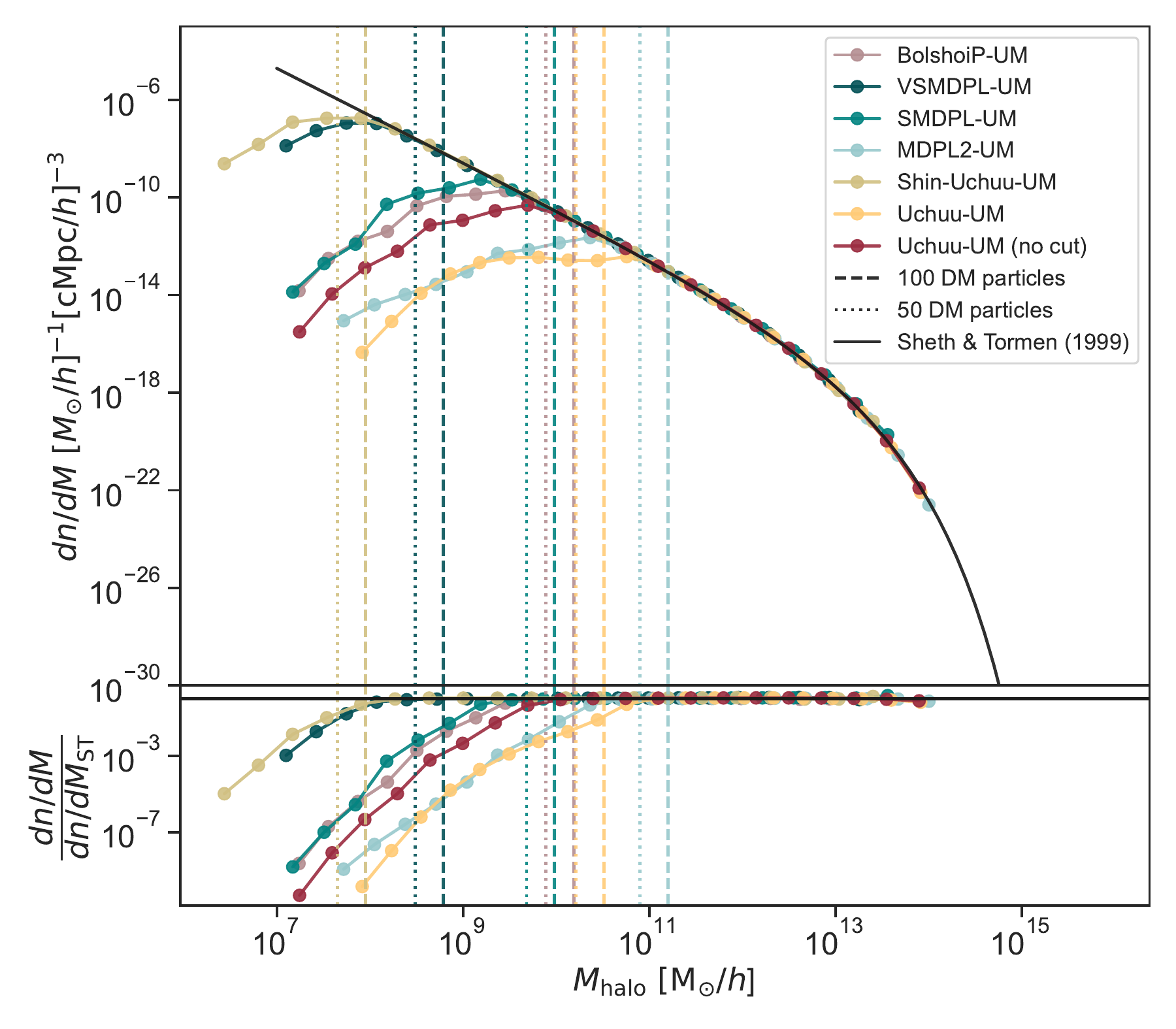}
\caption{Halo mass function comparison for \texttt{BolshoiPl-UM} (Bolshoi series), \texttt{MDPL2-UM}, \texttt{SMDPL-UM}, and \texttt{VSMDPL-UM} (MultiDark Planck series), and \texttt{Uchuu-UM} and \texttt{Shin-Uchuu-UM} (Uchuu series). The solid black curve denotes the \protect\cite{1999MNRAS.308..119S} halo mass function. Dashed vertical lines indicate the mass corresponding to $100$ dark matter particles for each catalog, and dotted vertical lines show the mass of $50$ particles (both color-coded with the same color used for the catalog). The lower panel presents the ratio of each simulation's mass function to the \protect\cite{1999MNRAS.308..119S} baseline, highlighting how resolution affects halo abundance in different $N$-body runs. The yellow curve is the publicly available \texttt{Uchuu-UM} catalog which include a cut of $M_* > 10^{8}\, \,M_{\odot}$.}
\label{fig:app_hmf}
\end{figure}

In Sec.~\ref{sec:initial-phase}, we adopt the \texttt{SMDPL} simulation as our baseline for building a realistic LAE catalog. A crucial motivation behind this choice is the simulation's mass resolution. This is because when halos are composed of too few particles, their star formation rates become poorly determined, and in extreme cases, low-mass halos could disappear entirely even though they can host faint LAEs. To quantify these resolution limits, we compare some widely used $N$-body plus \texttt{UniverseMachine} catalogs, each derived from a different simulation with different dark matter particle mass resolution (summarized in Table~\ref{Table:app_sims_res}). These include \texttt{BolshoiPl-UM} (Bolshoi series), \texttt{MDPL2-UM}, \texttt{SMDPL-UM}, and \texttt{VSMDPL-UM} (MultiDark Planck series), as well as \texttt{Uchuu-UM} and \texttt{Shin-Uchuu-UM} (Uchuu series). Fig.~\ref{fig:app_hmf} compares their halo mass functions against the analytic \citet{1999MNRAS.308..119S} prediction. Vertical lines show the halo masses corresponding to 100 and 50 dark matter particles color-coded for each catalog. In our fiducial \texttt{SMDPL-UM} case, 100 particles correspond to a halo mass of roughly $9\times 10^{9}\, \text{M}_\odot/h$, well below the typical LAE host halo mass $\gtrsim 10^{11}\,\rm{M}_{\odot}$ at $2\lesssim z\lesssim 3$ \citep{Ouchi:2020AR} ensuring that we could resolve the halos that matter for our LAE sample.\\

\begin{table}[H]
  
\refstepcounter{table}\label{Table:app_sims_res}

  {\small\textsc{Table}~\thetable.\textemdash\ }%
  {\footnotesize
    A summary of six distinct \texttt{Nbody–UniverseMachine} catalogs,
    each with its own box size, particle count, and mass resolution.
    \texttt{SMDPL-UM}, the focus of this work, provides an optimal
    trade-off between moderate volume ($400\,\mathrm{cMpc}/h$) and
    comparatively high mass resolution ($9.63\times10^{7}\,\mathrm{M_\odot}/h$),
    enabling it to sufficiently resolve the halos that typically host
    LAEs at $z\sim2\!-\!3$.
  }\medskip

  \centering
  \scalebox{1}{%
    \begin{tabular}{c|c|c|c}
      \textsc{Nbody-UniverseMachine} catalogs & Boxsize & Number of particles & Mass Resolution \\ 
      \hline\hline
      \textsc{BolshoiPl-UM}  & $250\,\,\mathrm{cMpc}/h$  & $2048^3$   & $1.55\times10^{8}\,\,\mathrm{M_\odot}/h$ \\
      \hline
      \textsc{VSMDPL-UM}     & $160\,\,\mathrm{cMpc}/h$  & $3840^3$   & $6.17\times10^{6}\,\,\mathrm{M_\odot}/h$ \\
      \textsc{SMDPL-UM}      & $400\,\,\mathrm{cMpc}/h$  & $3840^3$   & $9.63\times10^{7}\,\,\mathrm{M_\odot}/h$ \\
      \textsc{MDPL2-UM}      & $1000\,\,\mathrm{cMpc}/h$ & $3840^3$   & $1.51\times10^{9}\,\,\mathrm{M_\odot}/h$ \\
      \hline
      \textsc{Shin-Uchuu-UM} & $140\,\,\mathrm{cMpc}/h$  & $6400^3$   & $8.97\times10^{5}\,\,\mathrm{M_\odot}/h$ \\
      \textsc{Uchuu-UM}      & $2000\,\,\mathrm{cMpc}/h$ & $12800^3$  & $3.27\times10^{8}\,\,\mathrm{M_\odot}/h$ \\
    \end{tabular}%
  }
\end{table}

\subsection{A.2. Redshift evolution of the model UV luminosity function}
\label{sec:app-UVLF-zevol}
\begin{figure}  
\centering
\includegraphics[scale=0.42]{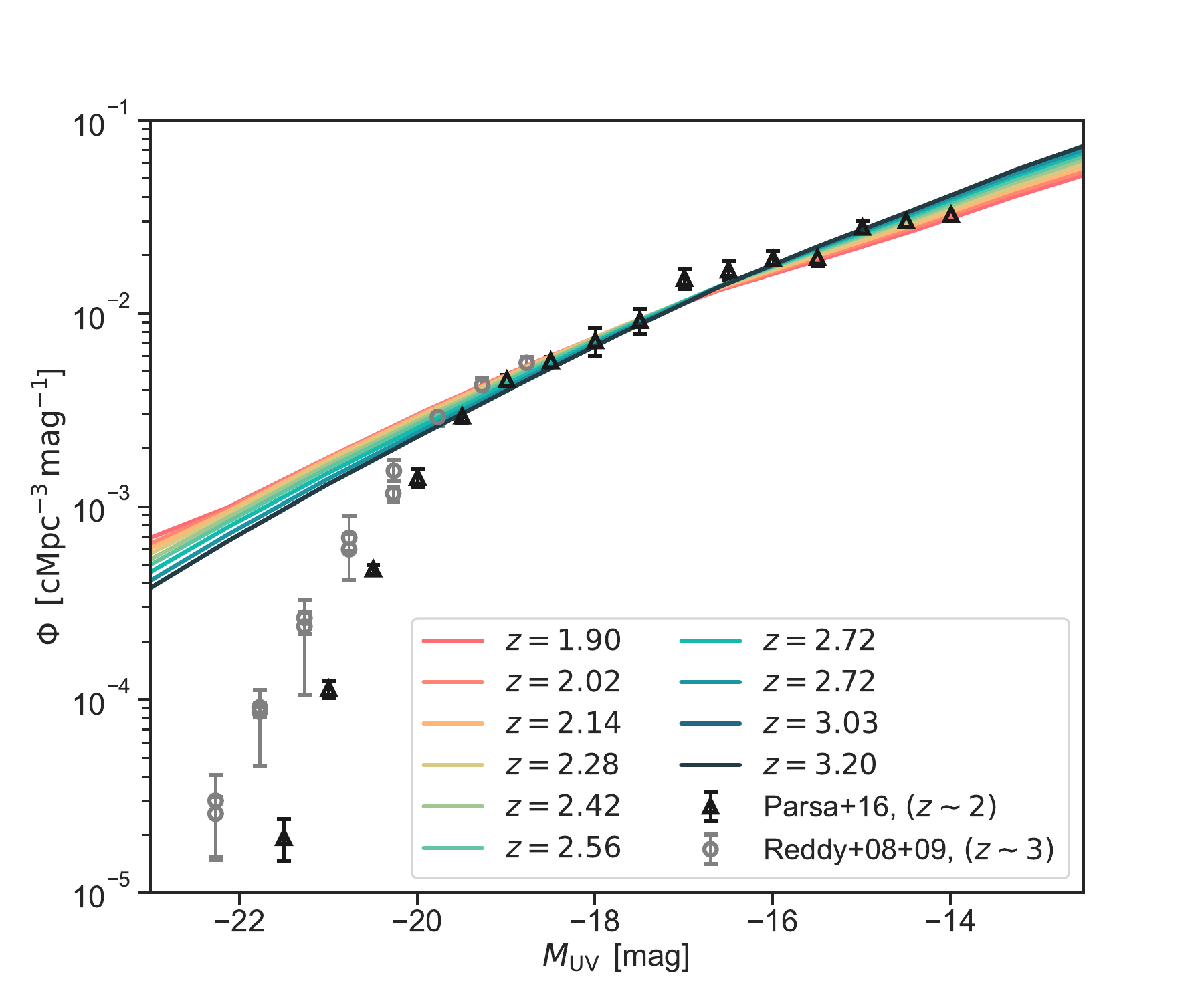}
\caption{Dust-free rest-frame UV luminosity functions predicted by the \texttt{UniverseMachine} model after applying (Eq.~\ref{eq:SFR_obs}). Colored curves show 10 snapshots from $z=1.9$ (red) to $z=3.2$ (dark blue, see legends for exact redshifts). Black triangles and gray circles are the measurements from  \cite{10.1093/mnras/stv2857} and \cite{Reddy_2008, Reddy_2009}. The model consistently reproduces the observed faint-end across all redshifts with minimal evolution ($\lesssim\,0.1 \,\text{dex}$) confirming that dust parameters calibrated at $z=2.42$ and $z=3.03$ suffice for the full $z\sim 2\!-\!3$ range.}
\label{fig:app_MUV_z}
\end{figure}

After recalibrating the SFR in \texttt{UniverseMachine} (Eq.~\ref{eq:SFR_obs} in Sec.\ref{sec:halos}), we confirm our model reproduces the observed UV luminosity function consistently across the redshift range $1.9 \leq z \leq 3.2$. Fig.~\ref{fig:app_MUV_z} presents the predicted dust-free UV luminosity functions for 10 snapshots within this redshift interval, compared to observational data from \cite{10.1093/mnras/stv2857} at $z\sim2$ and \cite{Reddy_2008, Reddy_2009} at $z\sim3$. It shows the model successfully matches the observed UV luminosity function at magnitudes fainter than $M_{\rm UV}\sim -20$. This agreement, with variation of less than $\sim 0.1 \,\rm dex$, holds across the entire redshift range and demonstrate that the weak redshift evolution justifies our calibration strategy: The adjustments made at $z=2.42$ and $z=3.03$, sufficiently characterize the galaxy population across $1.9 \leq z \leq 3.2$ without requiring redshift-dependent corrections. While this appendix focuses on the dust-free faint end, we emphasize that the empirical dust model from \citet{2020MNRAS.492.5167V}  corrects the bright end overprediction in our main analysis.

\subsection{A.3. Orphan Galaxies and their Impact on LAEs}
\label{sec:app-orphans}

In the \texttt{UniverseMachine} galaxy-halo connection model, ``orphan'' galaxies represent satellite galaxies whose host subhalos have fallen below resolution limits of the simulation and are no longer tracked by the halo finder. These orphans originate from subhalos in earlier snapshots but lack an associate dark matter structure in the final halo catalog. To evaluate their impact on our model predictions for LAE populations, we compare model outputs constructed with and without orphans at redshifts $z\sim2$ and $z\sim3$. We define the orphan fraction as: $f_{\rm orphan} \equiv N_{\rm orphan} / N_{\rm LAE}$, where $N_{\rm orphan}$ is the number of orphan LAEs and $N_{\rm LAE}$ is the total LAE population. At $z = 3.03$, $f_{\rm orphan} = 0.02$, rising to $0.03$ at $z = 2.42$. While subdominant, we systematically quantify their influence to ensure robustness.\\

\begin{figure}
\centering
\includegraphics[scale=0.36]{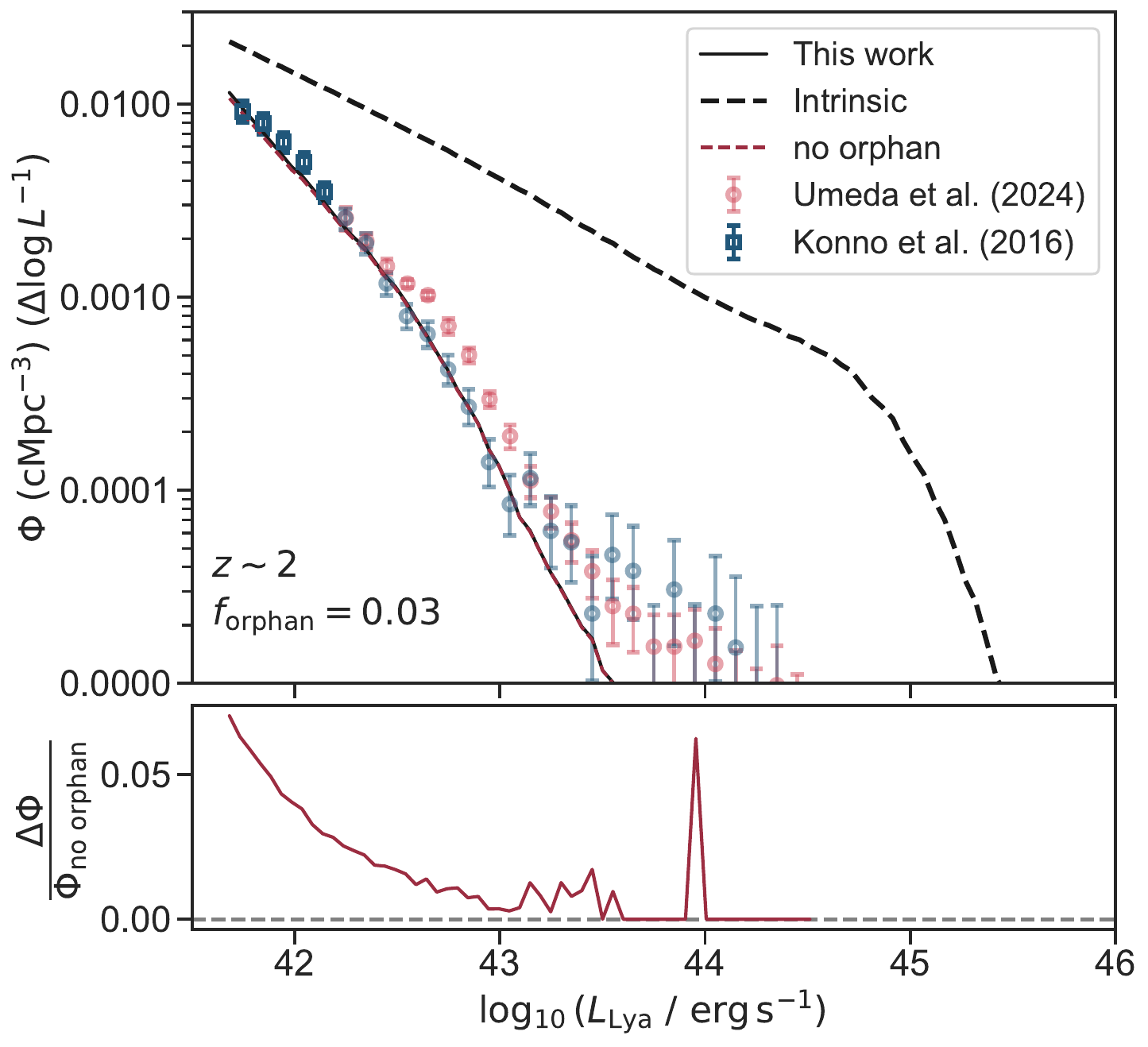}
\includegraphics[scale=0.36]{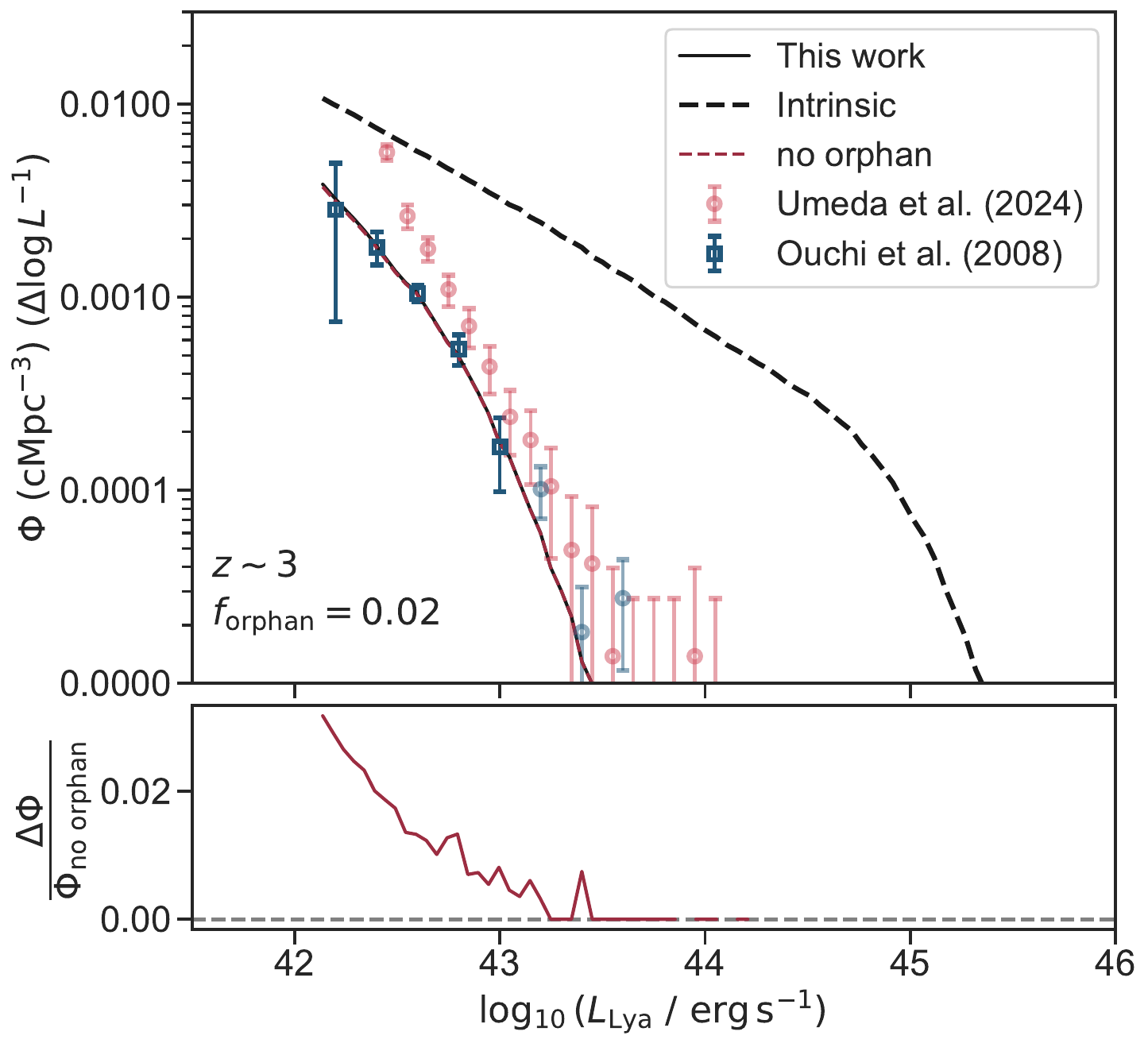}
\caption{Impact of orphan galaxies on the Ly$\alpha$ luminosity function at $z\sim2$ (\textit{left}) and $z\sim3$ (\textit{right}). Black solid and red dashed curves show LAE samples with and without orphans, respectively. The inclusion of orphans boosts the faint end ($\log_{10}\, (L_{\rm Ly\alpha}/\rm erg\,s^{-1}) \lesssim 42.5$) by $\sim5\%$, while leaving the bright end unchanged. Lower panels quantify the relative difference ($(\Phi_{\rm w\,orphan} - \Phi_{\rm no\,orphan}) /{\Phi_{\rm no\,orphan}}$), demonstrating how orphan removal suppresses faint LAE counts. This analysis extends the main luminosity function results from Fig.~\ref{fig:LF} by isolating the specific contribution of orphan galaxies to the LAE population.}
\label{fig:app_LF_no_orphan}
\end{figure}

\begin{figure}
\centering
\includegraphics[scale=0.37]{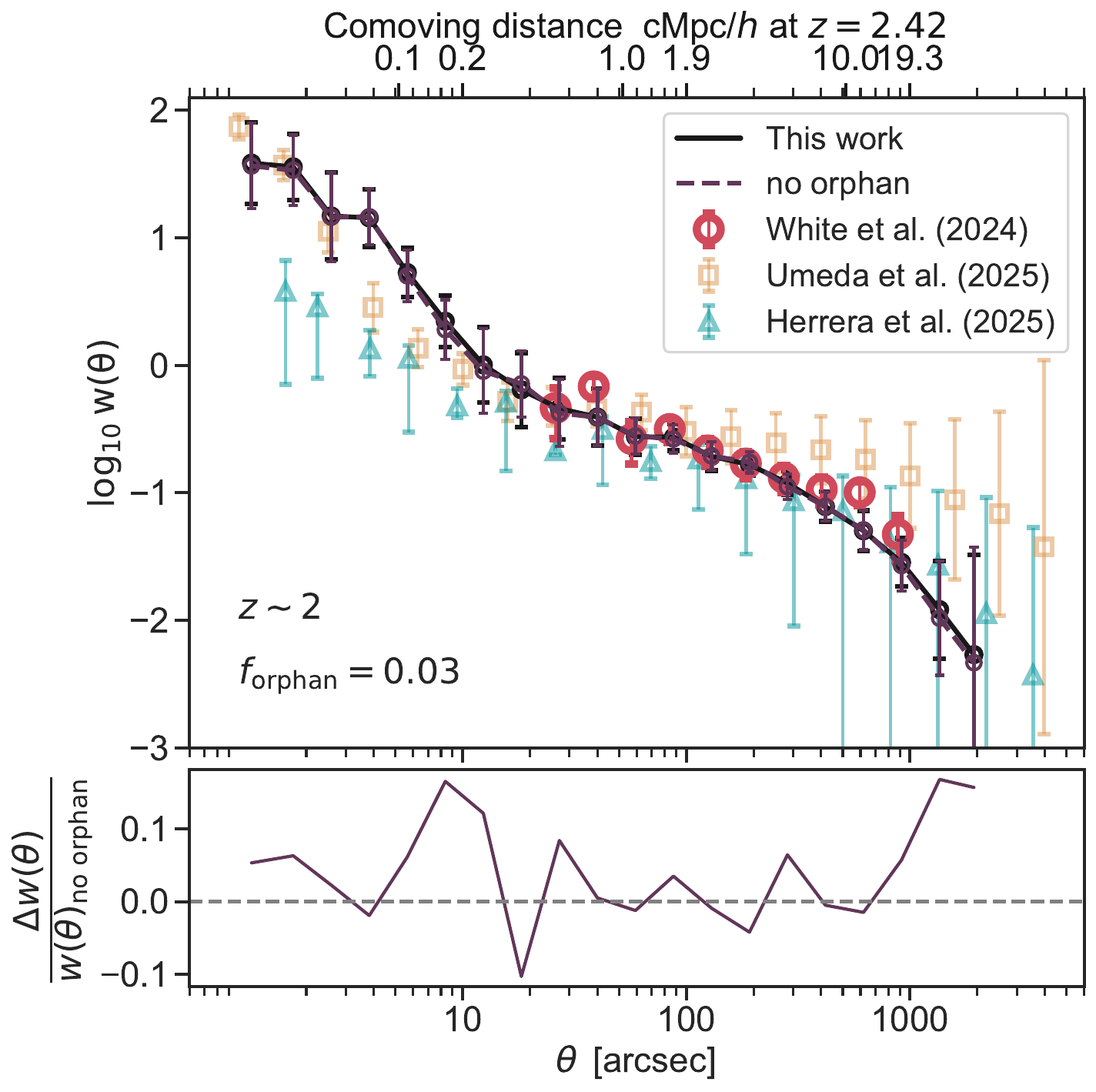}
\includegraphics[scale=0.37]{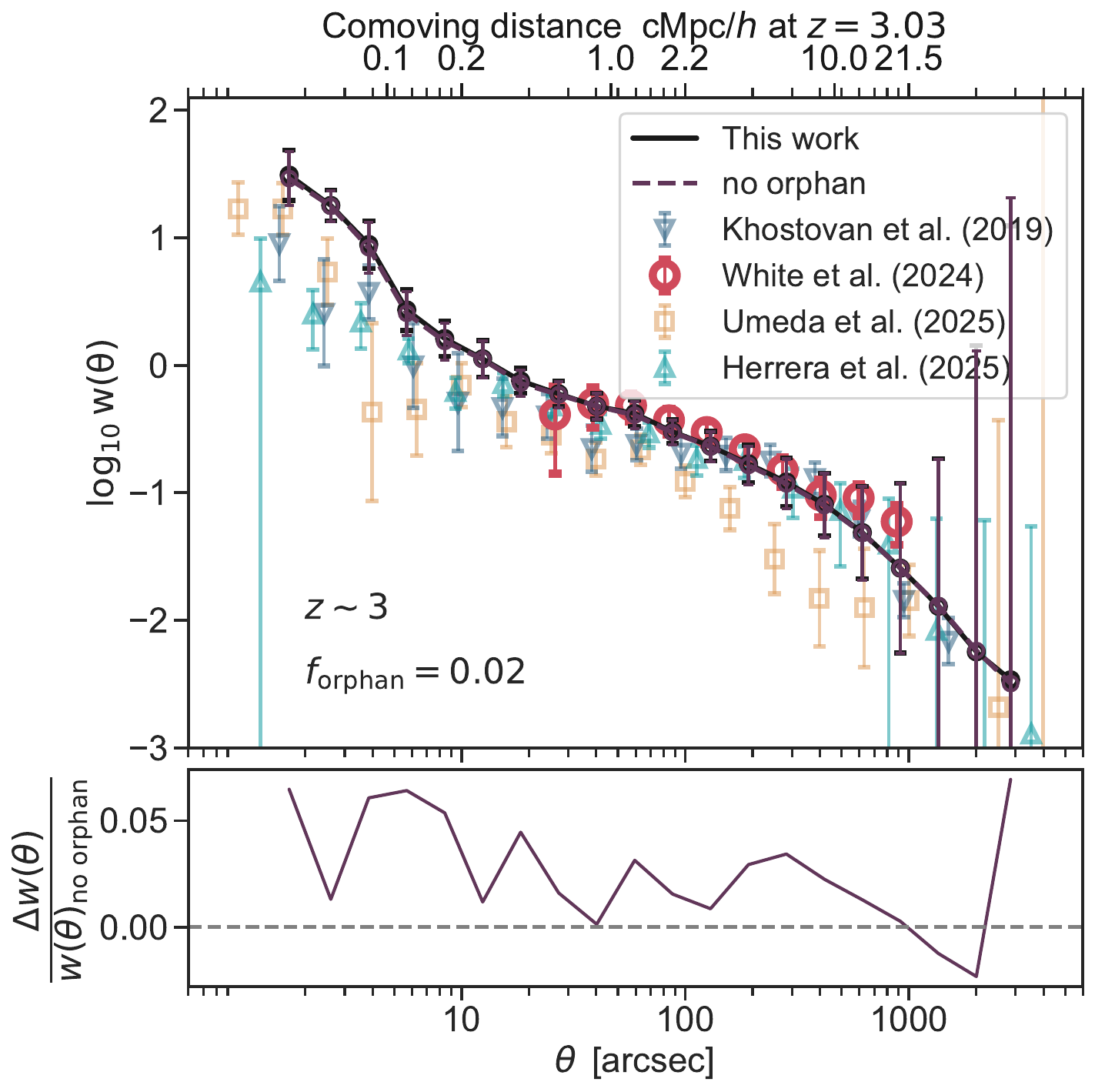}
\caption{Impact of orphan galaxies on LAE clustering at $z\sim2$ (\textit{left}) and $z\sim3$ (\textit{right}). Black solid and red dashed curves show the angular correlation function, $w(\theta)$, for samples with and without orphans, respectively. Orphans enhance small-scale clustering ($\theta \lesssim 10''$) by up to $\sim 10\%$ by contributing to intra-halo satellite pairs, while leaving large-scale clustering ($\theta \gtrsim 100''$) unaffected. Lower panels quantify the relative difference $\big((w(\theta)_{\rm w\,orphan} - w(\theta)_{\rm no\,orphan}) /{w(\theta)_{\rm no\,orphan}}\big)$, confirming their exclusive impact on the 1-halo regime. These results demonstrate how orphan satellites, despite constituting only $2\!-\!3\%$ of the LAE population, disproportionately affect small-scale clustering due to their spatial concentration within halos.}
\label{fig:app_CF_no_orphan}
\end{figure}

\begin{figure}
\centering
\includegraphics[scale=0.38]{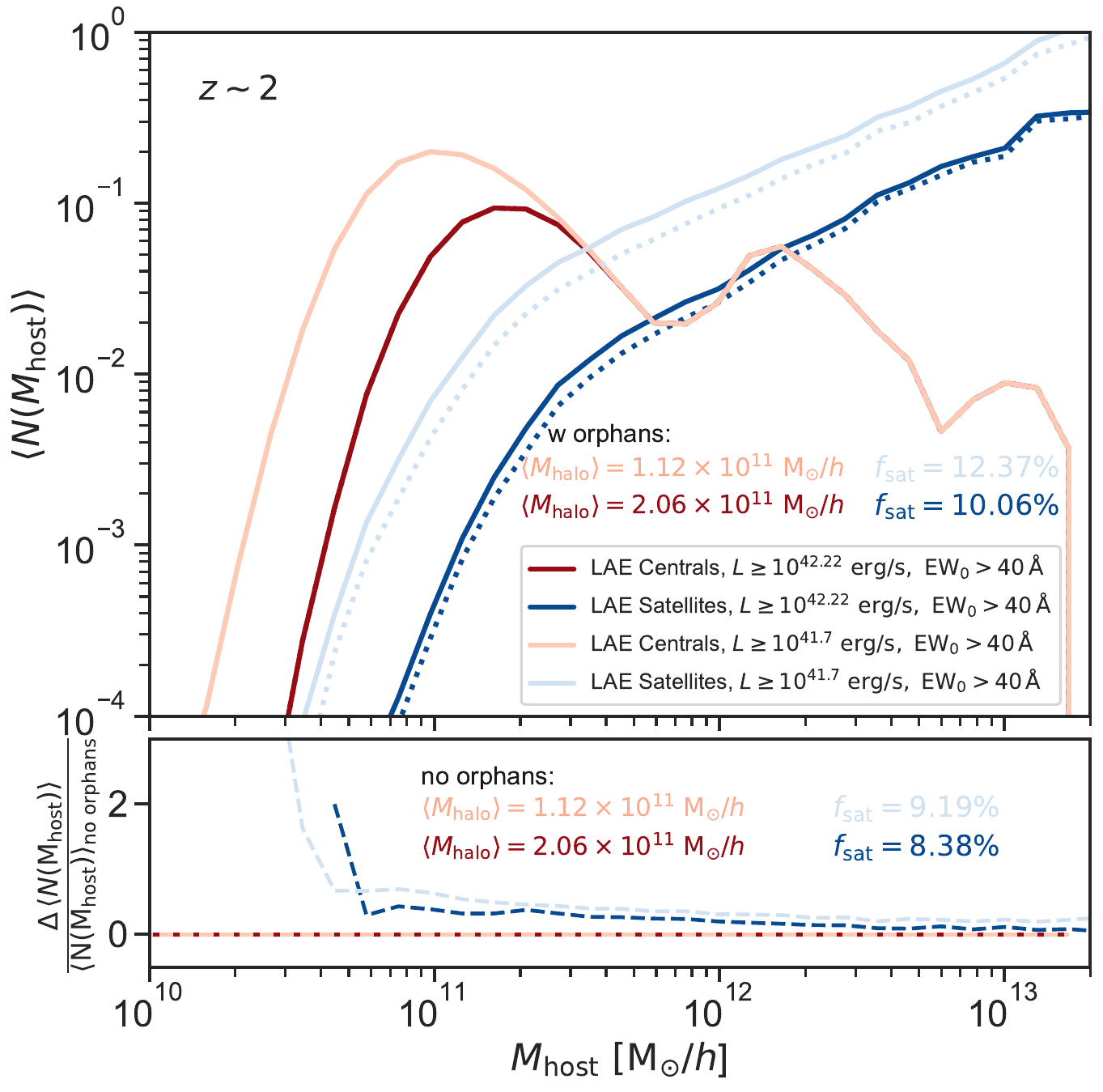}
\includegraphics[scale=0.38]{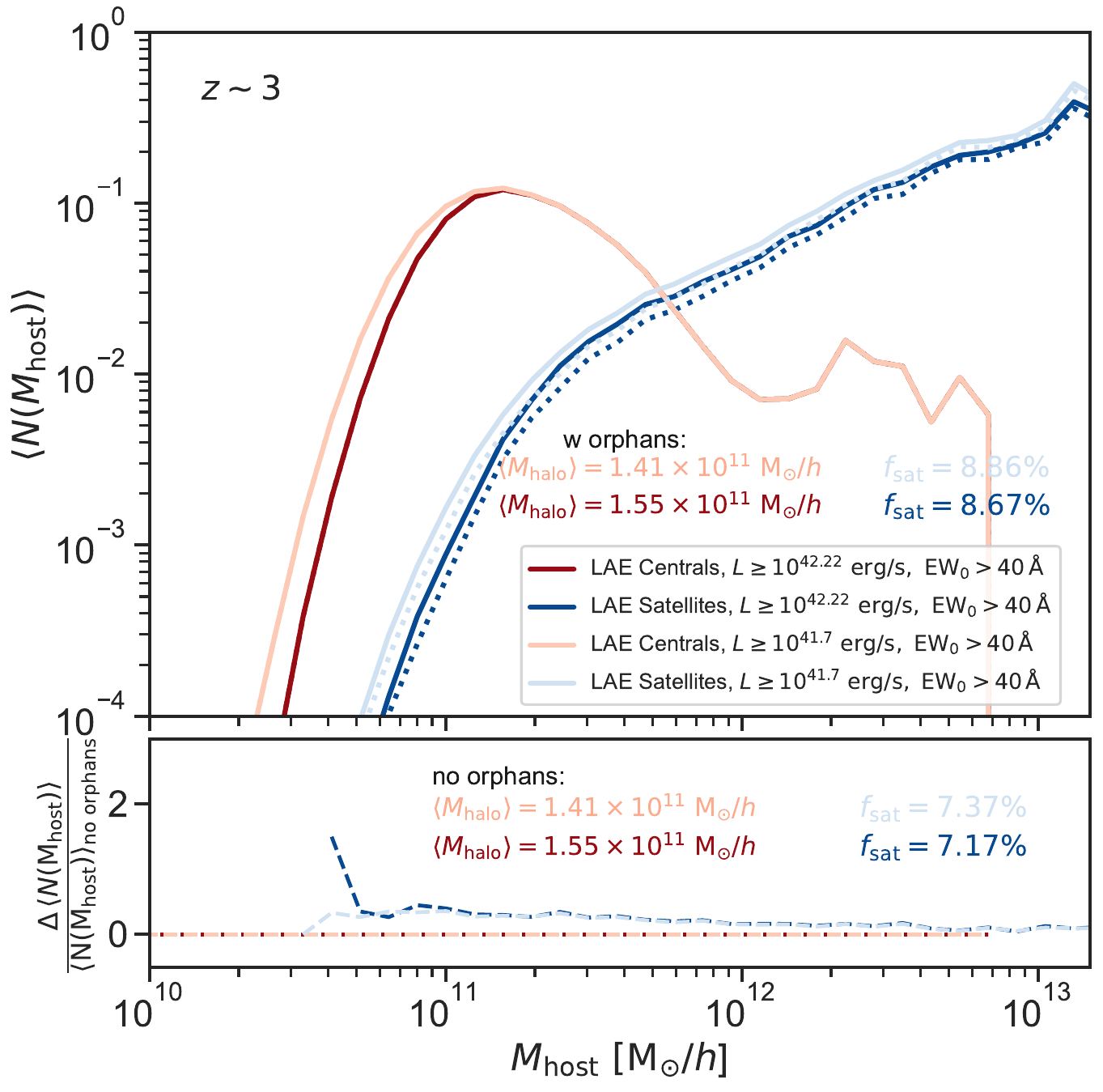}
\caption{Halo Occupation Distribution of LAEs at redshifts $z\sim2$ (\textit{left}) and $z\sim3$ (\textit{right}), comparing samples with (solid lines) and without (dotted lines) orphan galaxies. Central LAEs are shown in red and satellite LAEs in blue. The darker curves represent LAEs with a luminosity cut of $\log_{10}\, (L_{\rm Ly\alpha}/\rm erg\,s^{-1}) > 42.22$, while the lighter curves correspond to a fainter cut of $\rm log_{10}\, (L_{\rm Ly\alpha}/\rm erg\,s^{-1}) > 41.7$. The lower panels quantify the relative difference in HOD between the two samples, $((\langle N(\rm M_{\rm host})\rangle_{\rm w\,orphan} - \langle N(\rm M_{\rm host})\rangle_{\rm no\,orphan}) /{\langle N(\rm M_{\rm host})\rangle_{\rm no\,orphan}})$, highlighting how excluding orphan galaxies, which are exclusively satellites reduces the satellite occupation by up to $\sim 20\%$, particularly in lower mass halos ($M_{\rm host} \lesssim 10^{11}\,M_{\odot}/h$). This removal of orphans leaves the central population of LAEs unchanged also does not affect the mean halo mass of the sample. The results underscore the minor but systematic role of orphans in shaping the satellite fraction and clustering of faint LAEs.}
\label{fig:app_HOD_no_orphan}
\end{figure}

To evaluate their impact, we have computed the effects of excluding orphans from our LAE sample on all key diagnostics. For conciseness, we present results only for the Ly$\alpha$ luminosity function (Fig.~\ref{fig:app_LF_no_orphan}), angular correlation function (Fig.~\ref{fig:app_CF_no_orphan}) and halo occupation distributions (Fig.~\ref{fig:app_HOD_no_orphan}), while discussing the other diagnostics in the main text of this appendix.\\

Beginning with the luminosity function, we find the most noticeable orphan effect appear at the faint end. Fig.~\ref{fig:app_LF_no_orphan} shows that including orphans produces a modest but systematic enhancement of $\sim5\%$ at most for galaxies with $\mathrm{log}_{10}\, (L_{\rm Ly\alpha}/\rm erg\,s^{-1}) \lesssim 42.5$. This boost reflects their contribution as predominantly low-luminosity satellites that would otherwise be absent from the sample. The effect is consistent across both redshifts, though slightly more pronounced at $z\sim2$ where the orphan fraction is slightly higher ($\sim3\%$ at $z\sim2$ vs. $\sim2\%$ at $z\sim3$), suggesting a mild redshift dependence in their impact on the faint-end luminosity function.\\

The orphan population's role as satellites become even more apparent when examining their impact on clustering. Fig.~\ref{fig:app_CF_no_orphan} shows that orphans enhance the 1-halo term of the angular correlation function by up to $\sim 10\%$ at small scales ($\theta \lesssim 10''$), while leaving the 2-halo term unchanged. This scale-dependent effect occurs because orphans contribute exclusively to intra-halo pair counts as unresolved satellites, with their $\sim3\%$ population fraction at $z\sim2$ translating to a proportionally larger effect on small-scale clustering due to their concentrated spatial distribution within halos. The unchanged 2-halo term confirms that orphans do not affect the large-scale bias of the LAE population, consistent with their nature as subdominant satellite systems.\\

The scale-dependent clustering patterns in Fig.~\ref{fig:app_CF_no_orphan}, find their physical explanation in the halo occupation statistics. Fig.~\ref{fig:app_HOD_no_orphan} illustrates that orphans contribute exclusively to the satellite population, accounting for up to $\sim 20\%$ of satellites in low-mass halos $M_{\rm halo} \lesssim 10^{11}\, \mathrm{M_{\rm \odot}}/h$. Their concentrated spatial distribution within these halos naturally explains both the enhanced small-scale clustering seen in Fig.~\ref{fig:app_CF_no_orphan} and the $\sim 20\%$ increase in satellite fraction when orphans are included ($f_{\mathrm{sat}}$ values in blue in Fig.~\ref{fig:app_HOD_no_orphan}). This halo occupation signature simultaneously clarifies why removing orphans reduces the faint-end luminosity function (Fig.~\ref{fig:app_LF_no_orphan}). The missing satellites preferentially reside in lower-mass halos hosting fainter LAEs. The complete lack of orphan contribution to central galaxies further confirms they represent a distinct population from normal satellites with resolved subhalos, consistent with their origin as numerically unresolved systems.\\ 

Beyond the luminosity function and clustering, we assessed how orphan galaxies influence additional LAE observables. Their impact is consistently minor. The relationship between halo mass and Ly$\alpha$ luminosity shifts by less than 0.005 dex. The rest-frame equivalent width distribution, including its exponential slope and high EW tail, shows no statistically significant changes, with differences confined to the $\pm 1\sigma$ uncertainty of the best-fit model. Orphans reduce the Ly$\alpha$ emitter fraction $X_{\mathrm{Ly\alpha}}$ by at most $\sim5\%$ for the faintest UV magnitudes, though this falls below typical Poisson errors. Similarly, the median Ly$\alpha$ escape fraction $f^{\rm obs}_{\rm esc,Ly\alpha}$, shifts by $< 0.02$ across stellar masses. Kinematic diagnostics like velocity offsets, $\Delta v_{\rm Ly\alpha}$, vary by less than $\lesssim 0.25 \, \rm kms^{-1}$, while the anti-correlation between Ly$\alpha$-to-UV escape ratios and $N_{\rm HI}$ remains indistinguishable (shifts $\lesssim 0.01$). The dust reddening trends are also unaffected, with median $f^{\rm obs}_{\rm esc,Ly\alpha}$ vs. $E(B-V)$ varying $\lesssim 0.005$, negligible compared to observational uncertainties. Collectively these tests confirm that orphans, while non-zero in their influence, introduce no systematic biases to the broader LAE population or its inferred physical properties.

\section{B. Scaling Relations - Evidence from Simulations}
\label{sec:app-scalerel}

Our model depends on three crucial scaling relations that link galaxies to their dark matter halos and gas properties. While these relations were initially motivated by observational studies, we demonstrate here that they are also supported by results from hydrodynamical simulations. This appendix shows that our choices for galaxy size, gas content and metallicities, all align well with what cosmological simulations predict. 
The agreement between observations and simulations suggests that our scaling relations reflect the key physics governing LAEs, particularly at redshifts where direct measurements are challenging.

\subsection{B.1. Galaxy size-halo size relation}

Our model connects stellar half-mass radius, $r_{1/2}$, to their halo virial radius, $R_{\rm vir}$ through a simple but physically motivated two-regime prescription:
for larger halos ($R_{\rm vir}> 113\,\, \rm pkpc$), the stellar half-mass-radius scales linearly with the virial radius as $r_{1/2}\approx 0.015\, R_{\rm vir}$, while for smaller halos, the relation flattens to account for the observed compactness of low-mass galaxies (see Fig.~\ref{fig:scaling_rels}, top left panel). This behavior aligns remarkably well with trends seen in cosmological simulations like IllustrisTNG (see \citet{2023MNRAS.520.1630K}), which shows a nearly linear size-halo relation with minimal evolution from $z=0$ to $z=3$. We also included a Gaussian scatter of 0.12 dex motivated by observational as well as simulation works. Our choice of slope, pivot and scatter (see sec.\ref{subsec:scale}) therefore reproduces the size-halo relation for both low- and high-mass halos seen in fully hydrodynamical simulations. As shown in top left panel of Fig.~\ref{fig:app_scale}, the consequence of our scaling relation is that for low-mass halos, the galaxy-size to halo-size ratio reduces by increasing  the halo mass, while at high-halo mass, the ratio remains approximately constant. This relation remains consistent with the results from \citet{2023MNRAS.520.1630K}. This agreement between simulations and observations strengthens our choice of this scaling relation.

\subsection{B.2. Stellar mass-neutral hydrogen mass relation}

When modeling how much neutral hydrogen surrounds galaxies, we start with a simple but effective trend. More massive galaxies tend to have proportionally less atomic gas relative to their stars. This matches what we see in local galaxy surveys like XGASS \citep{10.1093/mnras/sty089}. The relation follows a scatter of 0.1 dex as we have discussed in Sec.~\ref{subsec:scale}. What is particularly useful is that modern simulations like \texttt{IllustrisTNG} \citep{10.1093/mnras/stx3040, 10.1093/mnras/stx3112, 10.1093/mnras/stx3304, 10.1093/mnras/sty2206, 10.1093/mnras/sty618}, \texttt{SIMBA}  \citep{10.1093/mnras/stz937} and \texttt{EAGLE} \citep{10.1093/mnras/stu2058, 10.1093/mnras/stv725} show this relationship holds remarkably well back to $z\sim 2\!-\!3$ (see Fig.~\ref{fig:app_scale}, top right panel). While gas fractions were higher in the early universe, the fundamental connection between stellar mass and HI mass appears stable over time. This means our scaling relation likely reflects the true gas content of LAEs during their peak epoch, even if direct HI measurements at these redshifts remain challenging. The modest scatter we include accounts for the natural diversity in gas accretion histories while staying grounded in both observational and simulation results.

\subsection{B.3. Cold gas metallicity - stellar mass relation}

\begin{figure}
\centering
\includegraphics[scale=0.42]{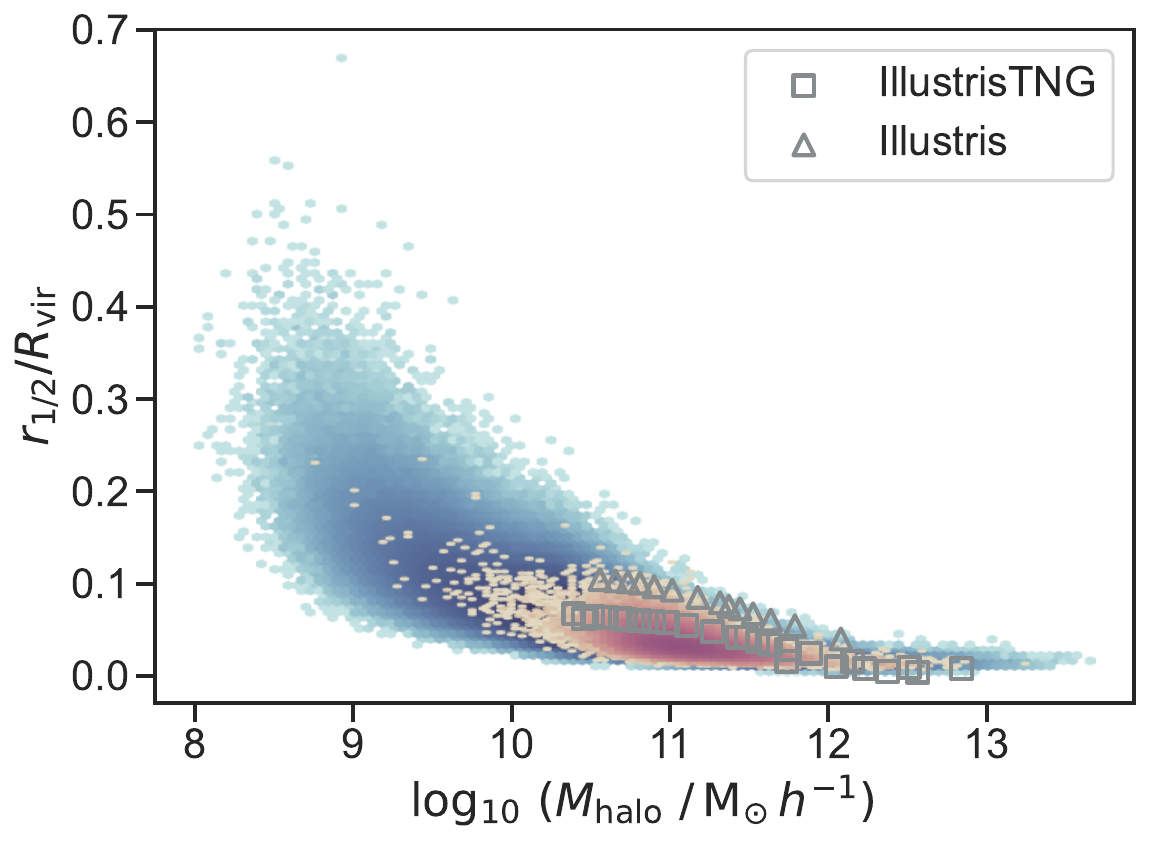}
\includegraphics[scale=0.42]{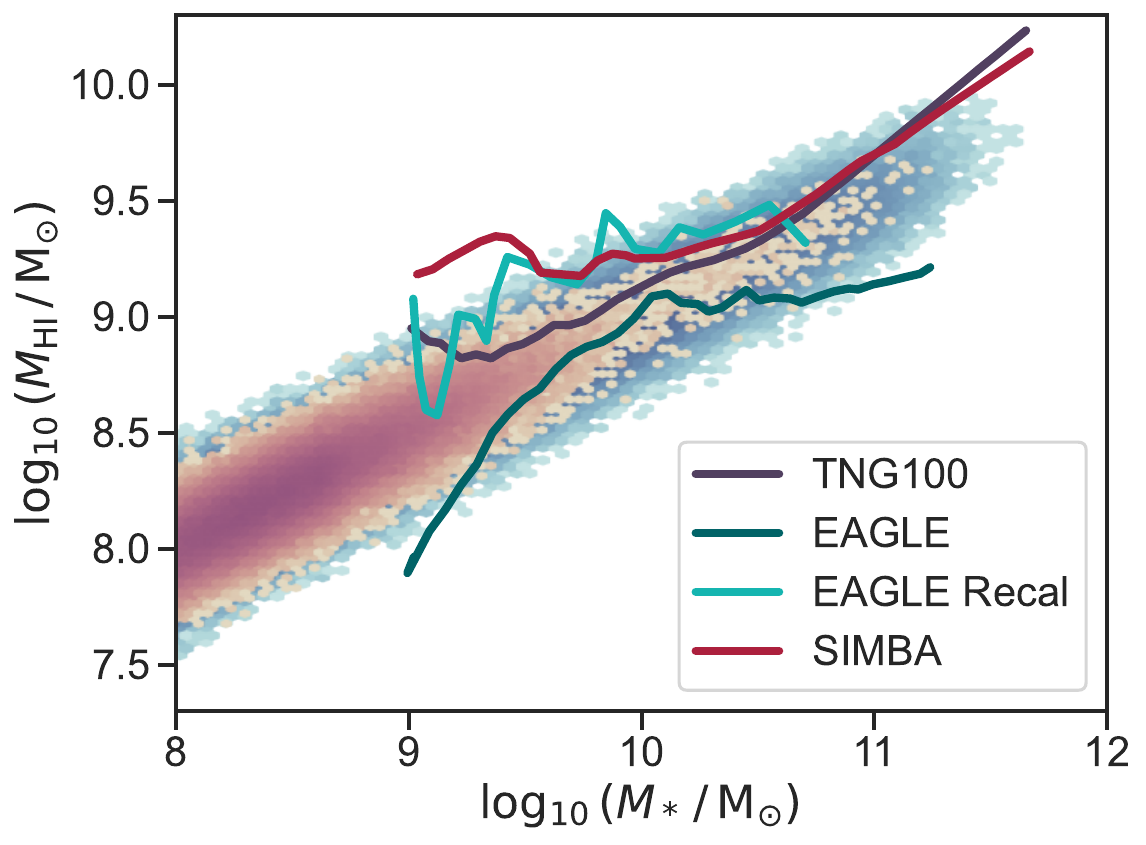}
\includegraphics[scale=0.42]{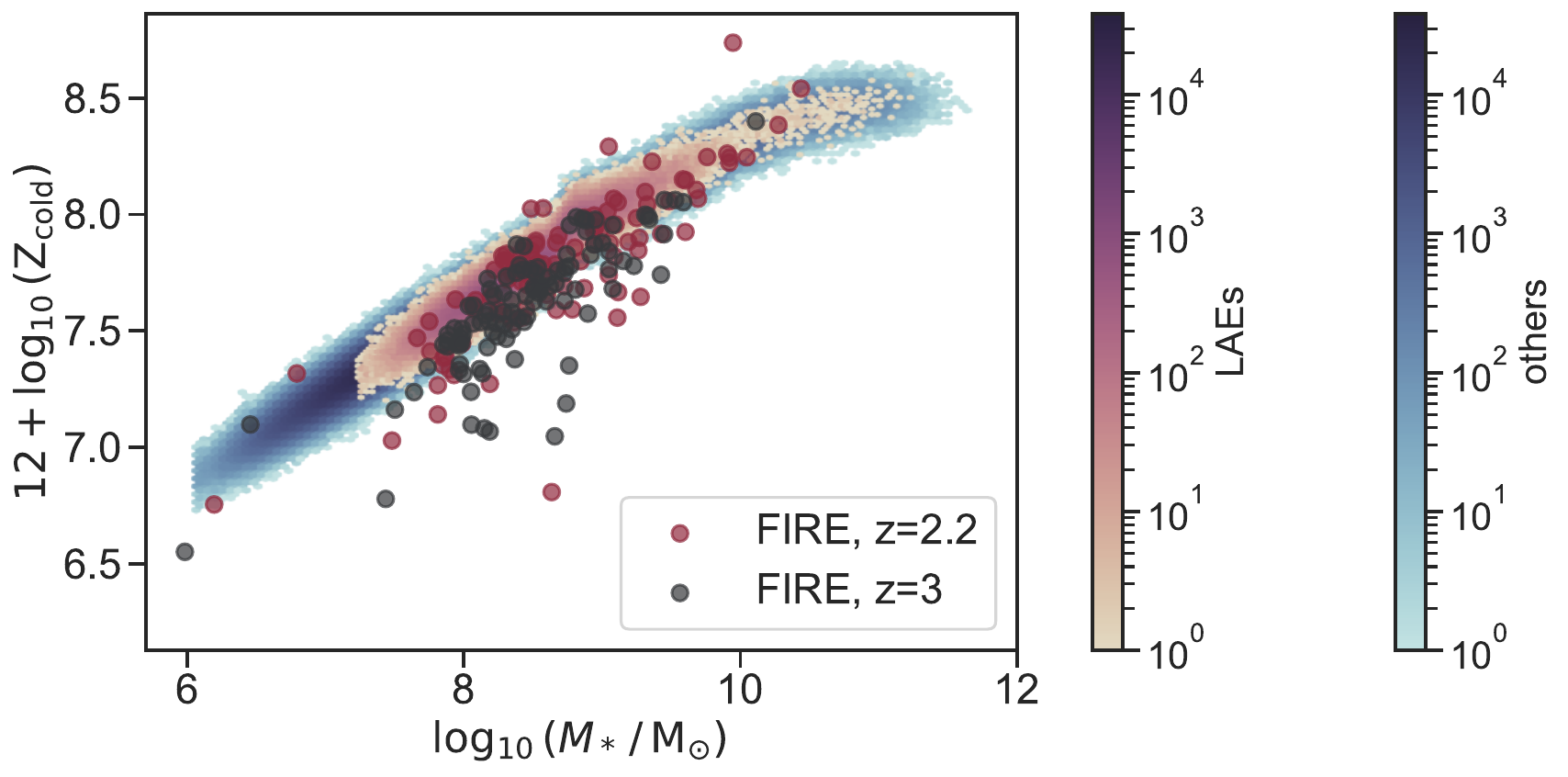}
\caption{Comparison of our adopted scaling relations with hydrodynamical simulation predictions. Same as Fig.~\ref{fig:scaling_rels}, the scatter points show our galaxy sample at $z\sim3$ where LAEs \big($ \mathrm{log}_{10}(L_{Lya}/\,\rm erg\,s^{-1}) \geq 41.7$ and $\mathrm{EW}_0 > 40 \,\, \text{\AA}$\big) are shown in pink while others are presented in blue. \textit{Top left:} Our galaxy size to halo virial radius ratio versus halo mass compared to Illustris and IllustrisTNG results \protect\citep{2023MNRAS.520.1630K}. \textit{Top right:} Neutral hydrogen mass ($M_{\rm HI}$), versus stellar mass $M_*$, showing our empirical relation against simulation predictions. \textit{Bottom:} Gas-phase metallicity, $12 + \rm log_{10}(Z_{\rm cold})$, versus stellar mass with our model compared to FIRE simulations \citep{2016MNRAS.456.2140M}.}
\label{fig:app_scale}
\end{figure}

For modeling the metal content of LAEs, we start with a simple but powerful trend. More massive galaxies tend to have more enriched gas. We adjusted the local mass-metallicity relation from \citet{2004ApJ...613..898T} to better match observations at $z\sim 2\!-\!3$ (see Sec.~\ref{subsec:scale}), lowering the overall normalization while keeping the same slope. This tweak makes physical sense, galaxies at cosmic noon were still building up their metal supplies through ongoing star formation. A key strength of this relation is its consistency across different approaches. While the absolute metallicity values in simulations can vary depending on feedback prescriptions, the fundamental trend we adopt aligns well with results from FIRE simulations \citep{2016MNRAS.456.2140M}. Our conservative 0.05 dex scatter accounts for the natural variations we see in both observations and simulations, some galaxies form stars more efficiently, while others lose metals more easily through outflows. For LAEs specifically, getting this metallicity right is important, because it directly impacts both dust attenuation and the radiative transfer of Ly$\alpha$ photons through their interstellar medium. (see Fig.~\ref{fig:app_scale} for our empirical adjusted relation with simulation predictions at $z \sim 2\!-\!3$ \citep{2016MNRAS.456.2140M}.)

\section{C. Calibration Against Different Observational Measurements}
\label{app:calibration}

To assess how sensitive our model is to observational inputs, we recalibrate it using alternative datasets. In Appendix~\ref{sec:app-umeda}, we examined the impact of luminosity function measurements by adopting the Ly$\alpha$ LF from \cite{Umeda:2025ApJS}, while maintaining our fiducial clustering calibration \citep{White:2024JC} at both redshifts. This test highlights how the faint-end slope of the luminosity function (especially at $z\sim3$, where the discrepancies between \cite{Umeda:2025ApJS} and \citet{Ouchi_2008} measurements are more pronounced) affects our parameter estimates. While the best-fit values shift slightly compared to our fiducial calibration, they remain consistent within uncertainties. In Appendix~\ref{sec:app-khostovan}, we investigate the small-scale clustering regime by fitting to the angular correlation function from \cite{Khostovan:2019mn} while keeping the \citet{Ouchi_2008} luminosity function calibration fixed at $z\sim3$. This analysis provides important constraints on satellite galaxy fraction and the 1-halo term that were less prominent in our fiducial clustering dataset choices. Together these tests offer insights into how different aspects of the observational datasets, to which our model is calibrated, influence our model parameters and predictions.

\subsection{C.1. Parameter constraints and model fitting with Umeda et al. (2024)}
\label{sec:app-umeda}

\begin{figure}
\includegraphics[scale=0.37]{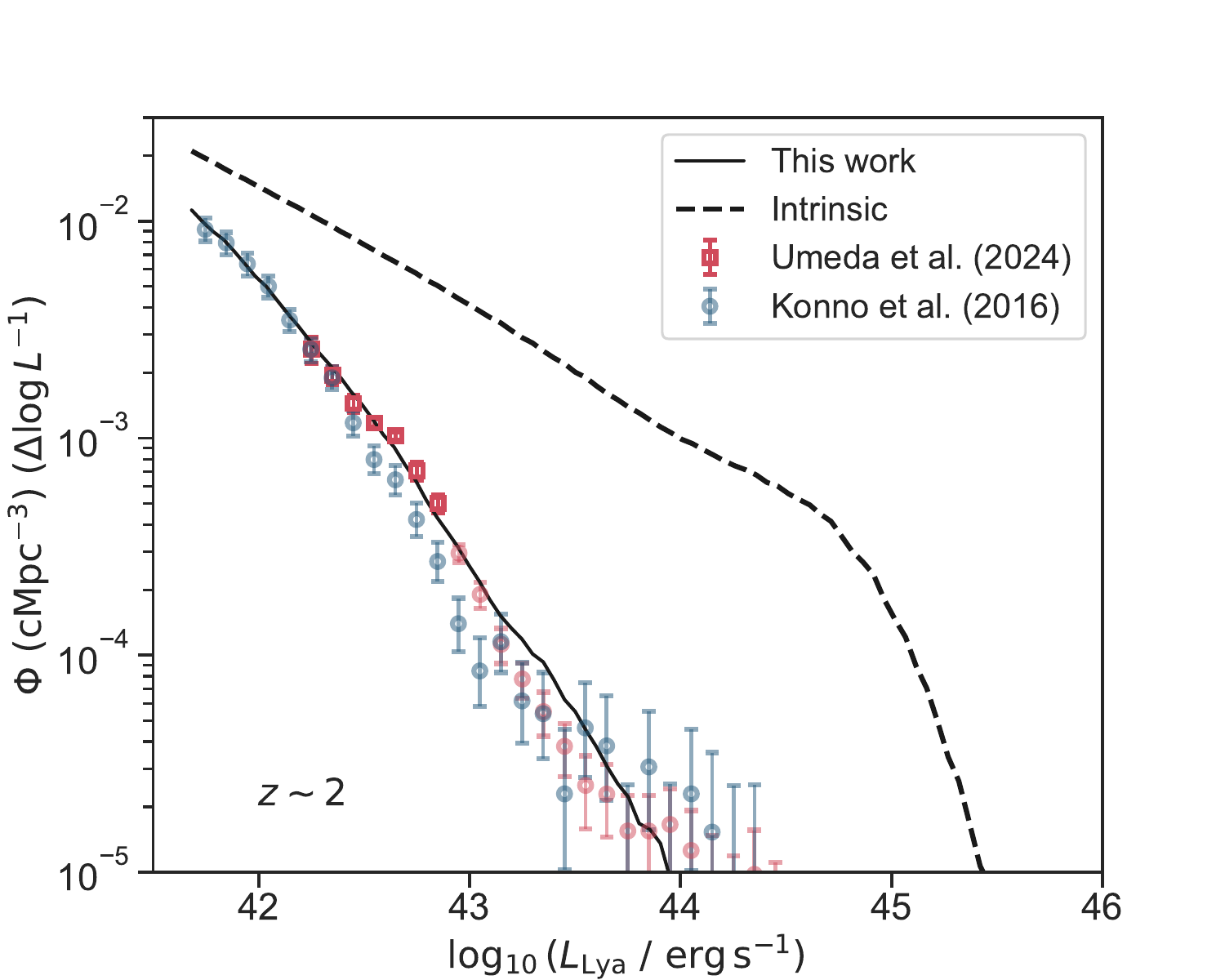}
\includegraphics[scale=0.37]{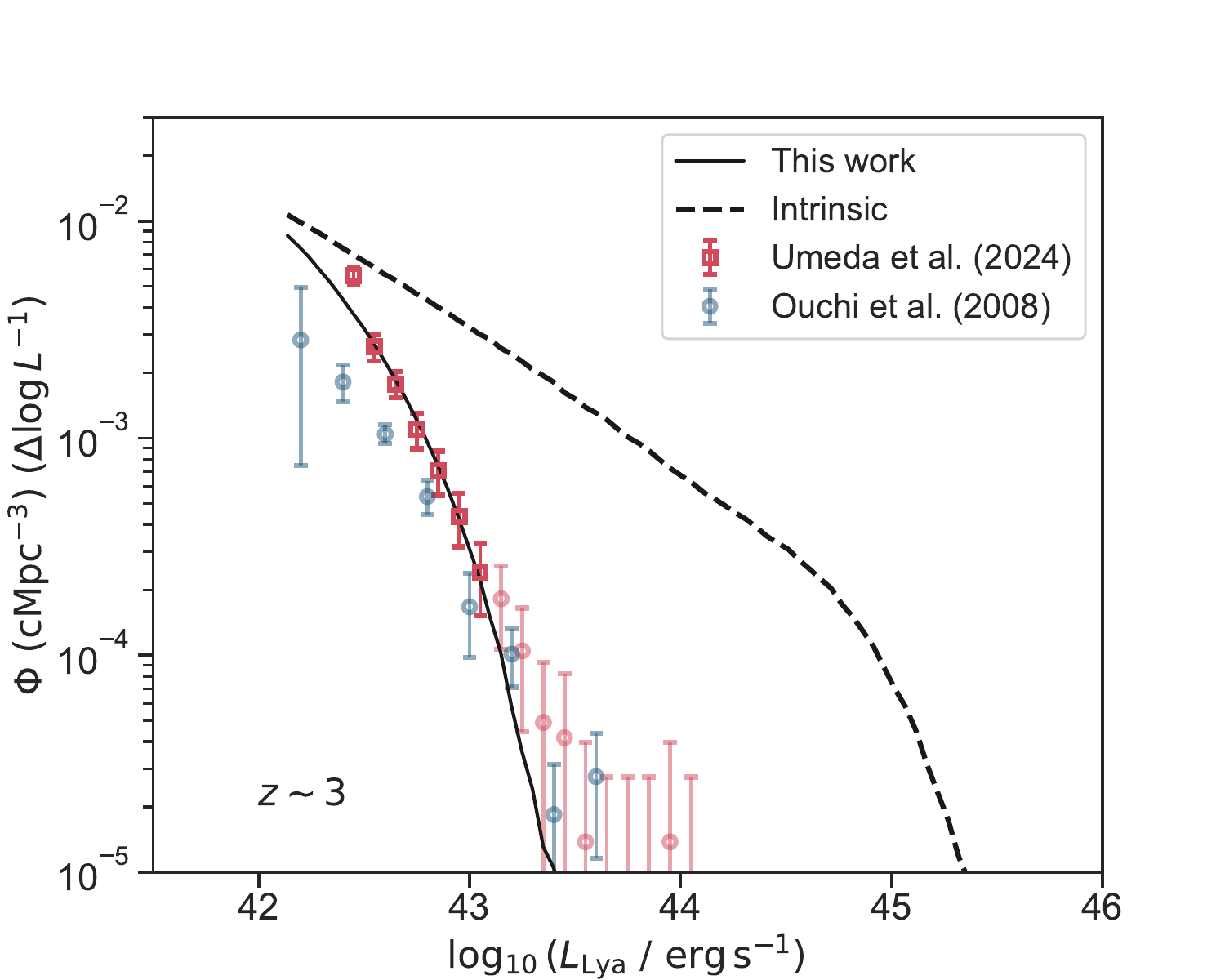}
\caption{Comparison of the Ly$\alpha$ luminosity function predictions from our model when calibrated to \cite{Umeda:2025ApJS} (solid line) against the intrinsic luminosity function (dashed line; before radiative transfer effects). The left panel shows results at $z\sim2$, and the right panel at $z\sim3$. The black line traces the red square data points (from \cite{Umeda:2025ApJS}) at both redshifts. For reference we overplot measurements from  \citet{Konno_2016} at $z\sim2$ and \citet{Ouchi_2008} at $z\sim3$ in faint blue circles. The fainter red data points are not used in our fitting procedure to avoid contamination from AGNs.}
\label{fig:app_umeda_fits}
\end{figure}

In Sec.~\ref{sec:result}, we calibrated our model parameters, $(\alpha, \beta, \gamma)$, using the luminosity functions from \citet{Konno_2016} at $z\sim 2$ and \citet{Ouchi_2008} at $z\sim 3$, finding that a single set of parameters could describe both redshits within uncertainties. To test the sensitivity of our results to the luminosity function measurements, we now recalibrate the model using \cite{Umeda:2025ApJS} luminosity function measurements, while keeping our fiducial clustering calibration fixed to \cite{White:2024JC} at both redshifts. At $z\sim2$, luminosity function measurements from \cite{Umeda:2025ApJS} are brighter than those from \citet{Konno_2016} in the faint end, though the two studies yield comparable slopes. At $z\sim3$, \citet{Umeda:2025ApJS} luminosity function is steeper than \citet{Ouchi_2008} at faint end. The best-fit model for this test is shown in Fig.~\ref{fig:app_umeda_fits} in solid black line. The bright red data points present the data used for the fit.  Our calibration to \citet{Umeda:2025ApJS}, requires adjustments to our free parameters. At $z\sim2$, $(\alpha, \beta, \gamma) = (2.000^{+0.966}_{-1.000}, 0.080^{+0.020}_{-0.039}, 70.0^{+15.78}_{-10.0})$, while at $z\sim3$, $(\alpha, \beta, \gamma) = (0.100^{+0.151}_{-0.080}, 0.030^{+0.011}_{-0.030}, 10.0^{+1.222}_{-0.782})$. Unlike our fiducial calibration, these solutions show no overlap between redshifts. The shifts arise because of the following reasons. The steeper luminosity function as $z\sim3$ demands lower $\alpha$ (weaker outflow) and $\gamma$ (less dust), while the $z\sim2$, data require stronger outflow and dust attenuation. Moreover, the sample selection also may probe different LAE populations especially at low luminosities. Also narrow-band filters as well as the cosmic variance could affect the slope of the measurements.

\subsection{C.2. Angular correlation function fit to Khostovan et al. (2019)}
\label{sec:app-khostovan}

\begin{figure}
\centering
\includegraphics[scale=0.42]{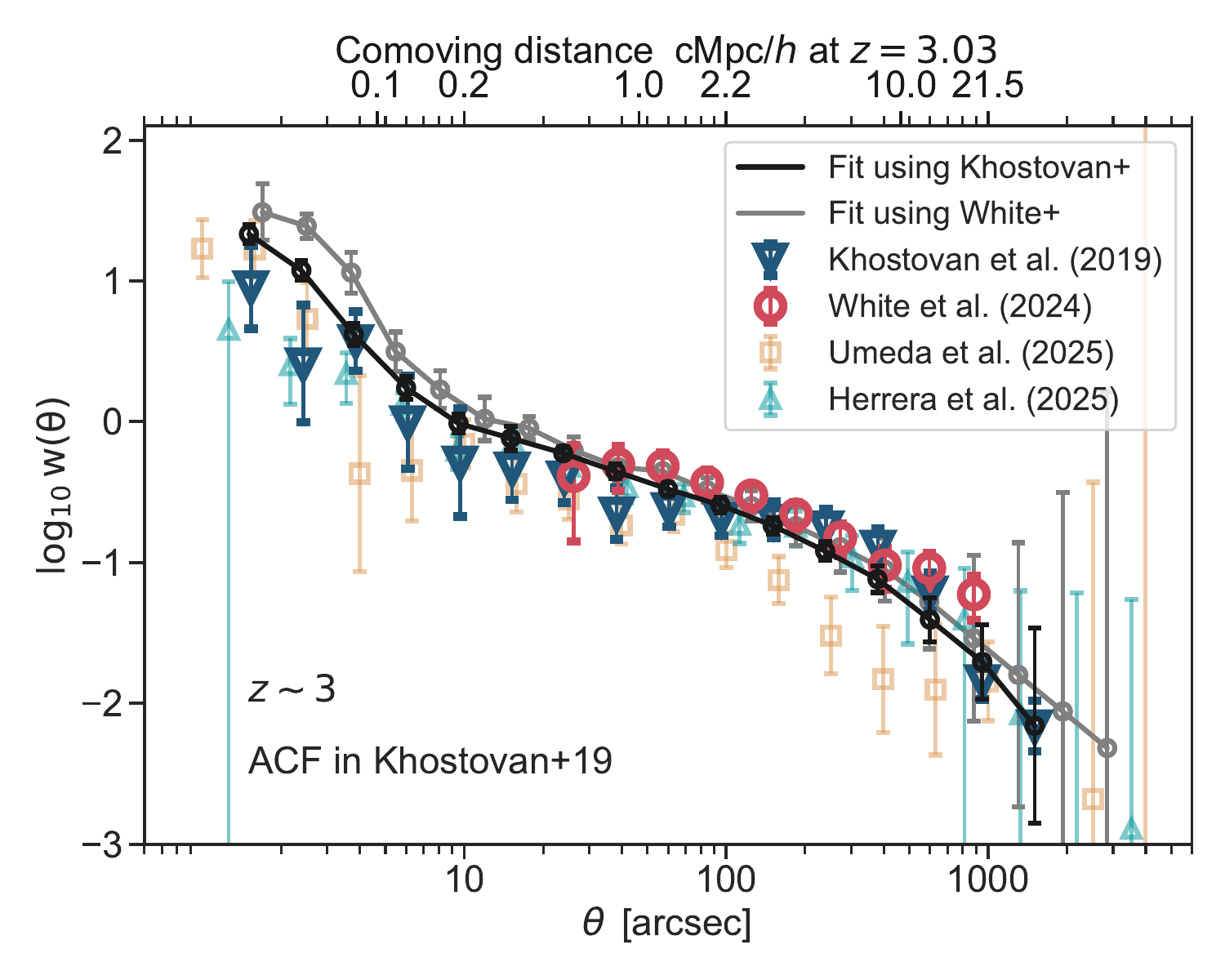}
\caption{Angular correlation function, $w(\theta)$, of LAEs at $z\sim3$. The gray curve shows our fiducial model calibrated to \cite{White:2024JC}, while the black curve shows the model fit to \cite{Khostovan:2019mn}. The two curves diverge most strongly at small angular separation, where the 1-halo term dominates. Because the \cite{White:2024JC} measurements do not probe those small scales, the 1-halo contribution in our fiducial model is not well constrained.}
\label{fig:app_khostovan_fits}
\end{figure}

To investigate the model's adaptability to different clustering constraints, we recalibrated the angular correlation function using the measurements from \citet{Khostovan:2019mn} at $z\sim3$ (while keeping the luminosity function fixed to \cite{Ouchi_2008}). The \citet{Khostovan:2019mn} measurement is based on 1198 LAEs detected over a $1.38\, \rm deg^{2}$ field using the NB497 filter, extend to smaller angular scales $(\theta \approx 1^{''})$ than our fiducial dataset \citep{White:2024JC} and provide valuable constraints on the 1-halo regime. Although their sample has a higher estimated contamination fraction ($\sim 15\%$), the median Ly$\alpha$ luminosity $\big( \log_{10}(L_{\rm Ly\alpha}\,/{\rm erg\,s^{-1}}) = 42.2\big)$ is also higher than in our fiducial sample. Fig.~\ref{fig:app_khostovan_fits} compares the results: the gray curve shows our fiducial model (calibrated to \cite{White:2024JC}; red circles), while the black curve shows the model recalibrated to \citet{Khostovan:2019mn} (blue triangles). Recalibrating the parameters successfully reproduces the \citet{Khostovan:2019mn} measurements, primarily by adjusting the satellite fraction to match their higher luminosity threshold, which naturally excludes fainter satellites. This demonstrates the framework's flexibility to adapt to different observational datasets through parameter adjustments informed by their specific selection functions.\\

\section{D. Impact of Free Parameters on Ly$\alpha$ Summary Statistics}
\label{app:free-params}

In this appendix, we systematically explore how variations in our three free parameters, $\alpha$ (outflow velocity scaling), $\beta$ (neutral hydrogen column density scaling) and $\gamma$ (dust attenuation scaling), shape the observable properties of LAEs. These parameters were carefully constrained during our model calibration to reproduce key observational diagnostics: the luminosity function and the angular correlation function. By isolating the influence of each free parameter we gain insights into the physical process governing Ly$\alpha$ emission and escape, and how these processes shape the observational properties of LAEs. We can directly see which observables are most sensitive to changes in outflow velocity, neutral hydrogen column density and dust optical depth. Appendix~\ref{sec:app-LF-abc} shows the effects on the luminosity function, while Appendix~\ref{sec:app-CF-abc} explores this effect on the angular correlation function. This analysis not only validates the robustness of our model but also highlights the interplay between the parameters and the observables they govern.

\subsection{D.1. Impact of free parameters on the Luminosity Function}
\label{sec:app-LF-abc}

The luminosity function of LAEs serves as a critical diagnostics for understanding how Ly$\alpha$ photons escape from galaxies and propagate through the ISM and IGM. In our model this escape process is governed by three key physical parameters: $\alpha$, which scales the outflow velocity of neutral hydrogen gas, $\beta$, which determines the column density of the neutral hydrogen, and $\gamma$, which sets the optical depth of dust attenuation. By systematically varying these parameters while holding others fixed, we can dissect their individual effects on the predicted luminosity function.\\

\begin{figure}
\centering
\includegraphics[scale=0.23]{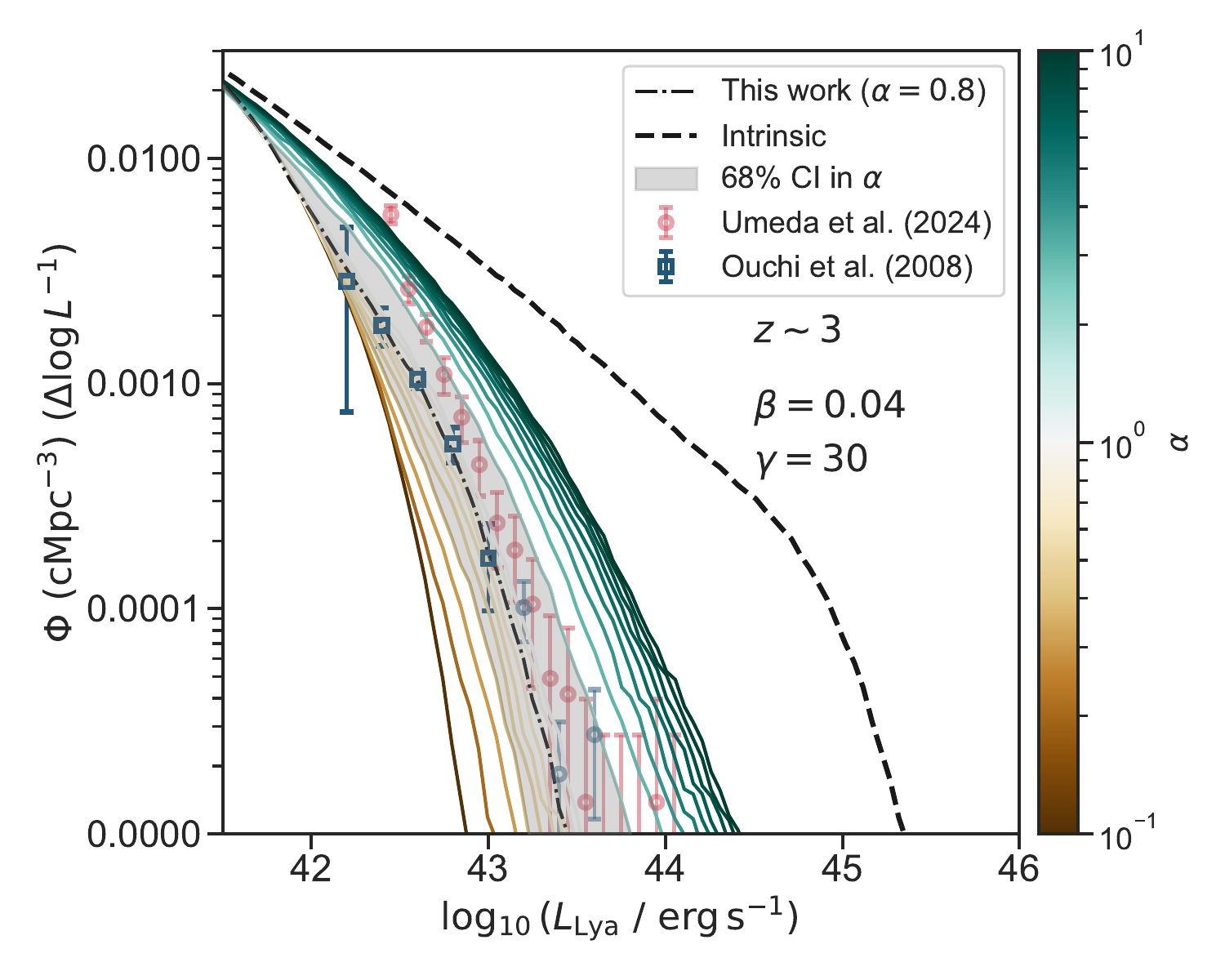}
\includegraphics[scale=0.23]{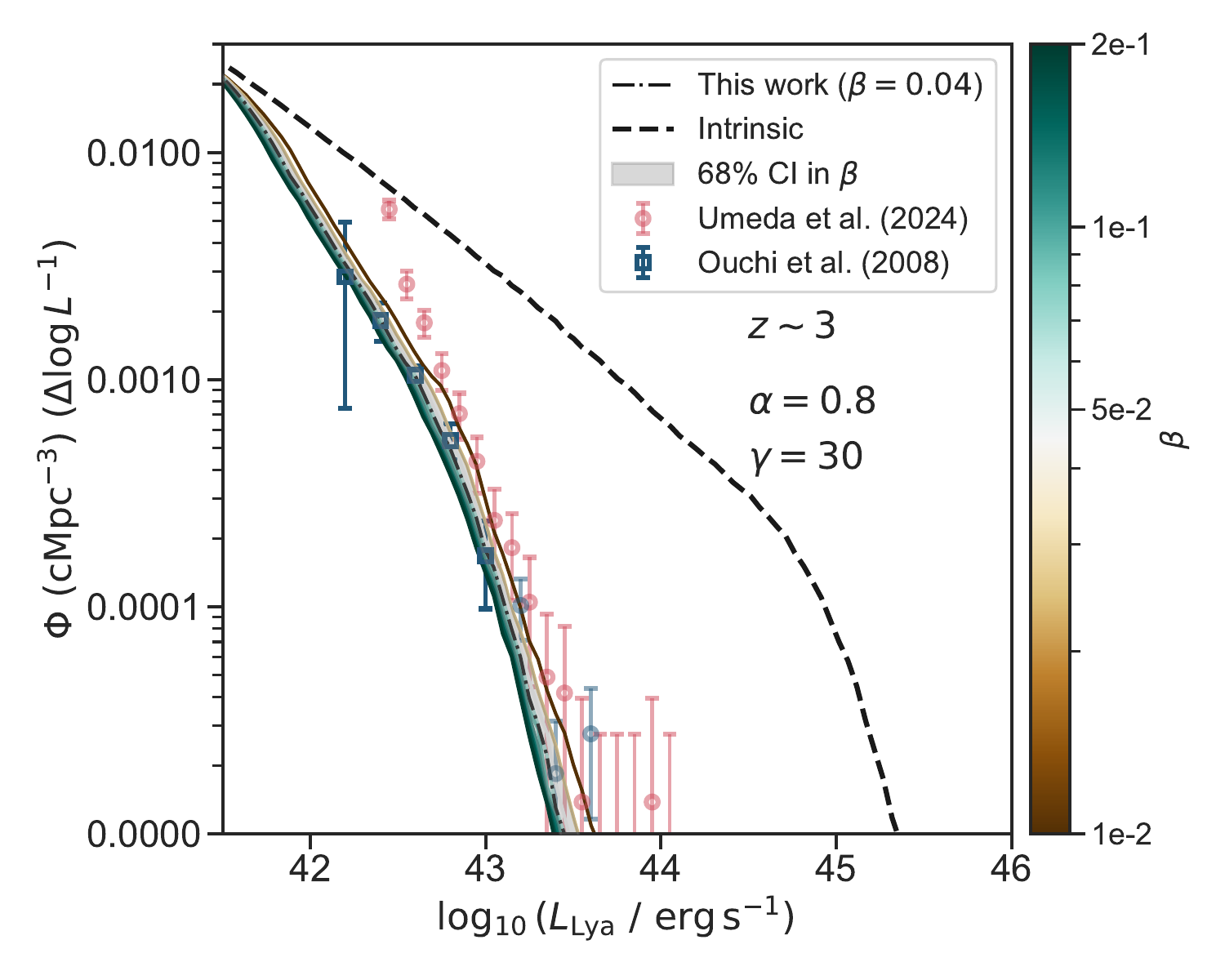}
\includegraphics[scale=0.23]{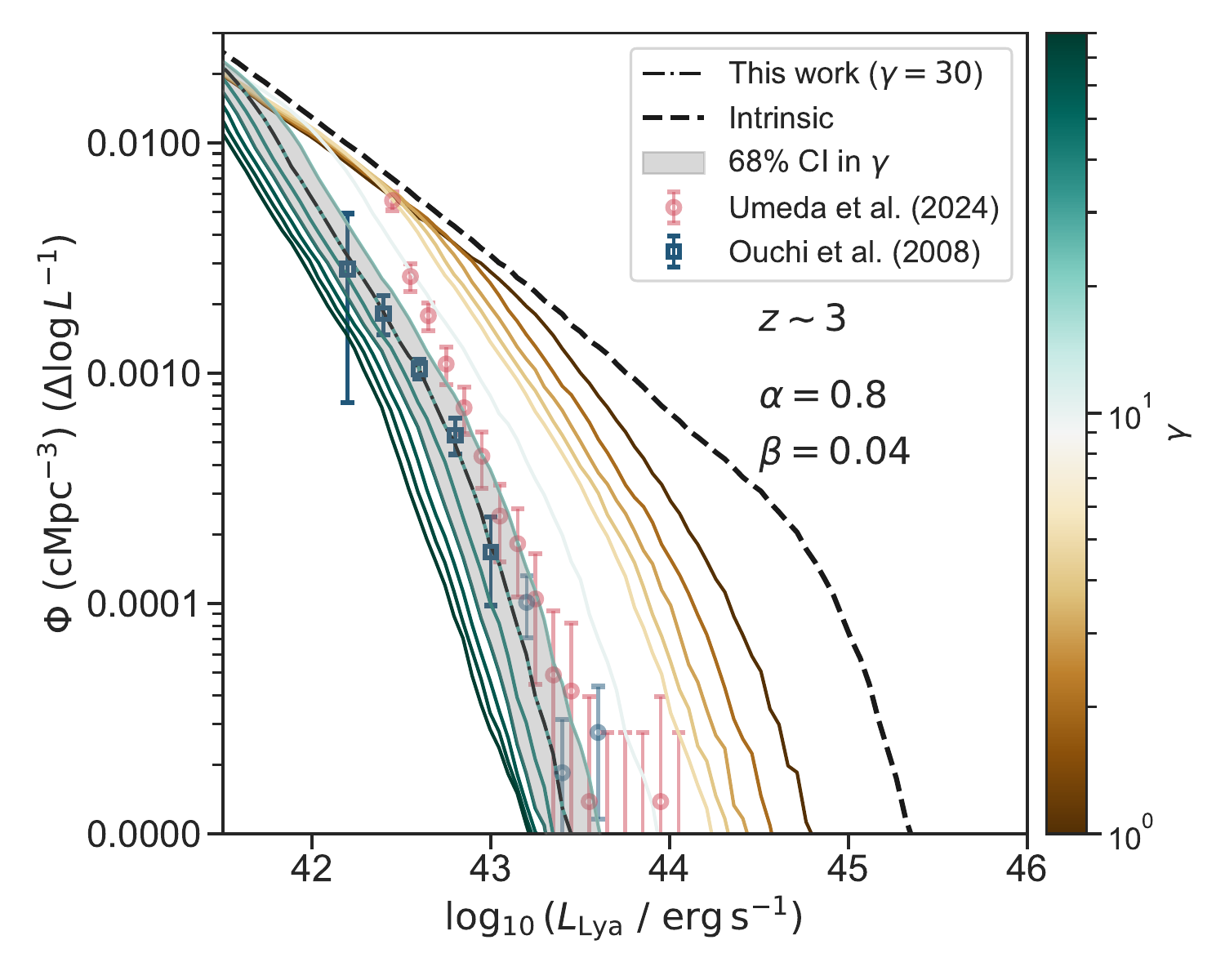}
\includegraphics[scale=0.23]{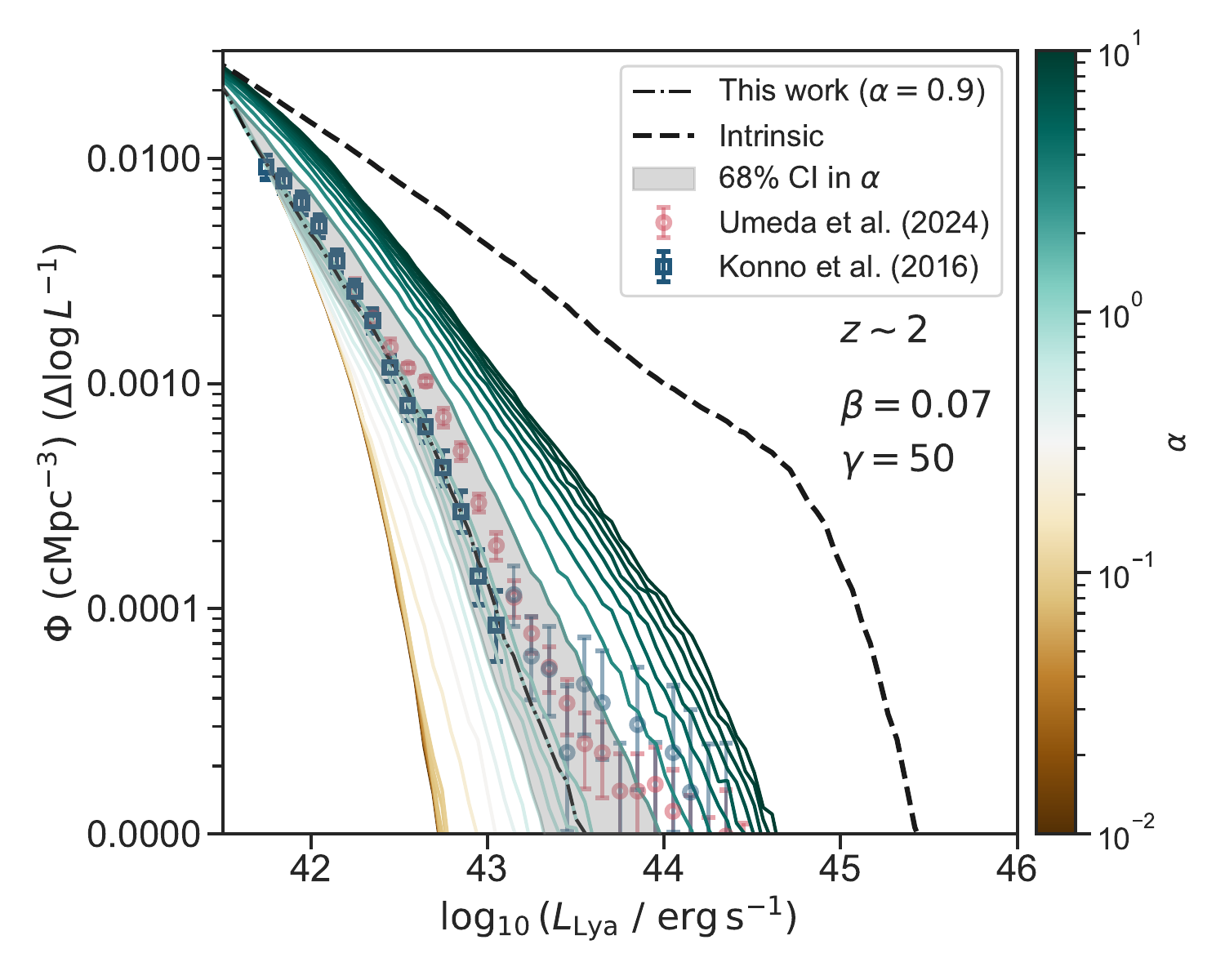}
\includegraphics[scale=0.23]{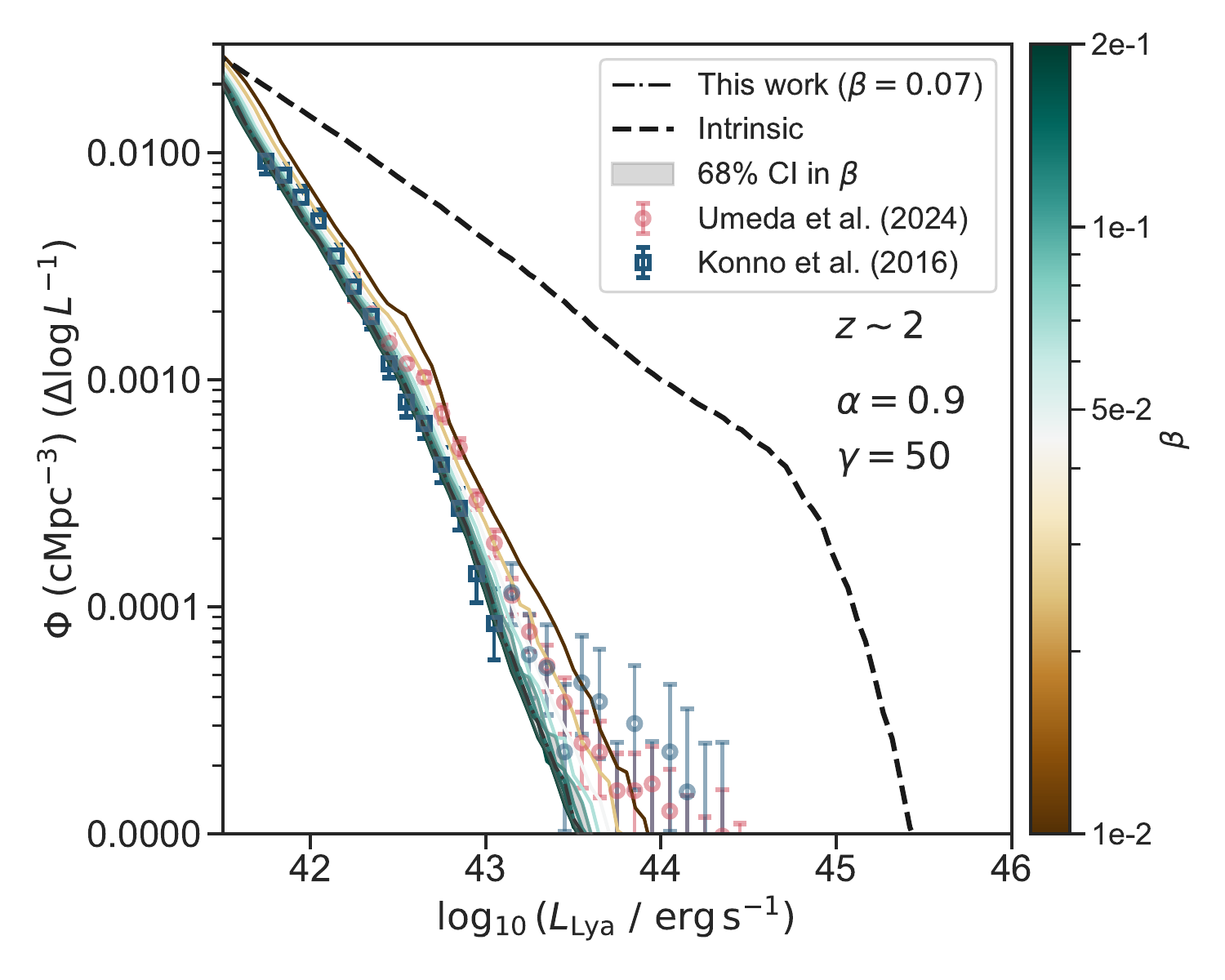}
\includegraphics[scale=0.23]{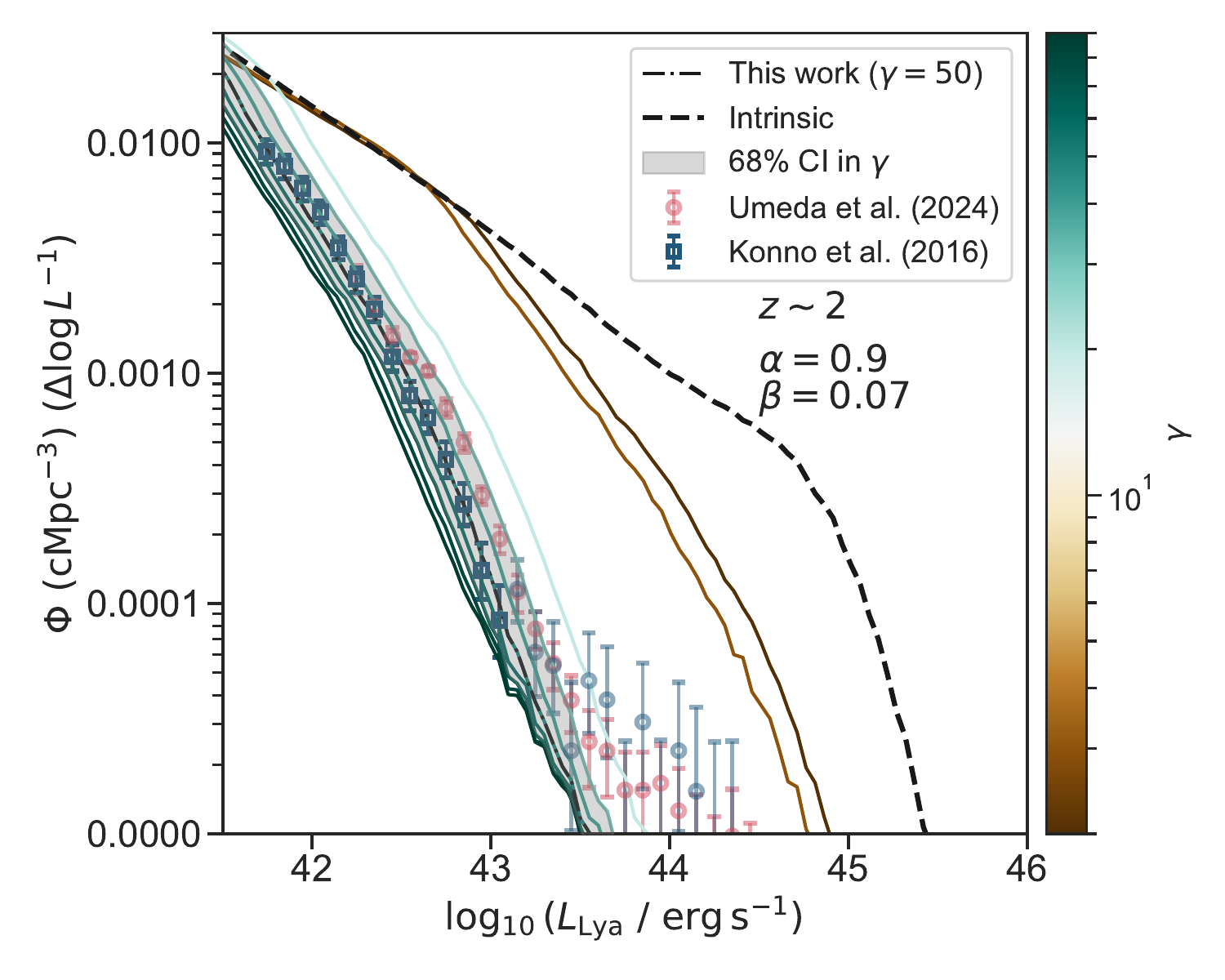}
\caption{Sensitivity of the Ly$\alpha$ luminosity function to variations in the model parameters, $\alpha$ (outflow velocity scaling), $\beta$ (HI column density scaling), and $\gamma$ (dust attenuation scaling). Each column shows the effect of varying one parameter (with values indicated in the legend), while fixing the others to their best-fit values derived in Sec.~\ref{sec:fit}. Upper panels show the impact of the free parameters on LF at $z\sim3$, while lower panels present the impacts at $z\sim2$. This analysis illustrates how increasing $\alpha$ primarily boosts the bright end as stronger outflows enhance Ly$\alpha$ escape in high-SFR galaxies, higher $\beta$ uniformly suppresses Ly$\alpha$ escape across all luminosities and increasing $\gamma$ causing a near parallel downward shift in normalization. Observational data from \cite{Konno_2016} at $z\sim2$ and \cite{Ouchi_2008} at $z\sim3$ are over-plotted, showing the consistency of the best-fit model with constraints across redshifts.}
\label{fig:app_LF_evol}
\end{figure}

Fig~\ref{fig:app_LF_evol} demonstrates how each parameter leaves its unique imprint on the Ly$\alpha$ luminosity function. The parameter $\alpha$, controlling the expansion velocity of gas shells, primarily affect the bright end of the luminosity function. As shown in the left-most panels of Fig~\ref{fig:app_LF_evol}, increasing $\alpha$ significantly boosts the number density of the most luminous LAEs. This is physically intuitive as galaxies with stronger outflows (higher $\alpha$), allow more Ly$\alpha$ photons to escape because the Doppler shift from the expanding shell reduces the number of resonant scatterings. At the bright end, where the galaxies typically have higher star formation rates and thus more intrinsic Ly$\alpha$ emission, even small changes in $\alpha$, lead to noticeable differences in the number of detectable LAEs. When we increase $\alpha$, we are essentially opening escape routes for photons in the most luminous galaxies, causing the bright end of the luminosity function to rise. In contrast, the faint end of the luminosity function is dominated by low-mass galaxies with smaller $V_{\rm exp}$. These galaxies are less affected by changes in $\alpha$ because their outflows are weaker (due to the lower sSFR or smaller sizes) so their escape fractions remain low. As $\alpha$ increases, the bright end rises significantly while the faint end stays relatively unchanged, resulting in a flatter overall slope of the luminosity function.\\

The parameter $\beta$, has a broad impact across all luminosities, uniformly suppressing the number of LAEs as it increases. This parameter scales the column density of neutral hydrogen ($N_{\rm HI}$), which determines how efficiently the ISM/CGM can trap Ly$\alpha$ photons through resonant scattering. Higher $\beta$ values correspond to denser neutral hydrogen gas, increasing the likelihood of resonant scattering and trapping Ly$\alpha$ photons before they escape. As seen in the middle panels of Fig~\ref{fig:app_LF_evol}, increasing $\beta$ shifts the entire luminosity function downward. Unlike $\alpha$, which selectively affects LAEs with strong outflows (due to the impact from SFR), $\beta$'s impact is more uniform because HI mass and size vary less dramatically with luminosity rather than SFR. However a subtle mass dependent may arise. Low-mass LAEs tend to have higher $M_{\rm HI}/M_*$ ratios (Fig.~\ref{fig:scaling_rels}, top right), so their $N_{\rm HI}$, and thus $\beta$'s suppression could be slightly more pronounced at the faint end.\\

The parameter $\gamma$, scales the dust optical depth ($\tau_a$), which depends on the product of the cold gas phase metallicity and the HI column density (see Eq.~\ref{eq:tau_a}). Since dust absorption destroys Ly$\alpha$ photons irrespective of the LAE's intrinsic luminosity or SFR, increasing $\gamma$ suppresses the Ly$\alpha$ escape fraction across all luminosities. As shown in the right-most panels of Fig~\ref{fig:app_LF_evol}, increasing $\gamma$, results in a nearly parallel downward shift of the entire luminosity function, changing its overall normalization rather than its shape. This uniform suppression occurs because $\tau_a$ is tied to dust-to-gas ratio (via $Z_{\rm cold}$) and $N_{\rm HI}$, both of which vary less steeply with galaxy mass than outflow velocities or SFR. However subtle deviations may occur. High-mass LAEs tend to have higher metallicities (Fig.~\ref{fig:scaling_rels}, bottom panel), so their dust attenuation (and thus $\gamma$'s impact) could be slightly more pronounced at bright end.\\

\begin{figure}
\centering
\includegraphics[scale=0.3]{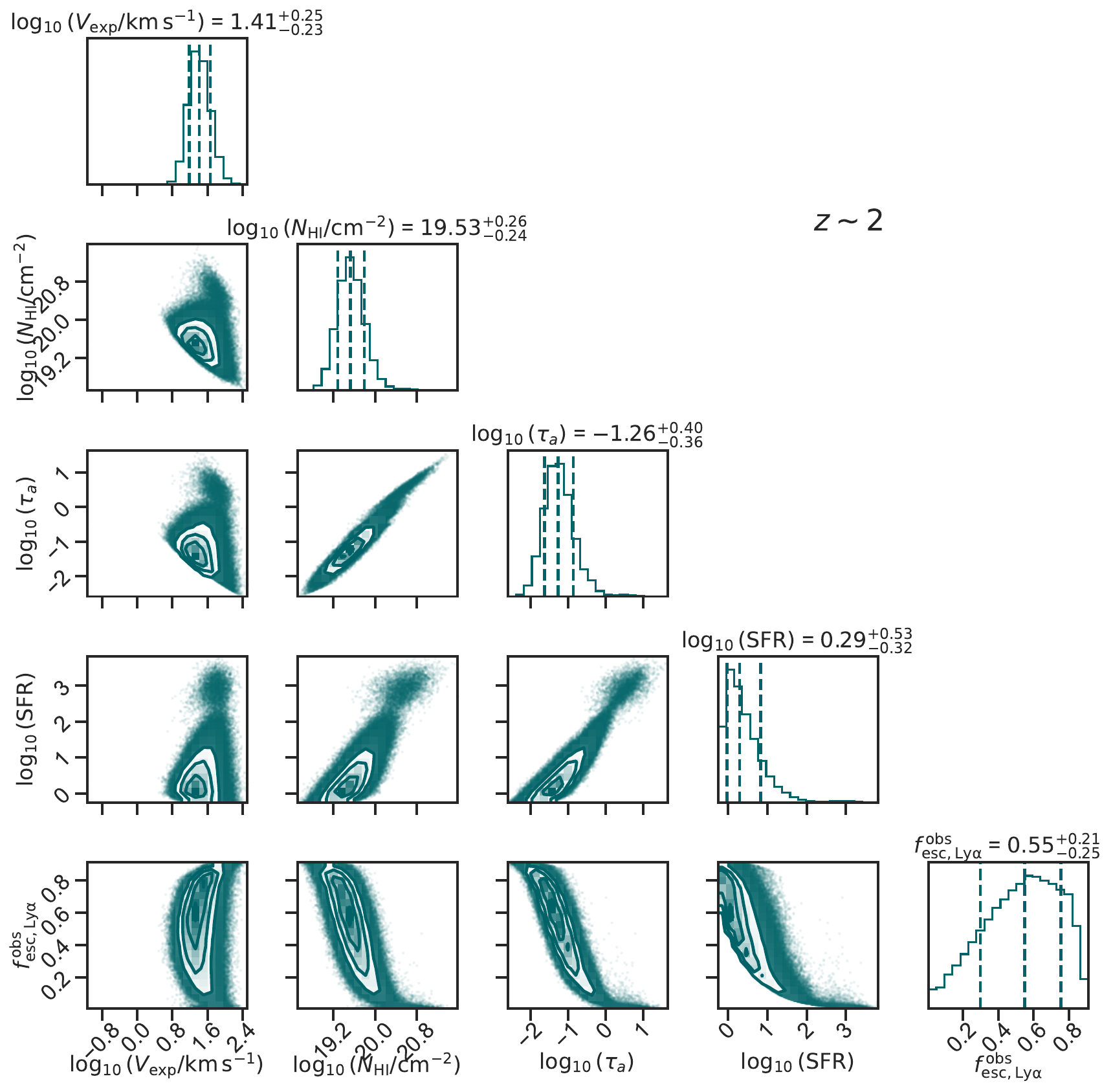}
\includegraphics[scale=0.3]{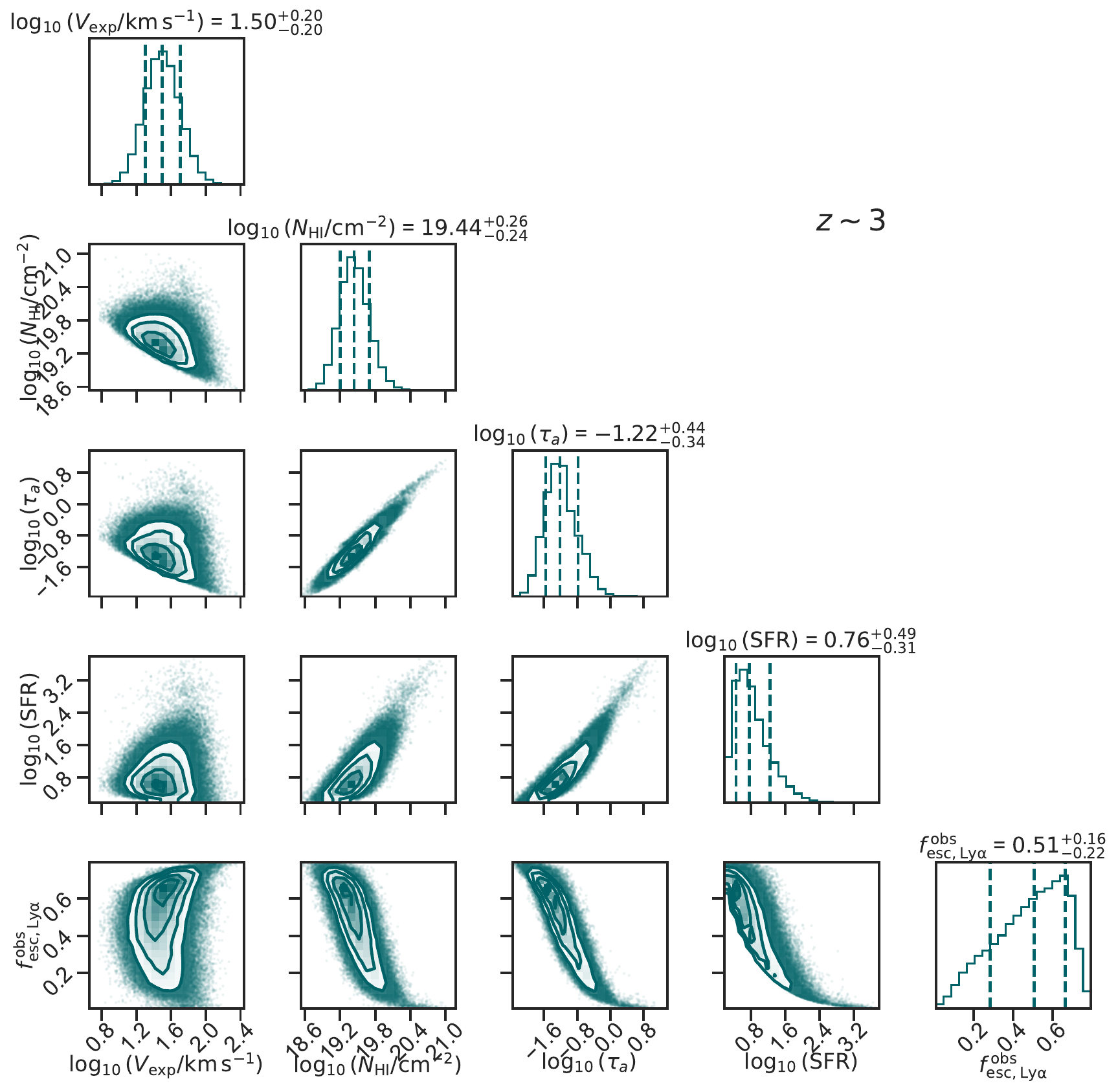}
\caption{Corner plot showing the joint distribution and the correlation between key radiative transfer parameters at $z\sim2$ (\textit{left}) and $z\sim3$ (\textit{right}). Each 1D histogram along the diagonal shows the distribution for one parameter, expansion velocity, $\rm log_{10}(v_{\rm exp}/km\,s^{-1})$, neutral hydrogen column density, $\rm log_{10}(N_{HI}/cm^{-1})$, dust absorption optical depth, $\rm log_{10}\,\tau_a$, star formation rate, $\rm log_{10}\, SFR$ and the resulting observed Ly$\alpha$ escape fraction, $f^{\,\rm obs}_{\rm esc,Ly\alpha}$ with vertical dashed lines marking the median and $68\%$ confidence intervals of the distribution. Off-diagonal panels show the joint distribution 2D credible regions at $68\%$, $95.4\%$ and $99.7\%$ levels illustrating the correlations between each pair of parameters.}
\label{fig:app-tiangle}
\end{figure}

Our model's effectiveness stems from the distinct yet interconnected roles of $\alpha$, $\beta$ and $\gamma$ in shaping Ly$\alpha$ escape fraction. Fig.~\ref{fig:app-tiangle} illustrates these relationships through joint distributions of key physical properties. The $V_{\rm exp}-f^{\,\rm obs}_{\rm esc,Ly\alpha}$ panel shows a clear trend: galaxies require sufficiently high outflow velocities (high $\alpha$) to achieve large escape fractions, but even strong outflows can not guarantee Ly$\alpha$ escape if HI column densities ($N_{\rm HI}$) or dust optical depths ($\tau_a$) are too high. This explains why $\alpha$ primarily influences the bright end of the luminosity function. Only galaxies with both high SFRs (generating abundance Ly$\alpha$ photons) and powerful outflows (enabling their escape) dominate this regime. The $N_{\rm HI}-f^{\,\rm obs}_{\rm esc,Ly\alpha}$ panel shows $\beta$'s broad impact. Higher $\beta$ increases $N_{\rm HI}$, enhancing resonance trapping and reducing escape across all luminosities. Low-mass galaxies with their higher $M_{\rm HI}/M_*$ ratios (Fig.~\ref{fig:scaling_rels}), experience slightly stronger suppression. Meanwhile, $\gamma$'s dust attenuation scale ($\tau_a \propto Z_{\rm cold}\, N_{\rm HI}$) uniformly lowers $f^{\,\rm obs}_{\rm esc,Ly\alpha}$, blocking photons independent of galaxy luminosity. This results in a near parallel shift of the luminosity function.\\

Understanding these distinct parameter effects is crucial for our calibration process. The sensitivity of the  bright-end slope of the luminosity function to $\alpha$, combined with the broad suppression in normalization from $\beta$, and the overall normalization controlled by $\gamma$ allows us to constrain each parameter independently. This multi-dimensional fitting approach ensures our model captures the complex physics governing Ly$\alpha$ escape while remaining constraint by observations. By isolating these effects (Figs.~\ref{fig:app_LF_evol} and \ref{fig:app-tiangle}) our model captures the essential physics governing Ly$\alpha$ visibility while remaining empirically grounded.\\

\begin{figure}
\centering
\includegraphics[scale=0.23]{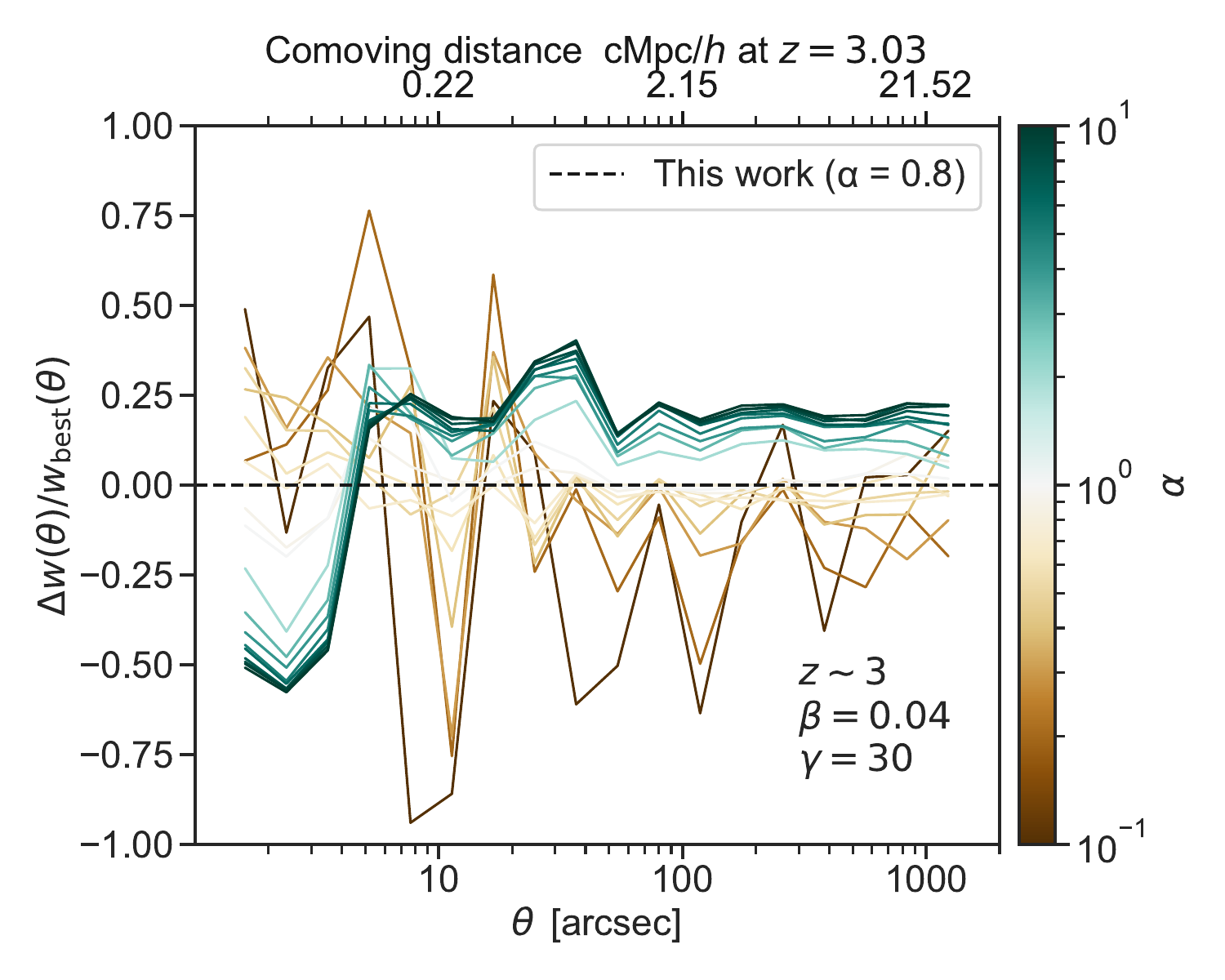}
\includegraphics[scale=0.23]{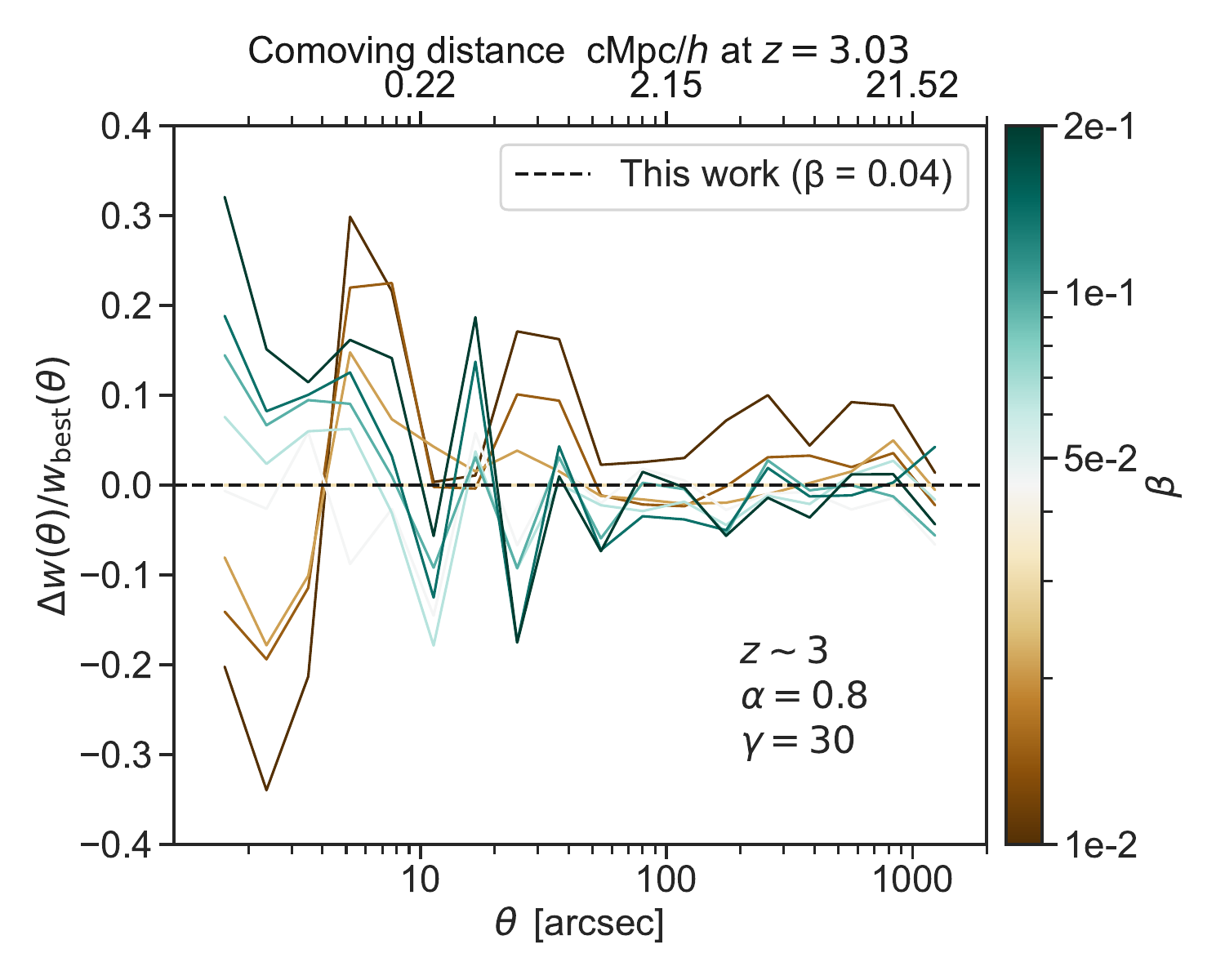}
\includegraphics[scale=0.23]{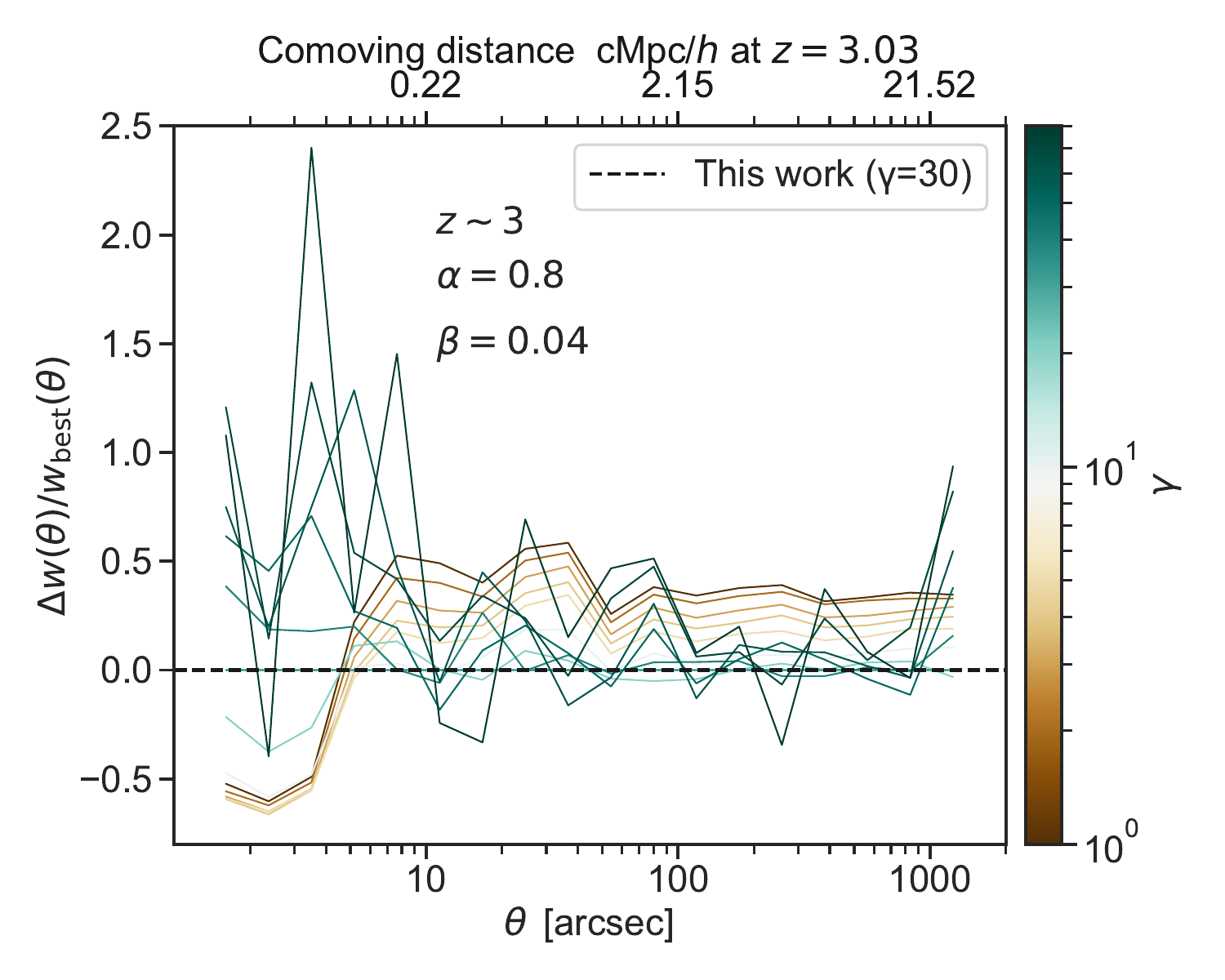}
\includegraphics[scale=0.23]{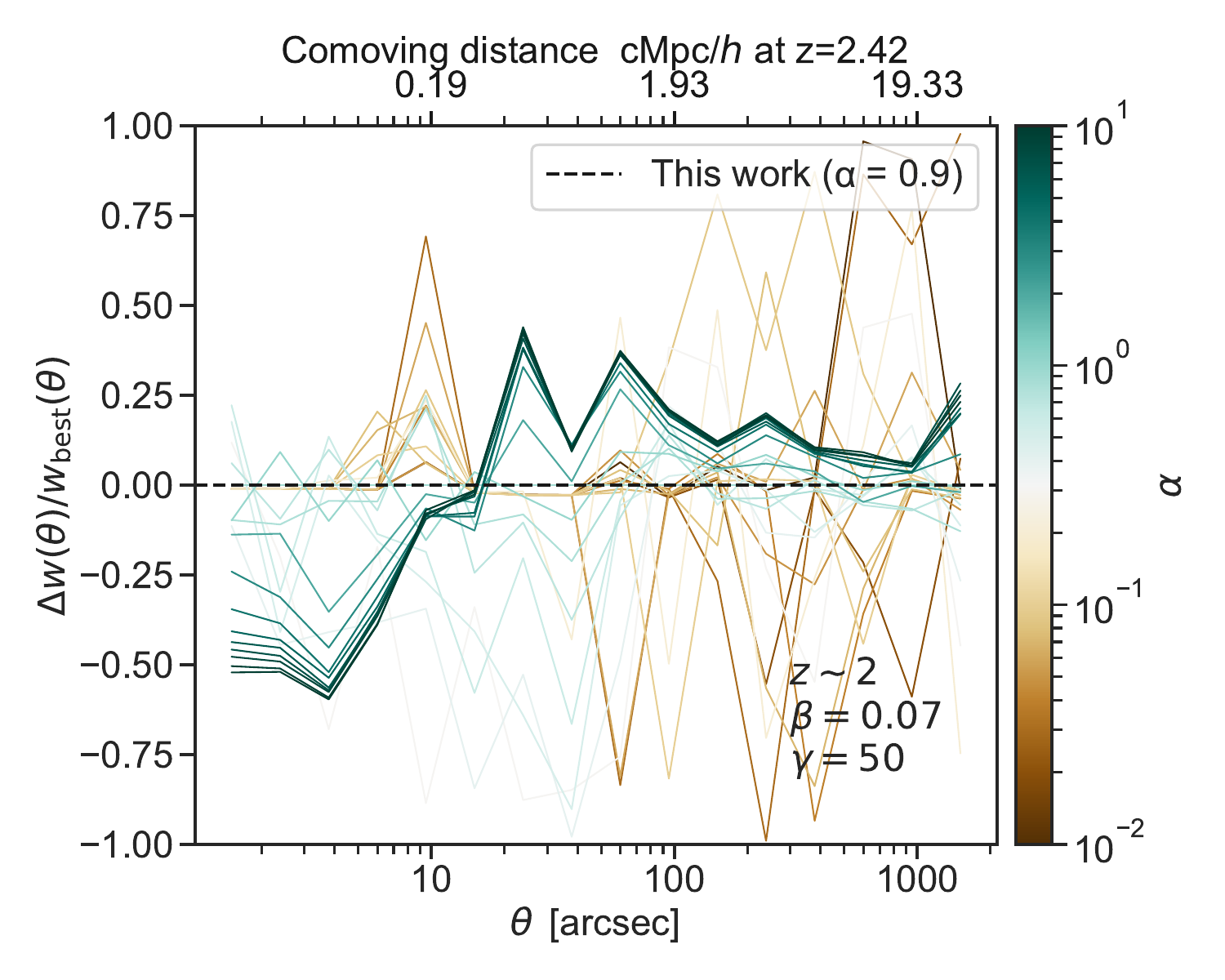}
\includegraphics[scale=0.23]{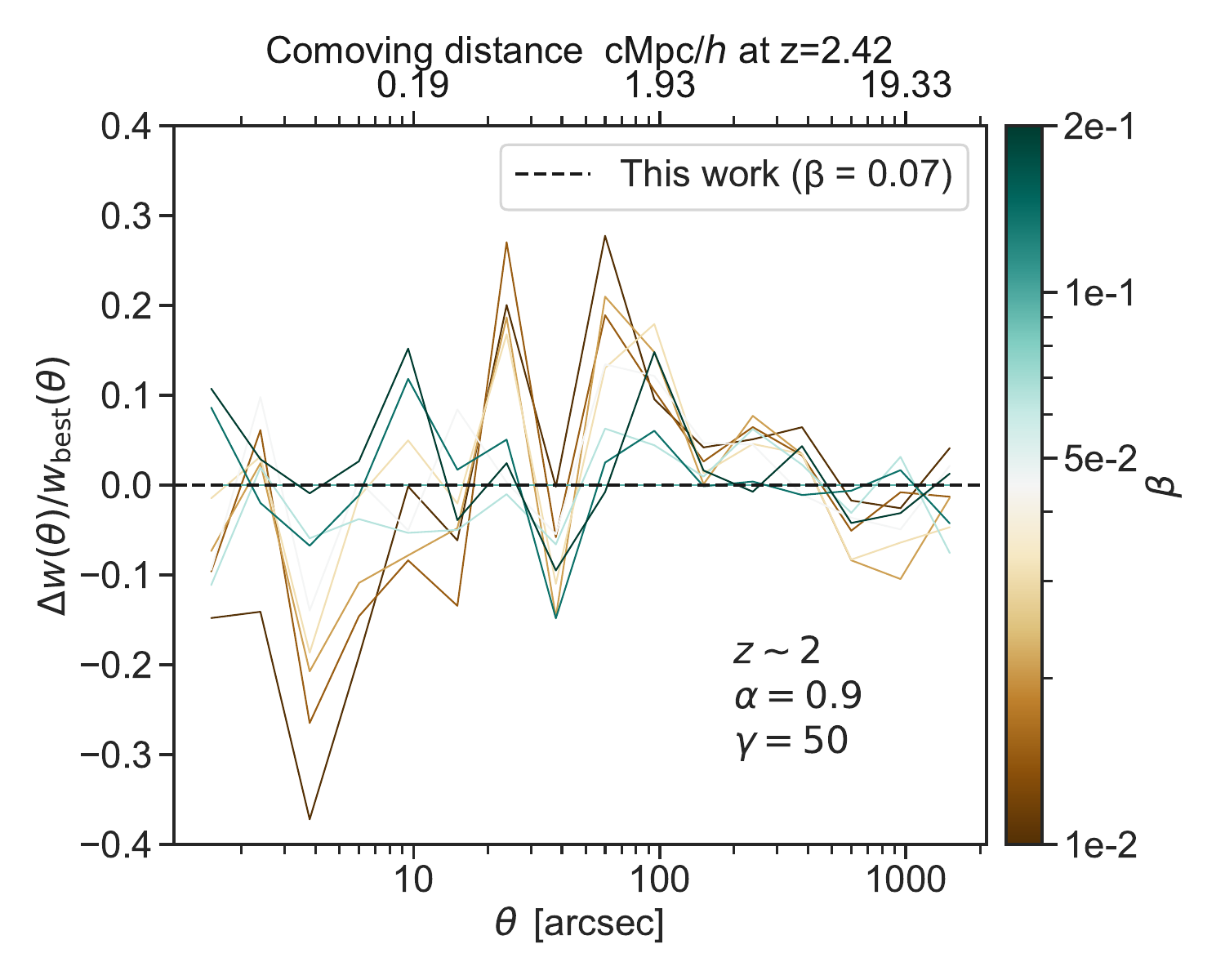}
\includegraphics[scale=0.23]{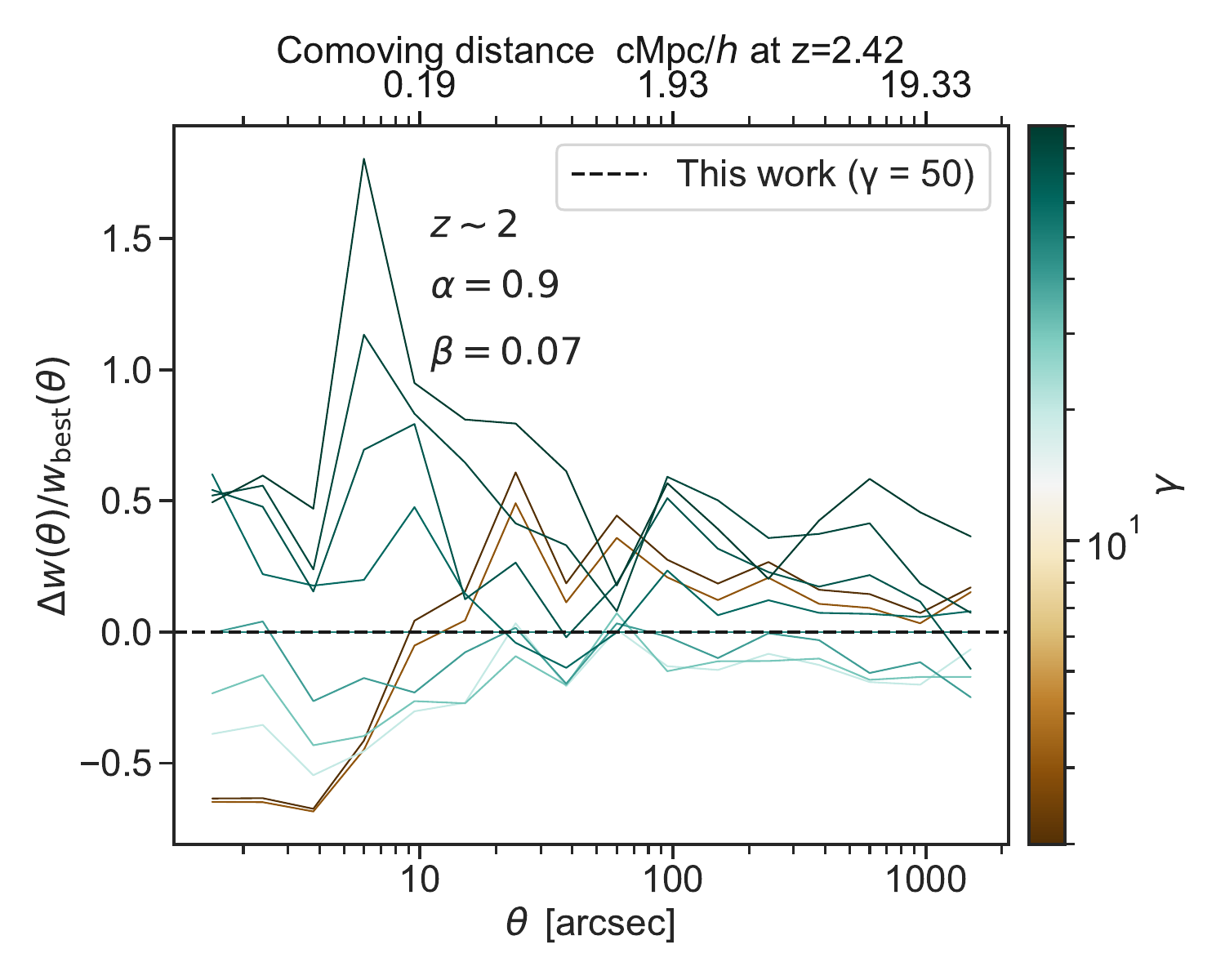}
\caption{Scale dependent response of the LAE angular correlation function, $w(\theta)$, to variation in radiative transfer parameters. \textit{Left:} Outflow velocity scaling, $\alpha$, enhances large-scale clustering while slightly suppressing small scales. \textit{Middle and right:} Both HI column density scaling, $\beta$ and dust attenuation scaling $\gamma$, show overall similar trends: increasing small-scale clustering, while weakening large scale correlations.}
\label{fig:app_CF_evol}
\end{figure}

\subsection{D.2. Impact of free parameters on the angular correlation function}
\label{sec:app-CF-abc}

The angular correlation function, $w(\theta)$, encodes how LAEs cluster across different scales. This clustering is sensitive to the underlying halo occupation of LAEs, which in turn depends on the radiative transfer processes governed by our three free parameters $\alpha$ (outflow velocity scaling), $\beta$ (neutral hydrogen column density scaling), and $\gamma$ (dust attenuation scaling). To understand their individual effects, we systematically vary each parameter, while holding the others fixed at their best-fit values and analyze the resulting change in $w(\theta)$.\\

The expansion velocity of gas shells, scaled by $\alpha$, determines how efficiently Ly$\alpha$ photons escape resonant scattering. Higher $\alpha$ values leads to stronger outflows which shift photons to the wings of the line profile, reducing interactions with neutral hydrogen. This enhances the detectability of central galaxies in massive halos, where star formation rates are high. As shown in Fig.~\ref{fig:app_CF_evol} (left-most panels), increasing $\alpha$ produces two opposite trends. At large scales ($\theta \gtrsim 100^{''}$), the clustering amplitude rises because more central galaxies in massive halos become detectable, amplifying the two-halo term. These systems are highly biased and dominate pair-counts at large separations. At small scales ($\theta \lesssim 10^{''}$), the signal weakens slightly. While outflows aid escape, satellite galaxies, even with boosted $f^{\,\rm obs}_{\rm esc,Ly\alpha}$, often remain below the detection threshold due to their lower intrinsic luminosities. This reduces the contribution of satellite-satellite and central-satellite pairs to the 1-halo term, in other words, it reduces the satellite fraction. This contrast highlights a key feature of LAE clustering. Bright centrals drive the large-scale signal, while the faint satellite population shapes small-scale correlations.\\

The parameter $\beta$ controls the neutral hydrogen column density, $N_{\rm HI}$, which sets the optical depth for Ly$\alpha$ scattering. Higher $\beta$ values increases $N_{\rm HI}$, suppressing Ly$\alpha$ escape across all galaxies but with important scale-dependent effect on clustering. Fig.~\ref{fig:app_CF_evol} (middle panels), illustrate this impact. At small scales ($\theta \lesssim 10^{''}$), higher $\beta$ removes LAEs from low-mass halos (where gas fractions are typically higher, see Fig.~\ref{fig:scaling_rels}). The surviving population becomes dominated by satellites in massive halos, where deeper gravitational potential well maintain dense gas reservoirs. This boosts the 1-halo term, as these satellites cluster tightly around their centrals. Conversely, at large scales ($\theta \gtrsim 100^{''}$), higher $\beta$ reduces the overall LAE number density by suppressing escape in lower-mass centrals (which have higher gas fraction and lower intrinsic luminosities). This weakens the 2-halo term, as fewer biased central galaxies contribute to large-scale pair counts.\\

The dust optical depth scaling parameter $\gamma$ (where $\tau_a \propto \gamma \, Z_{\rm cold}\, N_{\rm HI}$; Eq.~\ref{eq:tau_a}), exhibits the same scale dependent trend as $\beta$. At small scales  ($\theta \lesssim 10^{''}$), higher $\gamma$ strengthens clustering because dust preferentially extinguishes LAEs in low-mass halos (where galaxies have lower metallicities; Fig.~\ref{fig:scaling_rels}, but insufficient SFR to compensate for increased attenuation). Meanwhile, satellites in massive systems survive due to deeper gravitational potentials that retain both high dust and abundant satellite populations. This creates a 
surviving sample dominates by clustered satellites, boosting the 1-halo term. Conversely, at large scales ($\theta \gtrsim 100^{''}$), higher $\gamma$ reduces clustering amplitude because the suppression of low-mass centrals (with high gas fraction but low luminosity) decreases the overall LAE number density. While the most massive centrals remain detectable, their scarcity weakens the 2-halo term, as fewer bias tracers contribute to large-scale pairs (see Fig.~\ref{fig:app_CF_evol} right-most panels).\\

The scale-dependent signatures of $\alpha$, $\beta$ and $\gamma$ in the angular correlation function show how different physical processes control LAE visibility across cosmic environments. While outflows ($\alpha$) enhance central galaxy detection, both HI scattering ($\beta$) and dust attenuation ($\gamma$) exhibit similar trends (increasing small-scale clustering through satellite preservation in massive halos, while weakening large-scale correlations through number density suppression). Although our fiducial model (calibrated to \cite{White:2024JC}) primarily constrain the 2-halo regime, lacking strong small-scale (1-halo term) constraints. However, our tests in this appendix shows the following overall trend: increasing the Ly$\alpha$ escape fraction, whether via stronger outflows (higher $\alpha$), reduced HI coulmn density (lower $\beta$) or weaker dust (lower $\gamma$), systematically decreases the satellite fraction. This occurs because higher escape fractions preferentially increases the detectability of luminous central galaxies, while fainter satellites remain more likely to fall below the luminosity threshold. Our results shows the complex behavior between galaxy formation physics and observable LAE populations. It highlights how clustering measurements when combined with luminosity function, offer constraints on the gas and dust environments shaping Ly$\alpha$ emission across cosmic time.

\bibliographystyle{aasjournal}
\bibliography{bibliography}

\end{document}